\documentclass[12pt]{article}
\usepackage[dvips]{graphics}
\begin{document}

\vspace{2cm}

\begin{center}

\Large{Theory of the Elementary Particles}

\vspace{0.75cm}

\large{E.L. Koschmieder}

\vspace{0.5cm}

\normalsize

{Center for Statistical Mechanics\\
The University of Texas at Austin, Austin TX 78712, USA\\
e-mail: koschmieder@utexas.edu}       

\end{center}

\bigskip

\noindent
{We show that the measured rest masses of the stable mesons and baryons 
are, in a very good approximation, integer multiples of the mass of the 
$\pi^0$ or $\pi^\pm$ mesons. The integer multiple rule is a summary of
experimental facts.
We use lattice theory in order to determine the rest masses of the 
stable mesons and baryons and their spin. The masses of the particles
so determined agree, within percents, with the measured masses
of the particles, following the integer multiple rule. And with the 
same concept we determine the masses of the leptons\,: the muon, 
the electron and the masses of the electron neutrino, the muon neutrino
and the tau neutrino. It turns out that the mass of the electron neutrino 
is equal to the mass of the muon neutrino times the fine structure 
constant. Only photons, neutrinos, charge and the weak nuclear force
are needed to explain the masses of the elementary particles.}

\bigskip

\hspace{2.5cm}Contents

\smallskip

\noindent
Introduction\\
1 The Spectrum of the Particle Masses.\\
2 Standing Waves in a Cubic Lattice.\\
3 The Masses of the $\gamma$-branch Particles.\\
4 The Rest Mass of the $\pi^0$ Meson.\\
5 The neutrino-branch Particles.\\
6 The Masses of the $\nu$-branch Particles.\\
7 The Weak Force.\\
8 The Strong Force.\\
9 The Rest Mass of the Muon.\\
10 The Neutrino Masses.\\
11 Neutrinos in the Electron.\\
12 The Magnetic Moment of the Electron.\\
13 The Ratios m($\mu^\pm$)/m(e$^\pm$) and m($\pi^\pm$)/m(e$^\pm$).\\
14 The Spin of the $\gamma$-branch Particles.\\
15 The Spin of the $\nu$-branch Particles.\\
 Conclusions. - Acknowledgements - References - Appendices

\section*{Introduction}

The rest masses of the elementary particles, the so-called ``stable" mesons, 
baryons and leptons,\\

\hspace{1cm} the $\pi^0$,\,$\eta$,\,$\Lambda$,\,$\Sigma^{\pm,0}$,\,$\Xi^{-,0}$,
\,$\Omega^-,$ \,$\Lambda_c^+$,\,$\Sigma_c^{+,0}$,\,$\Xi_c^{+,0}$, and $\Omega_c^0$, 
\hspace{1cm}\\

\hspace{2cm} and the $\pi^\pm,$\,K$^{\pm,0}$, p, n, D$^{\pm,0}$, and D$_s^\pm$,\\

\hspace{2cm} as well as the e$^\pm$, $\mu^\pm$, and $\tau^\pm$ particles\\

\noindent
have been measured with great accuracy, usually to the sixth
decimal or better, but have not been explained so far. That means 
that neither the mass of the fundamental proton nor the mass of the 
fundamental electron have been explained. The quarks, which have been 
introduced by Gell-Mann [1] more than fifty years ago, are said  
to explain the elementary particles. But the standard model 
does not explain neither the mass, nor the charge, nor the spin 
of the mesons, baryons and leptons. Mass, charge and spin are 
the fundamental properties of the particles. The measured values 
of the properties of the particles are in the Review of 
Particle Physics [2]. There are many other attempts to explain the 
elementary particles or only one of the particles, too many 
to list them here. For example Skyrme [3] has proposed a 
unified theory of the mesons and baryons, and El Naschie has 
proposed a topological theory for high 
energy particles and the spectrum of the quarks [4]\,-\,[7].

   The need for the present investigation has been expressed by  
Feynman [8] as follows: ``There remains one especially unsatisfactory 
feature: the observed masses of the particles, m. There is no theory 
that adequately explains these numbers. We use the numbers in all 
our theories, but we do not understand them - what they are, or where 
they come from. I believe that from a fundamental point of view, this is 
a very interesting and serious problem". Today, thirty
years later, we still cannot explain the masses, the charge and the 
spin of the particles. It is time to try something different.   

\section {The spectrum of the particle masses}

As we have done before [9] we will focus attention on the so-called 
``stable" mesons and baryons 
whose masses are reproduced with other data in Tables 1 and 2.
It is obvious that any attempt to explain the masses of the mesons and 
baryons should begin with the particles that are affected by the fewest 
parameters. These are certainly the particles without isospin (I = 0) and 
without spin (J = 0), but also with strangeness S = 0, and charm C = 0. 
Looking at the particles with I,J,S,C = 0 it is startling to find that their
masses are quite close to integer multiples of the mass of the 
$\pi^0$\,meson. According to the Review of Particle Physics it is 
m($\eta$) = (1.0140 $\pm$ 0.0003)\,$\cdot$\,4m($\pi^0$), and the mass of 
the resonance $\eta^{\,\prime}$ is m($\eta^{\,\prime}$) = (1.0137 
$\pm$ 0.00015)\,$\cdot$\,7m($\pi^0$). Three particles seem hardly to be 
sufficient to establish a rule. However, if we look a little further we 
find that m($\Lambda$) = 1.0332\,$\cdot$\,8m($\pi^0$) or m($\Lambda$) = 
1.0190\,$\cdot$\,2m($\eta$). We note that the $\Lambda$ baryon has spin 
1/2, not spin 0 as the $\pi^0$ and $\eta$ mesons. Nevertheless, the mass of 
$\Lambda$ is close to 8m($\pi^0$). Furthermore we have m($\Sigma^0$) = 
0.9817\,$\cdot$\,9m($\pi^0$), m($\Xi^0$) = 0.9742\,$\cdot$\,10m($\pi^0$), 
m$(\Omega^-)$ = 1.0325\,$\cdot$\,12m($\pi^0)$ = 
1.0183\,$\cdot$\,3m($\eta$), ($\Omega^-$ is charged 
and has spin 3/2). Finally the masses of the charmed baryons 
are m($\Lambda_c^+$) = 0.99645\,$\cdot$\,17m($\pi^0$) = 
1.024\,$\cdot$\,2m($\Lambda$), m($\Sigma_c^0$) = 
1.00995\,$\cdot$\,18m($\pi^0$) = 1.0287\,$\cdot$\,2m($\Sigma^0$), 
m($\Xi_c^0$) = 1.0170\,$\cdot$\,18m($\pi^0$), and m($\Omega_c^0$) = 
0.99925\,$\cdot$\,20m($\pi^0$) = 1.0249\,$\cdot$\,2m($\Xi^0$).
 
    \begin{table}\caption{The ratios m/m($\pi^0$) of the particles of the 
$\gamma$-branch.} 
    \begin{tabular}{lllllcl}\\
\hline\hline\\
 & m/m($\pi^0$) & multiples & decays & fraction & spin & 
mode
 \\
 & & & & (\%) & &\\
[0.5ex]\hline
\\
$\pi^0$ & 1.0000 & 1.0000\,\,$\cdot$\,\,$\pi^0$ & $\gamma\gamma$ & 98.823 
& 0 & (1.)\\
 & & & e$^+$e$^-\gamma$ & \,\,\,1.174 & &\\
 & & & e$^+$e$^-$e$^+$e$^-$ & \,\,\,3.34$\cdot$10$^{-5}$ & &\\
\\
$\eta$ & 4.0563 & 1.0141\,$\cdot$\,\,4$\pi^0$ & $\gamma\gamma$ & 39.31 & 0 
& 
(2.)\\
 & & & 3$\pi^0$ & 32.57 & &\\
 & & & $\pi^+\pi^-\pi^0$ & 22.74 & &\\
 & & & $\pi^+\pi^-\gamma$ & \,\,\,4.60 & &\\
\\
$\Lambda$ & 8.26575 & 1.0332\,$\cdot$\,\,8$\pi^0$ & p$\pi^-$ & 63.9 & 
$\frac{1}{2}$ 
& 2$\ast$(2.)\\
 & & 1.0190\,$\cdot$\,\,2$\eta$ & n$\pi^0$ & 35.8 & &\\
&&& n$\gamma$ & 1.75\,$\cdot$\,10$^{-3}$ & &\\
\\
$\Sigma^0$ & 8.8359 & 0.9817\,$\cdot$\,\,9$\pi^0$ & $\Lambda \gamma$ & 100 
& 
$\frac{1}{2}$ & 2$\ast (2.) + (1.)$\\
\\
$\Xi^0$ & 9.7412 & 0.9741\,$\cdot$\,10$\pi^0$ & $\Lambda\pi^0$ & 99.52 & 
$\frac{1}{2}$ & 
2$\ast(2.) + 2(1.)$\\
\\
$\Omega^-$ & 12.390 & 1.0326\,$\cdot$\,12$\pi^0$ & $\Lambda$K$^-$ & 67.8 
& 
$\frac{3}{2}$ & 3$\ast(2.)$\\
 & & 1.0183\,$\cdot$\,\,3$\eta$ & $\Xi^0\pi^-$ & 23.6 & &\\
 & & &  $\Xi^-\pi^0$ & \,\,\,8.6 & &\\
\\
$\Lambda_c^+$ & 16.939 & 0.99645\,$\cdot$\,17$\pi^0$ & many & & 
$\frac{1}{2}$ & 
2$\ast(2.) + (3.)$\\
 & & 0.9630\,$\cdot$\,17$\pi^\pm$\\
\\
$\Sigma_c^0$ & 18.179 & 1.0099\,$\cdot$\,18$\pi^0$ & $\Lambda_c^+\pi^-$ & 
$\approx$100 & $\frac{1}{2}$ & $\Lambda_c^+ + \pi^-$\\
&&1.0287\,$\cdot$\,\,2$\Sigma^0$\\
\\
$\Xi_c^0$ & 18.307 & 1.0170\,$\cdot$\,18$\pi^0$ & eleven & (seen) & 
$\frac{1}{2}$   & 
2$\ast(3.)$\\
\\
$\Omega_c^0$ & 19.985 & 0.99925\,$\cdot$\,20$\pi^0$ & seven & (seen) & 
$\frac{1}{2}$ & 
2$\ast(3.) + 2(1.)$\\
& & 1.0249\,$\cdot$\,\,2$\Xi^0$\\
& & 0.9854\,$\cdot$\,\,5$\eta$ & & &\\
[0.3cm]\hline\hline
\vspace{0.1cm}
\end{tabular}
\footnotemark{\footnotesize The modes apply to neutral particles only. 
The $\ast$ marks coupled modes.}
    
    \end{table}

   Now we have enough material to formulate the 
\emph{integer multiple rule} of the particle masses, 
according to which the masses of the 
$\eta$,\,$\Lambda$,\,$\Sigma^0$,\,$\Xi^0$,\,
$\Omega^-$,\,$\Lambda_c^+$,\,$\Sigma_c^0$,\,
$\Xi_c^0$, and $\Omega_c^0$ 
particles are, in a first approximation, integer multiples of the mass 
of the $\pi^0$\,meson, although some of the particles have spin, 
and may also have charge as well as strangeness and charm. A 
consequence of the integer multiple rule must be that the ratio of the 
mass of any meson or baryon listed above divided by the mass of another 
meson or baryon listed above is equal to the ratio of two integer numbers. 
And indeed, for example m($\eta$)/m($\pi^0$) is practically two times 
(exactly 0.9950$\,\cdot$\,2) the ratio m($\Lambda$)/m($\eta$). There is 
also the ratio m($\Omega^-$)/m($\Lambda$) = 0.9993\,$\cdot$\,3/2. 
We have furthermore e.g. the ratios m($\Lambda$)/m($\eta$) = 
1.019\,$\cdot$\,2, m($\Omega^-$)/m($\eta$) = 1.018\,$\cdot$\,3, 
m($\Lambda_c^+$)/m($\Lambda$) 
= 1.0247\,$\cdot$\,2, m($\Sigma_c^0$)/m($\Sigma^0$) = 
1.0287\,$\cdot$\,2,  m($\Omega_c^0$)/m($\Xi^0$) =\\
1.0249\,$\cdot$\,2, and m($\Omega_c^0$)/m($\eta$) = 0.9854\,$\cdot$\,5.

   We will call, for reasons to be explained soon, the particles 
listed above, which follow in a first approximation the integer 
multiple rule, the \emph{$\gamma$-branch} of the particle spectrum.  
The mass 
ratios of these particles are in Table\,1. The deviation of the mass ratios 
from exact integer multiples of m($\pi^0$) is at most 3.3\%, the average 
of the factors before the integer multiples of m($\pi^0$) of the nine 
$\gamma$-branch particles in Table\,1 is 1.0066 $\pm$ 0.0184. From a least 
square analysis follows that the masses of the ten particles on Table 1 
lie on a straight line given by the formula

\begin{equation} \mathrm{m}(N)/\mathrm{m}(\pi^0) = 1.0065\,N - 0.0043 
\qquad  N\,>\,1, 
\end{equation}     
where N is the integer number nearest to the actual ratio of the particle 
mass divided by m($\pi^0$). The correlation coefficient in Eq.(1) 
has the nearly perfect value R$^2$ = 0.999. Since the particle masses are 
unquestionable, and since the least square analysis is a routine method, 
Eq.(1) is unquestionable.
 
   The integer multiple rule applies to more than just the stable mesons 
and baryons. The integer multiple rule applies also to the $\gamma$-branch 
baryon resonances which have spin J = 1/2 and the meson resonances with 
I,J  $\leq$ 1, listed in the Review of Particle Physics,
 or in Table\,2. The $\Omega^-$\,baryon will not be considered because it 
has spin 3/2, but would not change the following equation significantly.
 If we consider all mesons and baryons 
of the $\gamma$-branch in Tables\,1 and 2, ``stable" or unstable,
then we obtain from a least square analysis the formula 

\begin{equation}    \mathrm{m}(N)/\mathrm{m}(\pi^0) = 0.999\,N + 
0.0867\qquad    N\,>\, 1,  
\end{equation}
with the correlation coefficient  0.9999. The line through the points is 
shown in Fig.\,1.

\begin{table}\caption{The particles following the integer multiple rule}
\begin{tabular}{lcll|lcll}\\
\hline\hline\\
&J&m/m($\pi^0$) & multiples & &J&m/m($\pi^0$) & multiples\\
[0.5ex]\hline\\

$\pi^0$ & 0&1.0000 & 1.0000\,$\cdot$\,1$\pi^0$ &  $\eta_c$(1S) & 0 & 
22.0809 & 1.0037\,$\cdot$\,22$\pi^0$\\

$\eta$ & 0 & 4.0563 & 1.0141\,$\cdot$\,4$\pi^0$ & J/$\psi$ & 1 & 22.9441 & 
0.9976\,$\cdot$23$\pi^0$\\

$\eta^{\,\prime}$(958) & 0 &  7.0959 &1.0137\,$\cdot$\,7$\pi^0$ & 
$\chi_{c0}$(1P) & 0 & 25.2989 & 1.0119\,$\cdot$\,25$\pi^0$\\

$\eta$(1295) & 0 & 9.5868 & 0.9587\,$\cdot$\,10$\pi^0$ & $\chi_{c1}$(1P) & 
1 & 26.0094 & 1.0004\,$\cdot$\,26$\pi^0$\\ 

$\eta$(1405) & 0 & 10.445 & 1.0445\,$\cdot$\,10$\pi^0$ & $\eta_c$(2S) & 0 
& 26.9528 & 0.9983\,$\cdot$\,27$\pi^0$\\

$\eta$(1475) & 0 & 10.9352 & 0.9941\,$\cdot$\,11$\pi^0$ & $\psi$(2S) & 1 & 
27.3091 & 1.0115\,$\cdot$\,27$\pi^0$\\

& & & &  $\psi$(3770) & 1 & 27.9389 & 0.9978\,$\cdot$\,28$\pi^0$\\

$\Lambda$ & 1/2 & 8.2658 & 1.0332\,$\cdot$\,8$\pi^0$ & $\psi$(4040) & 1 & 
29.9237 & 0.9975\,$\cdot$\,30$\pi^0$\\ 

$\Lambda$(1405) & 1/2 & 10.4166 & 1.0417\,$\cdot$\,10$\pi^0$ & 
$\psi$(4191) & 1 & 31.0543 & 1.00175\,$\cdot$\,31$\pi^0$\\

$\Lambda$(1670) & 1/2 & 12.3725 & 1.0310\,$\cdot$\,12$\pi^0$ & 
$\psi$(4415) & 1 & 32.7538 & 0.9925\,$\cdot$\,33$\pi^0$\\

$\Lambda$(1800) & 1/2 & 13.335 & 1.0258\,$\cdot$\,13$\pi^0$ & & & &\\  

$\Sigma^0$ & 1/2 & 8.8359 & 0.9818\,$\cdot$\,9$\pi^0$ & B$^{\pm,0}$ & 
0 & 39.1116 & 1.0029\,$\cdot$\,39$\pi^0$\\
 
$\Sigma$(1660) & 1/2 & 12.2984 & 1.0249\,$\cdot$\,12$\pi^0$ &
B$^0_s$ & 0 & 39.7573 & 0.99393\,$\cdot$\,40$\pi^0$\\
 
$\Sigma$(1750) & 1/2 & 12.9652 & 0.9973\,$\cdot$\,13$\pi^0$  
& & & & \\

$\Xi^0$ & 1/2 & 9.7412 & 0.9741\,$\cdot$\,10$\pi^0$ & 
$\Upsilon$(1S) & 1 & 70.0884 & 1.0013\,$\cdot$\,70$\pi^0$\\

$\Omega^-$ & 3/2 & 12.3907 & 1.0326\,$\cdot$\,12$\pi^0$ &
$\chi_{b0}$(1P) & 0 & 73.0455 & 1.0006\,$\cdot$\,73$\pi^0$\\ 

$\Lambda_c^+$ & 1/2 & 16.9397 & 0.9965\,$\cdot$\,17$\pi^0$ & 
$\chi_{b1}$(1P) & 1 & 73.2926 & 1.0040\,$\cdot$\,73$\pi^0$\\

$\Lambda_c$(2593)$^+$ & 1/2 & 19.2285 & 1.0120\,$\cdot$\,19$\pi^0$ & 
$\Upsilon$(2S) & 1 & 74.259 & 1.0035\,$\cdot$\,74$\pi^0$\\

$\Sigma_c$(2455)$^0$ & 1/2 & 18.1792 & 1.00995\,$\cdot$\,18$\pi^0$ & 
$\chi_{b0}$(2P) & 0 & 75.8094 & 0.9975\,$\cdot$\,76$\pi^0$\\

$\Xi_c^0$ & 1/2 & 18.3069 & 1.01705\,$\cdot$\,18$\pi^0$ &
$\chi_{b1}$(2P) &1 & 75.9795 & 0.9997\,$\cdot$\,76$\pi^0$\\ 
 
$\Xi^{\prime0}_c$ & 1/2 & 19.0996 & 1.0052\,$\cdot$19\,$\pi^0$ &  
$\Upsilon$(3S) & 1 & 76.7185 & 0.9963\,$\cdot$\,77$\pi^0$\\

$\Xi_c$(2790) & 1/2 & 20.6843 & 0.9850\,$\cdot$\,21$\pi^0$ &
$\Upsilon$(4S) & 1 & 78.3795 & 1.0049\,$\cdot$\,78$\pi^0$\\  

$\Omega_c^0$ & 1/2 & 19.9849 & 0.9992\,$\cdot$\,20$\pi^0$ &
$\Upsilon$(10860) & 1 & 80.4954 & 1.0062\,$\cdot$\,80$\pi^0$\\ 

& & & & $\Upsilon$(11020) & 1 & 81.6364 & 0.9956\,$\cdot$\,82$\pi^0$\\
[0.4cm]\hline\hline
\end{tabular}
\end{table}

\begin{figure}
    \includegraphics{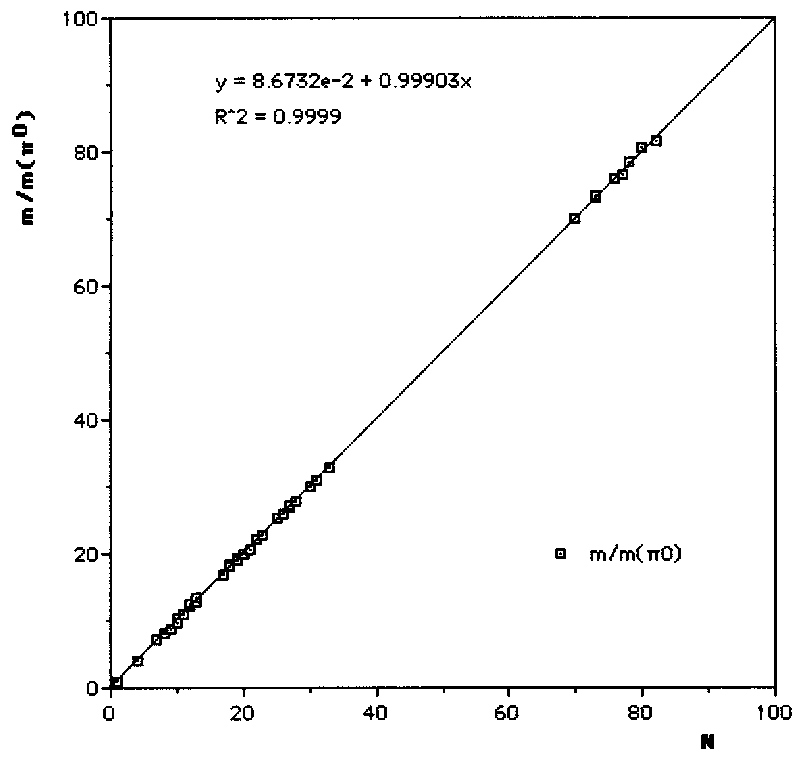}
    \begin{quote}
Fig.\,1: The mass of the mesons and baryons of the $\gamma$-branch,
  stable or unstable, with I\,$\leq$ 1,\,J\,$\leq$ 1 in units of 
m($\pi^0$) 
 as a function of the integer  N, demonstrating the integer multiple rule.
    \end{quote}
\end{figure}

   Only particles, stable or unstable, with J\,$\le$\,1  are listed. 
To be on the safe side, we use only particles which the Particle Data 
Group considers to be ``established". The $\Omega^-$ baryon 
with J = 3/2 is also given for comparison, but is not included 
in the least square analysis. In all there are 43 particles which 
follow the integer multiple rule. The line on Fig.\,1 is determined 
by Eq.(2).

   Fig.\,1 tells that 43 particles of the $\gamma$-branch  
of different spin and isospin, strangeness and charm; five I,J = 0,0 
$\eta$\,\,mesons, fifteen J = 1/2 baryons, two I = 0,1/2, J = 0 bottom 
mesons, ten I = 0, J = 0,1 c$\bar{\mathrm{c}}$ mesons, ten I = 0, 
J = 0,1 b$\bar{\mathrm{b}}$ mesons and the
$\pi^0$\,meson with I,J = 1,0, lie on a straight line with slope 
0.999. In other words they approximate the integer multiple rule 
very well. Spin 1/2 and spin 1 does not seem to affect the integer
multiple rule, i.e. the ratios of the particle masses, neither
does strangeness S $\neq$ 0 and charm C $\neq$ 0. Actually, spin 1/2 
is independent of the mass of the particle, as we will show at the end  
of Section 15.

   Searching for what else the  
$\pi^0$,\,$\eta$,\,$\Lambda$,\,$\Sigma^0$,\,$\Xi^0$,\,$\Omega^-$ 
particles have in common, we find that the principal decays (decays 
with a fraction\,$>$\,1\%) of these particles, 
as listed in Table\,1, involve primarily $\gamma$-rays, the 
characteristic case is $\pi^0 \rightarrow \gamma\gamma$  (98.8\%). We 
will later on discuss a possible explanation for the 1.174\% of the 
decays of $\pi^0$ which do not follow the $\gamma\gamma$ route but
decay via $\pi^0 \rightarrow$ e$^+$ + e$^-$ + $\gamma$. After the 
$\gamma$-rays the next most frequent 
decay product of the heavier particles of the $\gamma$-branch are 
$\pi^0$\,mesons, which again decay into $\gamma\gamma$. To describe 
the decays in another way, the principal decays of the particles listed 
above take place $\emph{always without the emission of neutrinos}$\,; 
see Table\,1. There the decays and the fractions of the principal decay 
modes are given. We cannot consider decays 
with fractions $<$\,1\%. We will refer to the particles whose masses are 
approximately integer multiples of the mass of the $\pi^0$\,meson,  
and which decay without the emission of neutrinos, as the 
$\gamma$-$\emph{branch}$ of the particle spectrum.

   To summarize the facts concerning the $\gamma$-branch. Within 
0.66\% on the average the masses of the stable particles of the 
$\gamma$-branch  in Table\,1 are integer multiples 
(namely 4,\,8,\,9,\,10,\,12, and even 17,\,18,\,20) of the mass of  
the $\pi^0$\,meson. 
It is improbable that nine particles have masses so close to integer 
multiples of m($\pi^0$), if there is no correlation between them and the 
$\pi^0$\,meson. It has, on the other hand, been argued that the integer 
multiple rule is a, quote, \emph{numerical coincidence}. But the probability 
that the mass ratios of nine particles of the $\gamma$-branch fall by 
coincidence on integer numbers between 1 and 20, instead on all 
possible numbers between 1 and 20 with two decimals  
after the period, is smaller than 10$^{-20}$, i.e.\,nonexistent. The 
integer multiple rule is not affected by more than 3\% by the spin, the 
isospin, the strangeness, and by charm. The integer multiple rule seems 
even to apply to the $\Omega^-$ and $\Lambda_c^+$ particles, although  
they are charged. In order for the integer multiple rule to be valid the 
deviation of the ratio m/m($\pi^0$) from an integer number must be smaller 
than 1/2N, where N is the integer number closest to the actual ratio 
m/m($\pi^0$). That means that the permissible deviation decreases rapidly 
with increased N. All particles of the $\gamma$-branch have deviations 
smaller than 1/2N.
 
   The remainder of the stable mesons and baryons are the  
$\pi^\pm$, K$^{\pm,0}$,\,p,\,\,n,\quad D$^{\pm,0}$, and D$_s^\pm$ 
particles, which make up the neutrino-branch ({$\nu$-branch}) of the 
particle spectrum. The ratios of their masses are given in Table\,3.
The characteristic particles of the $\nu$-branch are the $\pi^\pm$ mesons,
whose masses are nearly the same as the mass of the $\pi^0$ meson, 
m($\pi^\pm$) = 1.034\,$\cdot$\,m($\pi^0$).

    \begin{table}\caption{The ratios m/m($\pi^\pm)$ of the particles
of the $\nu$-branch} 
    \begin{tabular}{llllrcl}\\

\hline\hline\\
 & m/m($\pi^\pm$) & multiples & decays
 & fraction & spin & mode 
 \\
 & & & & (\%) & &\\
[0.5ex]\hline
\\
$\pi^\pm$ & 1.0000 & 1.0000\,\,$\cdot$\,\,$\pi^\pm$ & $\mu^+\nu_\mu$ & 
\,\,99.9877 & 0 & (1.)\\
 & & & e$^\pm$$\nu_e$($\bar{\nu_e}$) & \,\,\,\, 1.230$\cdot$10$^{-4}$\\
\\
K$^{\pm}$ & 3.53712 & 0.88428\,$\cdot$\,\,4$\pi^\pm$ & $\mu^+\nu_\mu$ & 
\,\,63.55 & 0 & 
(2.) + $\pi^0$\\
 & & & $\pi^\pm\pi^0$ & \,\,21.13 && \,\,\,\,(K$^\pm$)\\
 & & & $\pi^+\pi^-\pi^+$ & \,\,\,\,\,5.58 && (2.) + $\pi^\mp$\\
 & & & $\pi^0$ e$^+ \nu_e$ (K$_{e3}^+$)& \,\,\,\,\,4.87 & 
&\,\,\,\,(K$^0$,$\overline{{\mathrm{K}}^0}$)\\
 & & & $\pi^0\mu^+\nu_\mu$ (K$_{\mu3}^+$) & \,\,\,\,\,3.27 & &\\
 & & & e$^\pm$$\nu_e$($\bar{\nu_e}$) & \,\,\,\,\,\,\,\, 1.581$\cdot$10$^{-5}$\\
\\
n & 6.73185 & 0.8415\,$\cdot$\,\,8$\pi^\pm$ & p\,e$^-\overline{\nu}_e$ & 
100.\,\,\,\,\,\,\,\,  & $\frac{1}{2}$ & 2$\ast$(2.)\\
 & & 0.9516\,$\cdot$\,\,(K$^+$ + K$^-$)  & & & & \,\, + $2\pi^\pm$\\
 & &  0.9440\,$\cdot$\,\,(K$^0$ + $\overline{\mathrm{K}}^0$) & & & &\\
\\
D$^{\pm,0}$ & 13.395 & 0.8372\,$\cdot$\,16$\pi^\pm$ & e$^+$ anything & 
\,\,17.2 & 
0 & 2(2$\ast$(2.)\\
 & & 0.9468\,$\cdot$\,\,4K$^\pm$ & K$^-$ anything & \,\,24.2 & & \,\, + 
$2\pi^\pm$)\\
 & & 0.9956\,$\cdot$\,(p + $\bar{\mathrm{n}}$) & $\overline{\mathrm{K}}^0$ 
anything\\  
 & &                       & \,\,+\,K$^0$ anything & \,\,59\\
 & &                       & $\eta$ anything & $<$\,13\\
&&                          &K$^+$ anything &5.8\\
\\
D$^\pm_s$ & 14.104 & 0.8296\,$\cdot$\,17$\pi^\pm$ & K$^-$ anything & 
\,\,13 & 0 & 
body\\
 & & 0.9968\,$\cdot$\,\,4K$^\pm$ & $\overline{\mathrm{K}}^0$ 
 anything & & &  centered\\
 & & & \,\,+\,K$^0$ anything & \,\,39 & & cubic\\
 & & & K$^+$ anything & \,\,20\\
 & & & e$^+$ anything & 8\\
[0.3cm]\hline\hline

\vspace{0.1cm}
    \end{tabular}
    \footnotemark{\footnotesize Only the decays of the positively 
         charged particles are listed. The particles with negative 
         charges have conjugate charges of the listed  decays. The K$^0$
         particles are listed
         in Table 4, p.\,36. The $\ast$ marks coupled modes.}   
    \end{table}

The particles of the $\nu$-branch are in general charged, exempting the K$^0$ 
and D$^0$ mesons and the neutron n, in contrast to the particles of the 
$\gamma$-branch, which are in general neutral. It does not make a 
significant difference whether one considers the mass of a particular 
charged or neutral particle. The largest mass difference between 
charged and neutral particles is, after the $\pi$\,mesons (3.40\%), that
of the K\,mesons (0.81\%), and thereafter all mass differences between 
charged and neutral 
particles are $<$\,0.5\%. The integer multiple rule does not immediately 
apply to the masses of the $\nu$-branch particles if m($\pi^\pm$) (or 
m($\pi^0$)) is used as reference, because m(K$^\pm$) = 
0.8843\,$\cdot$\,4m($\pi^\pm$). 0.8843\,$\cdot$\,4 = 3.537 is far from 
integer. Since the masses 
of the $\pi^0$\,meson and the $\pi^\pm$\,mesons differ by only 
3.4\% it has been argued that the $\pi^\pm$\,mesons are, but for the
charge, the same type of particle as the $\pi^0$\,meson, and that 
therefore the $\pi^\pm$\,mesons cannot start a different particle branch. 
However, this argument is not supported 
by the completely different decays of the $\pi^0$\,mesons 
and the $\pi^\pm$\,mesons. The $\pi^0$\,meson decays almost 
exclusively into $\gamma\gamma$ (98.8\%),  whereas the 
$\pi^\pm$\,mesons decay practically exclusively into muons 
and neutrinos, as in $\pi^\pm$ $\rightarrow$  $\mu^\pm$  + 
$\nu_\mu(\bar{\nu}_\mu)$ (99.9877\%). Furthermore, the lifetimes of the 
$\pi^0$ and the $\pi^\pm$ mesons differ by nine orders of magnitude, 
being $\tau$($\pi^0$) = 8.4\,$\cdot$\,10$^{-17}$ sec versus 
$\tau$($\pi^\pm$) = 2.6\,$\cdot$\,10$^{-8}$ sec.

   If we make the $\pi^\pm$\,mesons the reference particles of the 
$\nu$-branch, then we must multiply the mass ratios m/m($\pi^\pm$) of  
the above listed particles with an average factor 0.848 $\pm$ 0.025, as 
follows from the mass ratios on Table\,3. 
The integer multiple rule may, however, apply 
directly if one makes m(K$^\pm$) the reference for masses larger than 
m(K$^\pm$). The mass of the neutron is 0.9516\,$\cdot$\,2m(K$^\pm$),
which is only a fair approximation to an integer multiple. There are,  
on the other 
hand, outright integer multiples in m(D$^\pm$) = 0.9954\,$\cdot$\,(m(p) + 
m($\bar\mathrm{n}$)), and in m(D$_s^\pm$) = 0.9968\,$\cdot$\,4m(K$^\pm$). 
A least square 
analysis of the masses of the $\nu$-branch particles in Table\,3 yields 
the formula

\begin{equation} \mathrm{m}(N)/0.853\mathrm{m}(\pi^\pm) = 1.000\,N + 
0.00575\qquad  N\,>\, 1 ,
\end{equation}
\noindent
with R$^2$ = 0.998. This means that the particles of the $\nu$-branch  
are integer multiples of m($\pi^\pm$) times the factor 0.853. One must, 
however, consider that the $\pi^\pm$\,mesons are not necessarily the 
perfect reference for all $\nu$-branch particles, because $\pi^\pm$ has  
isospin I = 1, whereas for example K$^\pm$ has I = 1/2 and S = $\pm$1   
and the neutron has also I = 1/2. Actually the 
factor 0.853 in Eq.(3) is only an average. The mass ratios indicate that 
this factor  decreases slowly with increased m(N).
 The existence of the factor and its decrease will be explained later.
 
   Contrary to the particles of the $\gamma$-branch, the $\nu$-branch 
particles decay preferentially with the emission of neutrinos, the 
foremost example is $\pi^\pm \rightarrow \mu^\pm$ + 
$\nu_\mu(\bar{\nu}_\mu)$ with a 
fraction of 99.9877\%. Neutrinos characterize the weak interaction. We 
will refer to the particles in Table\,3 as the $\emph{neutrino branch}$ 
($\nu$-branch) of the particle spectrum. We emphasize that a weak decay of 
the particles of the $\nu$-branch is by no means guaranteed. Although the 
neutron decays via n $\rightarrow$ p + e$^-$ + $\bar{\nu}_e$ in 885.7 sec 
(100\%), the proton is stable. There are, on the other hand, weak decays 
such as e.g. K$^+ \rightarrow \pi^+\pi^-\pi^+$ (5.59\%), but the 
subsequent decays of the $\pi^\pm$\,mesons lead to neutrinos and e$^\pm$. 

   To summarize the facts concerning the $\nu$-branch of the mesons 
and bary-ons. The masses of these particles seem to follow the integer 
multiple rule if one uses the $\pi^\pm$\,mesons as reference, however 
the mass ratios share a common factor 0.848 $\pm$ 0.025.

   To summarize what we have learned about the \emph{integer multiple rule}, 
which is not a theory, but a summary of experimental facts: 
In spite of differences in charge, spin, strangeness, and 
charm the masses of the ``stable" mesons and baryons of the 
$\gamma$-branch  are integer multiples of the mass 
of the $\pi^0$\,meson, within at most 3.3\% and on the average 
within 0.66\%. Correspondingly, the masses of the ``stable" particles 
of the $\nu$-branch are, after multiplication with a factor 0.848 
$\pm$ 0.025, integer multiples of the mass of the $\pi^\pm$\,mesons. 
The integer multiple rule has been anticipated 
much earlier by Nambu [10], who wrote in 1952 that ``some regularity 
[in the masses of the particles] might be found if the masses were 
measured in a unit of the order of the $\pi$-meson mass". A similar 
suggestion has been made by Fr\"ohlich [11]. The integer multiple rule 
suggests that the particles are the result of superpositions of modes 
and higher modes of a wave equation.

\section {Standing waves in a cubic lattice} 

 We will now study, as we 
have done in [12], whether the so-called ``stable" particles of the 
$\gamma$-branch cannot be described by the frequency spectrum of 
standing waves in a cubic lattice, which can accommodate automatically 
the Fourier frequency spectrum of an extreme short-time collision by 
which the particles are created. 
The investigation of the consequences of lattices for particle theory 
was initiated by Wilson [13] who studied a cubic fermion lattice. His 
study has developed over time into lattice QCD. 

   It will be necessary for the following to outline the most elementary 
aspects of the theory of lattice oscillations. The classic paper describing
lattice oscillations is from Born 
and v.\,Karman [14], henceforth referred to as B\&K. They looked at first 
at the oscillations of a one-dimensional chain of points with mass m, 
separated by a constant distance $\emph{a}$. This is the 
$\emph{monatomic}$ case, all lattice points have the same mass. B\&K assume 
that the forces exerted on each point of the chain originate only from the 
two neighboring points. These forces are opposed to and proportional to 
the displacements, as with elastic springs (Hooke's law). The equation of 
motion is in this case

\begin{equation} \mathrm{m}\ddot{u}_{n} = \alpha(u_{n+1} - u_n)
 - \alpha(u_n - u_{n-1})\,. \end{equation}
\noindent
The $u_n$ are the displacements of the mass points from their 
equilibrium position 
which are apart by the distance \emph{a}. The dots signify, as usual, 
differentiation with respect 
to time, $\alpha$ is a constant characterizing the force between the 
lattice points, and n is an integer number. For \emph{a} $\rightarrow$ 0 
Eq.(4) becomes the wave equation c$^2\partial^2$u/$\partial$x$^2$ = 
$\partial^2$u/$\partial$t$^2$ (B\&K).
 
   In order to solve Eq.(4) B\&K set 

\begin{equation} u_n = Ae^{i(\omega\,t\, +\, n\phi)}\,,
\end{equation}

\noindent
which is obviously a temporally and spatially periodic solution or 
describes \emph{standing waves}. n is an 
integer, with n\,$<$\,N, where N is the number of points in the chain. 
$\phi$ = 0 is the monochromatic case. We also consider higher modes,
 by replacing n$\phi$ in Eq.(5) by n$^\prime\phi$, where n$^\prime$
is n times an\,\,integer number $>$\,1. The wavelengths are then shorter 
by one over the number n$^\prime$/n. At n$\phi$ = $\pi$/2 are 
nodes, where for all times $\emph{t}$ the displacements are zero, as 
with standing waves f(x,t) = A\,cos($\omega$t)\,cos(n$\phi$) = 
A\,cos($\omega$t)\,cos(kx). 
If a displacement is repeated after n points we have n$\emph{a}$ = 
$\lambda$, where $\lambda$ is the wavelength, and $\emph{a}$ 
the lattice constant, and it must be n$\phi$ = 2$\pi$ according to (5). 
It follows that

\begin{equation} \lambda = 2\pi\emph{a}/\phi\,.
\end{equation}

\noindent 
Inserting (5) into (4) one obtains a continuous frequency spectrum of  
the standing waves as given by Eqs.(5) and (6) of B\&K

\begin{equation}  \omega = \pm\,2\sqrt{\alpha/\mathrm{m}}
\,\mathrm{sin}(\phi/2)\,. \end{equation}
\noindent
B\&K point out that there is not only a continuum 
of frequencies, but also a  \emph{maximal frequency} which 
is reached when $\phi$ = $\pi$, or at the 
minimum of the possible wavelengths $\lambda$ = 2$\emph{a}$. The 
boundary conditions are periodic, that means that $u_n$ = $u_{n + N}$, 
where N is the number of points in the chain. Born referred to the 
periodic boundary condition as a ``mathematical convenience". The number 
of normal modes must be equal to the number of particles in the lattice.
 
    Born's model of the crystals has been verified in great detail by X-ray
scattering and even in much more complicated cases by neutron scattering. 
The theory of lattice oscillations has been pursued in particular by 
Blackman [15], a summary of his and other studies is in [16]. 
Comprehensive reviews of the results of linear studies of lattice dynamics 
have been written by Born and Huang [17], by Maradudin et al. [18], and 
by Ghatak and Kothari [19]. A quantum mechanical description of atomic
lattices can be found in Born and Huang. Whether this description applies
to lattices in which the force has a range of 10$^{-16}$ cm remains to be
seen.

\section {The masses of the $\gamma$-branch particles}

   We will now assume, as seems to be quite natural, that  
\emph{the particles consist of the same particles into which they 
decay}, directly or ultimately. This assumption is fundamental
for the explanation of the particles. We know this from atoms, which 
consist of nuclei and electrons, and from nuclei, which consist of 
protons and neutrons. Quarks 
have never been found among the decay products of 
elementary particles. For the $\gamma$-branch particles 
our assumption means that they consist of photons. Photons 
and $\pi^0$\,mesons are the principal decay products of the 
$\gamma$-branch particles, the characteristic example 
is $\pi^0$ $\rightarrow$ $\gamma\gamma$ (98.82\%). Table\,1 
shows that there are decays of the $\gamma$-branch particles which 
lead to particles of the $\nu$-branch, in particular to pairs of 
$\pi^+$ and $\pi^-$ mesons.  It appears that this has to do with pair 
production in the $\gamma$-branch particles. Pair production 
is evident in the decay $\pi^0 \rightarrow$ e$^+$ + e$^- + 
\gamma$ (1.174\%), or in the $\pi^0$\,meson's third most frequent 
decay $\pi^0 \rightarrow$ e$^+$e$^-$e$^+$e$^-$ 
(3.34$\cdot10^{-5}$). Pair production requires the
presence of electromagnetic waves of high energy. Anyway, 
the explanation of the $\gamma$-branch particles must begin 
with the explanation of the most simple example of its kind, the 
$\pi^0$\,meson, which by all means seems to consist of photons.
We take 98.82\% for a good approximation of 100\%, and say that the 
$\pi^0$\,meson consists of photons. 
The composition of the particles of the $\gamma$-branch suggested here 
offers a direct route from the formation of a $\gamma$-branch particle, 
through its lifetime, to its decay products. Particles that are made of 
photons are necessarily neutral, as the majority of the particles of the 
$\gamma$-branch are.
 
   We also base our assumption that the particles of the $\gamma$-branch 
are made of photons on the circumstances of the formation of the 
$\gamma$-branch particles. The most simple and straightforward creation of 
a $\gamma$-branch particle are the reactions $\gamma$ + p $\rightarrow$ 
$\pi^0$ + p, or in the case that the spins of $\gamma$ and p are parallel 
\,\,$\gamma$ + p $\rightarrow$ $\pi^0$ + p + $\gamma^\prime$. A 
photon impinges on a proton and creates  a $\pi^0$\,meson. The 
considerations which follow apply as well for other photoproductions 
such as $\gamma$ + p $\rightarrow \eta$ + p or $\gamma$ + d 
$\rightarrow \pi^0$ + d and to the photoproduction of $\Lambda$  in
$\gamma$ + p $\rightarrow \Lambda$ + K$^+$, but also for the 
electroproductions e$^-$ + p $\rightarrow 
\pi^0$ + e$^-$ + p or e$^-$ + d $\rightarrow$   $\pi^0$  + e$^-$ + d,  
see Rekalo et al.\,\,[20]. The most simple example of the creation of a 
$\gamma$-branch particle by a strong interaction is the reaction 
p + p $\rightarrow$ p + p + $\pi^0$. The electromagnetic energy 
accumulated in a proton during its acceleration reappears as the 
$\pi^0$\,meson. 

   In $\gamma$ + p $\rightarrow \pi^0$ + p the pulse of the 
incoming electromagnetic wave is in 10$^{-23}$\,sec  converted 
into a continuum of electromagnetic waves with frequencies ranging  
from 10$^{23}$ sec$^{-1}$ to $\nu$ $\rightarrow$ $\infty$ according  
to Fourier analysis. There must be a cutoff frequency, 
otherwise the energy in the sum of the frequencies would exceed the 
energy of the incoming electromagnetic wave.  Conservation of momentum 
requires that all waves in the wave packet move with the velocity of light 
in the same direction, the direction of the incoming $\gamma$-ray. The wave
packet so created decays, according to experience, after 
8.4\,$\cdot$\,10$^{-17}$\,sec into two 
electromagnetic waves or $\gamma$-rays, $\pi^0$ $\rightarrow$ $\gamma\gamma$. 
It seems to be very 
unlikely that Fourier analysis does not hold for the case of an 
electromagnetic wave impinging on a proton. The question then arises of 
what happens to the electromagnetic waves in the timespan of 10$^{-16}$ 
seconds between the creation of the wave packet and its decay into two 
$\gamma$-rays\,? We will show that the electromagnetic waves 
can continue to exist for the 10$^{-16}$ seconds until the wave packet 
decays into two $\gamma$-rays. $\gamma$$\gamma$ means
experimentally that the $\pi^0$\,meson decays into two separate 
$\gamma$-rays moving with equal
energy in opposite direction, forward and backward with regard to the
incoming $\gamma$-ray, as conservation of momentum requires. We will 
show that the $\pi^0$\,meson is, during its lifetime, already filled with 
electromagnetic waves moving with equal energies in opposite direction, 
forward and backward.

   If the wave packet created by the collision of a $\gamma$-ray  
with a proton consists of electromagnetic 
waves, then the waves cannot be progressive because the
wave packet must have a \emph{rest mass}. The rest mass is 
the mass of a particle whose center of mass does not move. However 
\emph{standing electromagnetic waves}  have a rest mass. Standing 
electromagnetic waves are equivalent to a lattice, because in standing 
waves the waves travel back and forth between the nodes, just as 
lattice points oscillate between the nodes of the lattice oscillations. 
The oscillations in the lattice take care of the continuum of
frequencies of the Fourier spectrum of the collision which created the 
particle. So we assume that 
the very many photons in the wave packet are held together in a cubic
lattice. It is not unprecedented that photons have been 
considered to be building blocks of the elementary particles. Schwinger 
[21] has once studied an exact one-dimensional quantum electrodynamical 
model in which the photon acquired a mass $\sim$ e$^2$. 

   We will now investigate the standing waves in a cubic photon 
lattice. We assume that the lattice is held together by a weak force 
acting from one 
lattice point to its nearest neighbors. We assume that the range  
of this force is 10$^{-16}$\,cm, because the range of the weak nuclear  
force is on the order of 10$^{-16}$\,cm, as stated e.g.\,\,on p.25 of
Perkins [22]. We set the \emph{lattice constant} at
\begin{equation}
\emph{a} = 1\cdot 10^{-16}\,\mathrm{cm}\,, \end{equation}
as we have done originally in [23]. The lattice constant 
of a cubic lattice can be derived from lattice theory, 
see Appendix A. For the sake of simplicity we set the 
sidelength of the lattice at 10$^{-13}$\,cm, there are then 10$^{\,9}$ 
lattice points. The exact size of the nucleon is given in [2] or in [26]
and will be used later. As we will see the ratios of the masses of 
the $\gamma$-branch particles are independent of the sidelength of the 
lattice. Because it is the most simple case, we assume that a 
central force acts between the lattice points. We cannot consider 
charge, spin, strangeness or charm of the particles. The frequency 
equation for the waves in an isotropic monatomic cubic lattice with 
central forces is, in the one-dimensional case, given by Eq.(7). The 
direction of the waves is determined by the direction of the 
incoming $\gamma$-ray.

   According to Eq.(13) of B\&K the force constant $\alpha$ in Eq.(7) is

\begin{equation}   \alpha = \emph{a}\,(c_{11} - c_{12} - c_{44})\,,
\end{equation}

\noindent
where c$_{11}$, c$_{12}$ and c$_{44}$ are the elastic constants in 
continuum mechanics which applies in the limit $\emph{a}$ $\rightarrow$  
0. If we consider central forces then c$_{12}$ = c$_{44}$, which is the 
classical Cauchy relation. Isotropy requires that c$_{44}$ = 
(c$_{11}\,-\,$  c$_{12}$)/2.  
The waves are longitudinal. Transverse waves in a cubic lattice 
with central forces are not possible according to [19]. All frequencies 
that solve Eq.(7) come with either a plus or a minus sign which is, as we 
will see, important. The reference frequency in Eq.(7) is

\begin{equation}  \nu_0 = \sqrt{\alpha/4\pi^2\mathrm{m}} 
= \mathrm{c}/2\pi \emph{a}\,,
\end{equation}
\noindent
using Eq.(12) and f($\phi$) = $\phi$/2 from Eq.(13a).  c is the velocity of 
light.

   The \emph{consequence of the group velocity} has 
now to be considered. The group velocity is given by

\begin{equation}  \mathrm{c}_g = \frac{d\omega}{dk} = 
\emph{a}\sqrt{\frac{\alpha}{\mathrm{m}}}\cdot\frac{d\,f(\phi)}{d\phi}\,.
\end{equation}
\noindent
The group velocity of water waves in deep water is one-half of the 
phase velocity. We assume that the same relation applies to the group
velocity of photons in a photon lattice. In order to learn how this 
requirement affects the frequency distribution we have to know the value 
of $\sqrt{\alpha/\mathrm{m}}$ in a photon lattice. But we do not have 
information 
about what either $\alpha$ or m might be in this case. In the following 
we set $\emph{a}\sqrt{\alpha/\mathrm{m}}$ = c, which means, 
since $\emph{a} = 10^{-16}$\,cm, that $\sqrt{\alpha/\mathrm{m}}$ = 
3\,$\cdot$\,10$^{26}$ sec$^{-1}$, or that the corresponding period is 
$\tau$ = 1/3\,$\cdot$\,10$^{-26}$ sec, which is the time it takes for a 
wave to travel with the velocity of light over one lattice distance. With 

\begin{equation} \mathrm{c} = \emph{a}\sqrt{\alpha/\mathrm{m}}\,
\end{equation}

\noindent
the equation for the group velocity becomes

\begin{equation} \mathrm{c}_g = \mathrm{c}/2 = 
 \mathrm{c}\cdot d\,f(\phi)/d\phi\,.
\end{equation} 
\noindent
For photons in a photon lattice that means that\\ 

\hspace {3cm}df($\phi$)/d$\phi$ = 1/2,\quad or\,\, f($\phi$) =
 $\phi$/2 + $\phi_0$,\hspace{2cm} (13a)

\vspace{0.5cm}

\noindent
that is the first approximation of f($\phi$) in Eq.(7).\\
 
The frequencies of the oscillations are then given from Eq.(7) by

\begin{equation}\nu = \pm\,\nu_0(\,\phi + \phi_0\,)\,, \end{equation}
with $\nu_0$  = c/2$\pi$a from Eq.(10). 

   For the time being we will disregard $\phi_0$ in Eq.(14) because
$\phi_0$ = 0 when the boundary condition is periodic. The frequencies
of the spectrum in Eq.(14) 
must increase from $\nu$ = 0 at the origin $\phi$ = 0 with slope 
1 (in units of $\nu_0$) until the maximum is reached at $\phi = \pi$. 
The energy  contained in the oscillations  must be proportional to
the sum of all frequencies (Eq.15). The \emph{second mode}  of 
the lattice oscillations contains 4 times as much 
energy as  the basic mode, because the frequencies are twice the 
frequencies of the basic mode, and there are twice as many oscillations,
see Eq.(20a). 
Adding, by superposition, to the second mode different 
numbers of basic modes or of second modes will give exact 
integer multiples of the  energy of the basic mode. 

Now we understand the integer multiple 
rule of the particles of the $\gamma$-branch. The wavelenghts 
of standing waves can only be equal to the width of the container
divided by any integer number. That means that the frequencies 
of the waves are integer multiples of the basic frequency. There is,
in the framework of this theory, no alternative but $\emph{integer 
multiples}$ of the basic mode for the energy contained in the frequencies 
of the different modes or for superpositions of different modes. In other 
words, the masses of the different particles are integer multiples of 
the mass of the $\pi^0$\,meson, if there is no charge, spin, 
strangeness or charm.

   We remember that the measured masses in Table\,1, which 
incorporate different spins, isospins, strangeness and charm, spell 
out the integer multiple rule within on the average 0.65\% accuracy. It 
is worth noting that $\emph{there is no free parameter}$ if one takes
 the ratio of the energies contained in the frequency distributions of 
the different modes, because the factor $\sqrt{\alpha/\mathrm{m}}$ 
in Eq.(7) or $\nu_0$ in Eq.(14) cancels. This means, 
in particular, that the ratios of the frequency distributions, or the mass 
ratios, are independent of the mass of the photons at the lattice points, 
as well as of the magnitude of the force between the lattice points.

   It is obvious that the integer multiples of the sum of the frequencies 
in the particles are only a 
first approximation of the theory of lattice oscillations and of the mass 
ratios of the particles. The equation of motion in the lattice Eq.(4) does 
not apply in the eight corners of the cube, nor does it 
apply to the twelve edges nor, in particular, to the six sides of the 
cube. A cube with 10$^{\,9}$ lattice points is not precisely described by  
the periodic boundary condition we have used to derive Eq.(7), but  
is what is referred to as a microcrystal. 
   A phenomenological theory of the frequency distributions in 
microcrystals, considering in particular the surface energy, can be 
found  in Chapter 6 of Ghatak and Kothari [19]. The surface energy 
may account for the small deviations of the mass ratios of the mesons 
and baryons from the integer multiple rule of the oscillations in a cube. 
 
   Let us summarize our findings concerning the $\gamma$-branch. 
The particles of the $\gamma$-branch consist of standing 
electromagnetic waves. The $\pi^0$\,meson is the basic mode.  
The $\eta$ meson corresponds to the 
second mode, as is suggested by m($\eta$) $\approx$ 
4m($\pi^0$). The $\Lambda$ baryon corresponds to the superposition
of two second modes, as is suggested by m($\Lambda) \approx$ 
2m($\eta$). This superposition apparently results in the creation of 
spin 1/2. The two modes would then have to be coupled. The
$\Sigma^0$ and $\Xi^0$ baryons are superpositions 
of one or two basic modes on the $\Lambda$ baryon, as 
indicated by the decays $\Sigma^0$ $\rightarrow$ $\Lambda + \gamma$
(100\%) and $\Xi^0$ $\rightarrow$ $\Lambda + \pi^0$ (99.5\%). The  
$\Omega^-$ particle corresponds to the superposition of three coupled 
second modes as is suggested by m($\Omega^-$) $\approx$ 
3m($\eta$). This procedure apparently causes spin 3/2. The charmed 
$\Lambda_c^+$ baryon seems to be the first particle incorporating a 
third mode. $\Sigma_c^0$ is apparently the superposition of  
a negatively charged basic mode on $\Lambda_c^+$, as is 
suggested by the decay of $\Sigma_c^0$. The easiest explanation 
of $\Xi_c^0$ is that it is the superposition of two coupled 
third modes. The superposition of two modes of the same type is, 
as in the case of $\Lambda$, accompanied by spin 1/2. The $\Omega_c^0$ 
baryon is apparently the superposition of two basic modes on the $\Xi_c^0$ 
particle. All particles of the $\gamma$-branch are thus accounted 
for. The explanation of the charged $\gamma$-branch particles 
$\Sigma^\pm$  and $\Xi^-$ has been described in [67]. The modes of
the particles are listed in Table\,1. As mentioned already in
[12] all $\gamma$-branch particles with strangeness contain pairs of
second modes of $\pi^0$ or $\eta$ doublets, but for $\Omega^-$ which is
a triplet of $\eta$. All particles of the $\gamma$-branch with charm
contain a (3.) mode of the $\pi^0$\,meson.

   We have also found the $\gamma$-branch $\emph{antiparticles}$.
The rest masses of the antiparticles of the $\gamma$-branch consist also 
of standing electromagnetic waves. Their masses are the same as the 
masses of the normal particles and follow from the sum of the energies
h$\nu$ of the negative frequencies which solve Eq.(7) or Eq.(14). 
Antiparticles have always been associated with negative
energies. Following Dirac's argument for electrons and positrons, we
associate the masses with the negative frequency distributions with 
antiparticles. We emphasize that the existence of antiparticles is an 
automatic consequence of our theory. In this model of the particles
the \emph{rest mass} of a particle has an antiparticle.

   All particles of the $\gamma$-branch are unstable with lifetimes on 
the order of 10$^{-10}$ sec or shorter. Born [24] has shown that the 
oscillations in cubic lattices held together by central forces are 
unstable. It seems, however, to be possible that the particles can be 
unstable for reasons other than the instability of the lattice, which 
apparently causes the most frequent (electromagnetic) decay of the 
$\pi^0$\,meson $\pi^0 \rightarrow \gamma\gamma$ (98.82\%),  
or the most frequent (electromagnetic) decay of the 
$\eta$ meson $\eta \rightarrow \gamma\gamma$ (39.31\%).
Pair production seems to make it possible to understand the decay 
of the $\pi^0$\,meson $\pi^0 \rightarrow$ e$^-$ + e$^+ + \,\gamma$  
(1.174\%), or the decay $\pi^0$  $\rightarrow$ e$^+$e$^-$e$^+$e$^-$.
Since in our model the $\pi^0$\,meson consists of a 
multitude of electromagnetic waves it seems that pair 
production takes place within the $\pi^0$\,meson, and even more so 
in the higher modes of the $\gamma$-branch, where the electrons and 
positrons created by pair production tend to settle on mesons, as e.g. 
in $\eta \rightarrow \pi^+ +  \pi^- + \pi^0$ (22.74\%) or in the decay 
$\eta \rightarrow \pi^+ + \pi^- + \gamma$ (4.60\%), where the origin 
of the pair of charges is more apparent. Pair production is also evident 
in the decays $\eta \rightarrow$ e$^+$e$^-\gamma$ (0.7\%)  or 
$\eta \rightarrow$ e$^+$e$^-$e$^+$e$^-$ (6.9$\cdot10^{-5}$).

   Finally we must explain the reason for which the photon lattice or the 
$\gamma$-branch particles are limited in size to a particular value of 
about $10^{-13}$ cm, as 
the experiments show. Conventional lattice theory using the periodic 
boundary condition does not limit the size of a crystal, and in fact very 
large crystals exist. If, however, the lattice consists of standing 
electromagnetic waves the size of the lattice is limited by the radiation 
pressure. The lattice will necessarily break up at the latest when the 
outward directed radiation pressure is equal to the inward directed 
elastic force which holds the lattice together. For details we refer to 
[25].

\section {The rest mass of the $\pi^0$\,meson}

   So far we have studied the ratios of the masses of the particles. We 
will now determine the mass of the $\pi^0$\,meson in order to validate 
that the mass ratios link with the actual masses of the particles. The 
energy in the mass of the $\pi^0$\,meson, which does not have spin, is 

\vspace{0.5cm}

\hspace{1.5cm} m($\pi^0$)c$^2$ = 134.9766\,MeV =
 2.16258\,$\cdot\,10^{-4}$\,erg.\\

\noindent
The sum of the energies  E = h$\nu$ of the frequencies of the 
one-dimensional waves in $\pi^0$, Eq.(14), seems to be given by 
the equation

\begin{equation}\mathrm{E}_\nu = 
\mathrm{Nh}\nu_0\cdot\frac{1}{2\pi}\,\,\int\limits_{-\pi}^{\pi}f(\phi)d\phi\,.
\end{equation}
\noindent
N is the number of all lattice points and $\nu_0$  = c/2$\pi$a is from 
Eq.(10). The total energy of the frequencies in a cubic lattice is equal 
to the number N of the oscillations times the average of the energy of  
the individual frequencies. 

   In order to arrive at an exact value of N in Eq.(15) we have to use 
the correct value of the radius of the proton, for which we use
\begin{equation}
\mathrm{r}_p = (0.880\pm0.015)\cdot10^{-13}\,\mathrm{cm}\,,
\end{equation} 
\noindent
according to [26], or it is
r$_p$ = (0.883 $\pm$  0.014)\,$\cdot$\,10$^{-13}$\,cm according to [27].
The Review of Particle Physics gives for the charge radius of the proton
the value r$_p$ = (0.877 $\pm$ 0.007)$\,\cdot\,10^{-13}$ cm. When the size of 
the proton is measured by scattering large numbers of randomly oriented electrons 
on large numbers of randomly oriented protons, only a radius of the proton can
emerge. With \emph{a} = 10$^{-16}$\,cm, (Eq.(8)), it follows from Eq.(16) that 
the number of all lattice points in the cubic photon lattice is

\begin {equation}
\bullet\hspace{1cm}\mathrm{N} = \frac{4\pi\,\mathrm{r}^3_p}{3\,\emph{a}^{\,3}} =
 2.854\cdot10^{\,9} \cong 1\,418^{\,3}\,.
 \end{equation}
\noindent
N is fundamental for the explanation of the masses of the particles, because
N lattice points contribute to the mass of the particle.
The radius of the $\pi^\pm$\,mesons has also been measured [28] and after 
further analysis [29] was found to be 0.83\,$\cdot$\,10$^{-13}$\,cm. A much
earlier measurement [30] found r$_\pi$ at (0.86 $\pm$ 0.14)$\,\cdot\,10^{-13}$
cm. The Review of Particle Physics does not give a value for r$_\pi$.   
Within the uncertainty of the radii we have r$_p$ = r$_\pi$. 
And according to [2] or [31] the charge radius of $\Sigma^-$ is 
(0.78 $\pm$ 0.10)\,$\cdot$\,$10^{-13}$\,cm.

  If the oscillations are parallel to an axis,  
the group velocity is taken into account, that means if 
Eq.(14) applies, and the absolute values 
of the frequencies are taken, then the value of  
the integral in Eq.(15) is $\pi^2$. With N = 2.854\,$\cdot$\,10$^{\,9}$
and $\nu_0$ = c/2$\pi\emph{a}$  follows from Eq.(15) that the sum 
of the energy of the frequencies of the basic 
mode corrected for the group velocity is 
E$_{corr}$ = 2.836\,$\cdot$\,10$^{\,9}$\,erg. That means  
that the energy is 13.12\,$\cdot\,10^{12}$ times larger than
m($\pi^0$)c$^2$. 
This discrepancy is inevitable, because the basic frequency of the 
Fourier spectrum after a collision on the order of 
10$^{-23}$ sec duration is $\nu$ = 10$^{23}$ sec$^{-1}$, which means,  
when E = h$\nu$, that one basic frequency alone contains an energy of  
about 9\,m($\pi^0$)c$^2$.
 
   To eliminate this discrepancy we use, instead of the simple 
form E = h$\nu$, the complete quantum mechanical energy of a linear 
oscillator as given by Planck

\begin{equation} E = \frac{h\nu}{e^{h\nu/kT} -\,1}\,\,.
\end{equation}
\\
\noindent
This equation was already used by B\&K for the determination of the 
specific heat of cubic crystals or solids. Equation (18) calls into 
question the value of the temperature T in the interior of a particle. We 
determine T empirically with the formula for the internal energy of solids

\begin{equation} u = \frac{R\Theta}{e^{\Theta/T} -1}\,\,,
\end{equation}
\noindent
which is from Sommerfeld [32]. In this equation R = N\,$\cdot$\,k = 
2.854\,$\cdot\,10^{\,9}$\,k, 
where k is Boltzmann's constant, and $\Theta$ is the 
characteristic temperature introduced by Debye [33] for the explanation of 
the specific heat of solids. It is $\Theta = h\nu_m$/k, where $\nu_m$ is 
the maximal frequency. In the case of the oscillations making up the 
$\pi^0$\,meson the maximal frequency follows from Eq.(14) and is
$\nu_m$ = $\nu_0$$\pi$ = c/2a = 1.5\,$\cdot$\,10$^{\,26}$ 
sec$^{-1}$. $\Theta$ = 2$\pi$h$\nu_m$/k is then $\Theta$ = 
7.19\,$\cdot$\,10$^{\,15}$ K.
 
  In order to determine T we set the internal energy u equal to 
m$(\pi^0)$c$^2$ = 2.16\,258\,$\cdot$\,10$^{\,-4}$ erg, with 
1\,MeV = 1.60219\,$\cdot$\,10$^{-6}$ erg.  It then 
follows from Eq.(19) that $\Theta$/T = 30.20, 
or T = 2.38\,$\cdot$\,10$^{\,14}$\,K. That means that Planck's 
formula (18) changes Eq.(15) into 
\begin{equation}\mathrm{E}_\nu(\pi^0) =  
\frac{\mathrm{Nh} \nu_0}{(\mathrm{e^{h\nu/kT}}\,\mathrm{-}\,1)}
\cdot\frac{1}{2\pi}\int\limits_{-\pi}^{\pi}\phi\,d\phi\,. \end{equation}
This type of equation was already used by B\&K, their Eq.(47), for the
determination of the internal energy of cubic crystals. The energy in 
the second mode of $\pi^0$ is, with $\nu$ = $\nu_0$2$\phi$, and with twice 
as many oscillations in the second mode as in the first mode, given by

\begin{displaymath}
\hspace{2.0cm}
\mathrm{E}_{2\nu}(\pi^0) =  
\frac{\mathrm{2Nh} \nu_0}{(\mathrm{e^{h\nu/kT}}\,\mathrm{-}\,1)}
\cdot\frac{1}{2\pi}\int\limits_{-\pi}^{\pi}2\phi\,d\phi =
4\cdot\mathrm{E}_\nu(\pi^0)\,.\hspace{1.8cm}(20a)
\end{displaymath}
  
\noindent Eq.(20a) is theoretical proof of the \emph{integer multiple 
rule}. The energy 
in the mass of the second mode E$_{2\nu}$($\pi^0$) = m($\eta$)c$^2$ is 
four times the energy in the mass of the first mode E$_\nu$($\pi^0$) = 
m($\pi^0$)c$^2$, or m($\eta$)/m($\pi^0$) $\cong$ 4, as on Table\,1. 

The function
f(T) = (e$^{h\nu/kT}$ $-$ 1) in Eq.(20) introduces the term
\begin{equation} 
f(T) = (\mathrm{e}^{\Theta/T} - 1\,) \cong \mathrm{e}^{\,30.2} = 
(1.305\cdot10^{\,13}) \,,  \end{equation} 
into Eq.(15). In other words, if we determine the temperature T of the 
particle through Eq.(19), and correct Eq.(15) accordingly, then we arrive 
with Eq.(20) at the oscillation energy in the $\pi^0$\,meson, the sum of 
the energy in the frequencies of Eq.(14). It  is 
\begin{equation} \sum_1^N\mathrm{E}_\nu = 1.0866\cdot10^{-4}\,
\mathrm{erg} = 67.82\,\mathrm{MeV}\,,\end{equation}
whereas m($\pi^0)$c$^2$(exp) = 134.9766\,MeV.
The sum of the energies of N one-dimensional oscillations  
in the $\pi^0$\,meson lattice is 0.5024\,m($\pi^0$)c$^2$(\emph{exp}).

   If the electromagnetic waves in the $\pi^0$\,meson are cicular, 
as seems to be likely, then we must double the energy in Eq.(22), because
we have then 2N oscillations instead of N oscillations, and we 
find that in this model

\newpage
 
\emph{the rest mass of the $\pi^0$\,meson} is  
\begin{eqnarray}\mathrm{m}(\pi^0)\mathrm{c^2} 
=  \mathrm{E}_\nu(\pi^0)(theor) = 
2.1732\cdot10^{-4}\,\mathrm{erg} = 135.64\,\mathrm{MeV} \nonumber\\
 = 1.005\,\mathrm{m}(\pi^0)\mathrm{c^2}(exp)\,.\end{eqnarray}

\vspace{0.5cm} 

   The energy in the measured mass of the $\pi^0$\,meson and the energy 
in the sum of the standing waves agree fairly well, considering the 
uncertainties of the parameters involved. The theoretical mass of the 
$\eta$ meson is according to Eq.(20a) m($\eta$)(\emph{theor}) = 
4\,$\cdot$\,m($\pi^0$) =
542.56\,MeV = 0.990\,m($\eta$)(\emph{exp}), and the theoretical mass 
of the $\Lambda$\,baryon, the superposition of two $\eta$ mesons, is 
then m($\Lambda$)(\emph{theor}) = 8\,$\cdot$\,m($\pi^0$)
 = 1085.1\,MeV = 0.9726\,m($\Lambda$)(\emph{exp}).  

   To sum up: The $\pi^0$\,meson is formed when a $\gamma$-ray
collides with a proton, $\gamma$ + p $\rightarrow \pi^0$ + p. By the 
collision the incoming $\gamma$-ray is converted into a packet of standing 
electromagnetic waves, the $\pi^0$\,meson. After $10^{-16}$ seconds
the wave packet decays into two electromagnetic waves, 
$\pi^0 \rightarrow \gamma\gamma$. Only electromagnetic waves are present 
throughout the entire process. The energy in the rest mass 
of the $\pi^0$\,meson
and the other particles of the $\gamma$-branch is correctly given 
by the sum of the energy of standing electromagnetic waves in a cube,  
if the energy of the oscillations is determined by Planck's formula for 
the energy of a linear oscillator.

\begin{itemize}
\item
\emph{The $\pi^0$\,meson is like a cubic black body filled with\\  
standing electromagnetic waves}.
\end{itemize}
 
A black body of a given size and temperature can certainly contain the 
energy in the rest mass of the $\pi^0$\,meson, which is O($10^{-4})$
erg, if only the frequencies are sufficiently high. We know from 
Bose's work [34] that Planck's formula applies to a photon gas as well.
For all $\gamma$-branch particles we have found a simple mode of 
standing electromagnetic waves. Since the equation determining 
the frequency of the standing waves is quadratic it follows 
\emph{automatically} that for each positive frequency there  
is also a negative 
frequency of the same absolute value, that means that for each particle 
there exists also an \emph{antiparticle}. For the explanation of the 
stable mesons and baryons of the $\gamma$-branch $\emph{we use 
only photons, nothing else}$. This is a rather conservative
explanation of the $\pi^0$\,meson and the $\gamma$-branch 
particles. \emph{We do not use hypothetical particles}.

   From the frequency distributions of the standing waves follow 
the ratios of the masses of the particles which obey the integer 
multiple rule. It is important to note that in this theory the ratios 
of the masses of the $\gamma$-branch particles to the mass of the 
$\pi^0$\,meson $\emph{do not depend}$ on the sidelength of 
the lattice, neither do they depend on the strength 
of the force between the lattice points nor on the mass of the 
lattice points. The mass ratios are determined only by the spectra  
of the frequencies of the standing electromagnetic waves. 

\section {The neutrino branch particles}

The masses of the neutrino branch, the $\pi^\pm$, K$^{\pm,0}$, n, p, 
D$^{\pm,0}$ and D$^\pm_s$ particles, are integer multiples of the 
mass of the $\pi^\pm$\,mesons times a factor $0.85\,\pm\,0.02$, as 
we stated before. We assume, 
as appears to be quite natural, that the $\pi^\pm$\,mesons and the 
other particles of the neutrino branch \emph{consist of the 
same particles into which they decay}, that means in the case of the 
$\pi^\pm$\,mesons of muon neutrinos $\nu_\mu$, antimuon neutrinos 
$\bar{\nu}_\mu$, electron neutrinos $\nu_e$, anti-electron neutrinos 
$\bar{\nu}_e$ and of an electron or positron,  as exemplified by the 
decay sequence 
$\pi^\pm$  $\rightarrow$  $\mu^\pm$ + $\nu_\mu$($\bar{\nu}_\mu$),
\,$\mu^\pm$ $\rightarrow$ e$^\pm$ + $\bar{\nu}_\mu$($\nu_\mu$) 
+ $\nu_e$($\bar{\nu}_e$). The absence of an electron neutrino $\nu_e$
in the decay branches of $\pi^-$, or of an anti-electron neutrino 
$\bar{\nu}_e$  in the decay
branches of $\pi^+$, can be explained with the composition of 
the electron or positron, which will be discussed in Section 11. The   
existence of neutrinos and antineutrinos is unquestionable.
Since the particles of the $\nu$-branch decay through weak  
decays, we assume, as appears likewise to be natural, that 
\emph{the weak nuclear force holds the particles of the $\nu$-branch 
together}. This assumption has far reaching consequences, it is not
only fundamental for the explanation of the $\pi^\pm$\,mesons, but
leads also to the explanation of the $\mu^\pm$\,muons and 
ultimately to the explanation of the mass of the electron.  
The existence of the weak nuclear force is unquestionable. 

   Since the 
range of the weak interaction, which is about 10$^{-16}$ cm [22],
is only about a thousandth of the diameter of the particles, which is 
about 10$^{-13}$ cm, the weak force can hold particles together only 
if the particles have a lattice structure, just as macroscopic crystals 
are  held together by microscopic forces between atoms. In the 
absence of a central force which originates in the center of the particle 
and extends throughout the entire particle, as the Coulomb force does,
the configuration of a particle held together by the weak force is 
\emph{not spherical but cubic}, reflecting the very short range of the 
weak nuclear force. Nuclei are not spherical either. We will show  
that the energy in the rest mass of the $\nu$-branch particles is 
the energy in the oscillations of a cubic lattice consisting of electron 
neutrinos and muon neutrinos and their antiparticles, plus the energy 
in the masses of the neutrinos, plus a small part with the energy 
in the charges e$^\pm$ the particle carries.

   First it will be necessary to outline the basic aspects of diatomic lattice 
oscillations. In $\emph{diatomic}$ lattices the lattice points have 
alternately the masses m and M, as with the masses of the electron
neutrinos m($\nu_e$) and muon neutrinos m($\nu_\mu$).
The classic example of a diatomic lattice is the salt 
crystal with the masses of the Na and Cl atoms in the lattice points. 
The theory of diatomic harmonic lattice oscillations was started by 
Born and v.\,Karman [14]. They first discussed a diatomic chain. The 
equation of motions in the chain are according to Eq.(22) of B\&K

\begin{equation} \mathrm{m}\ddot{u}_{2n} = \alpha(u_{2n+1} +
 u_{2n-1} - 2u_{2n})\, ,\end{equation}

\begin{equation} \mathrm{M}\ddot{u}_{2n+1} = \alpha(u_{2n+2} +
 u_{2n} - 2u_{2n+1}) \,, \end{equation}
\noindent    
where the u$_n$ are the displacements, n an integer number and $\alpha$ a 
constant characterizing the force between the particles. Eqs.(24,25) are 
solved with

\begin{equation}u_{2n} = Ae^{i(\omega\,t\,+\,2n\phi)} ,\end{equation} 

\begin{equation}u_{2n+1} = 
Be^{i(\omega\,t\,+\,(2n+1)\phi)}\,,\end{equation} 

\noindent
where A and B are constants and $\phi$ is given by $\phi = 
2\pi\emph{a}/\lambda$ as in Eq.(6). $\emph{a}$ is the lattice constant as 
before and $\lambda$ the wavelength, $\lambda$ = n$\emph{a}$. The 
solutions of Eqs.(26,27) are obviously periodic in time and space and 
describe again standing waves. Using (26,27) to solve (24,25) leads to a 
secular equation from which, according to Eq.(24) of B\&K, the frequencies 
of the oscillations of the chain follow from 

\begin{equation}4\pi^2\nu^2_\pm  = \alpha/\mathrm{Mm}\cdot((\mathrm{M+m}) 
\pm\sqrt{(\mathrm{M-m})^2 + 
4\mathrm{mM}\mathrm{cos}^2\phi}\,)\,.\end{equation}    

Longitudinal and transverse waves are distinguished by the plus or minus 
sign in front of the square root in (28).

\section{The masses of the $\nu$-branch particles}

The characteristic case of the neutrino branch particles are the
$\pi^\pm$\,mesons which can be created in the process $\gamma$ + p
$\rightarrow \pi^-$ +  $\pi^+$  + p. A photon impinges on a proton and is
converted in $10^{-23}$ sec into a pair of particles of opposite charge. 
A simple example of the creation of a $\nu$-branch particle by strong
interaction is the case p + p $\rightarrow$  p + p + $\pi^-$  + $\pi^+$. 
Fourier analysis dictates that a continuum of frequencies must be in the 
collision products. The waves must be standing waves in order to be part 
of the rest mass of a particle. The $\pi^\pm$\,mesons decay via 
$\pi^\pm$ $\rightarrow$  $\mu^\pm$ + $\nu_\mu(\bar{\nu}_\mu)$ 
(99.98770\%) followed by 
$\mu^\pm \rightarrow$ e$^\pm$ + $\bar{\nu}_\mu(\nu_\mu)$ 
+ $\nu_e(\bar{\nu}_e)$ ($\approx$ 100\%). 
If the particles consist of the particles into which they decay, 
then the $\pi^\pm$\,mesons are made of neutrinos, antineutrinos
and e$^\pm$. Since neutrinos interact through the weak 
force which has a range of O(10$^{-16}$)\,cm according to p.25 of [22], 
and since the radius of the $\pi^\pm$\,mesons [30] is on the order of 
10$^{-13}$\,cm, \emph{the $\nu$-branch particles must have a lattice
with  N neutrinos}, N = 2.854\,$\cdot$\,10$^{\,9}$ being the same as in 
Eq.(17). It is not known with certainty that neutrinos actually have a 
mass as was originally suggested by Bethe [35] and Bahcall [36]
and what the values of m($\nu_e$) and m($\nu_\mu$) are. However, 
the results of the Super-Kamiokande [37] and the Sudbury [38] 
experiments indicate that the neutrinos have masses. Different masses 
of the electron neutrino, muon neutrino and tau neutrino guarantee that  
the three neutrino types are different. And, as the experiments show, 
they are different. Otherwise the neutrinos of the three types do not 
differ, they do not have charge and have the same spin.
   
The neutrino lattice must be \emph{diatomic}, meaning that the lattice 
points have alternately larger 
(m($\nu_\mu$)) and smaller (m($\nu_e$)) masses. We will retain the 
traditional term diatomic. \emph{The term neutrino lattice will refer to a 
lattice consisting of neutrinos and antineutrinos}. The lattice we consider  
is shown in Fig.\,2. Since the neutrinos have spin 1/2, the
 $\nu_\mu$,\,$\bar{\nu}_\mu$,\,$\nu_e$,\,$\bar{\nu}_e$ lattice is a 
four-Fermion lattice, which is required for the explanation of the weak 
decays. The first investigation of cubic Fermion lattices in context with 
the elementary particles was made by Wilson [13]. A neutrino lattice is 
electrically neutral. Since we do not know the interaction of the electron
with a neutrino lattice we cannot consider lattices with a charge.

    \begin{figure}[h]
    \hspace{3cm}
    \includegraphics{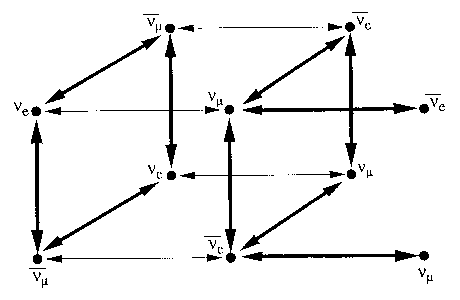}
    \vspace{-0.2cm}
    \begin{quote}
Fig.\,2: A cell in the neutral neutrino lattice of the $\pi^\pm$\,mesons. 
Bold lines mark the forces between neutrinos 
and antineutrinos. Thin lines mark the forces between  
either neutrinos only, or antineutrinos only.
    \end{quote}
    \end{figure}

The neutrino lattice oscillations take care of the continuum of 
frequencies which must, according
to Fourier analysis, be present after the high energy collision 
which created the particle. We will, for the sake of simplicity, first set 
the sidelength of the lattice at 10$^{-13}$\,cm, that means approximately 
equal to the size of the nucleon. The lattice then contains about 10$^{\,9}$ 
lattice points. The sidelength of the lattice does not enter Eq.(28) 
for the frequencies of diatomic oscillations. The calculation of the 
ratios of the masses  is consequently
independent of the size of the lattice, as was the case with the 
$\gamma$-branch. The size of the lattice can be explained with 
the pressure which the lattice oscillations exert on a crossection of 
the lattice. The pressure cannot exceed Young's modulus of the lattice.  
We require that the lattice is isotropic.

    \begin{figure}[h]
    \hspace{1.5cm}
    \includegraphics{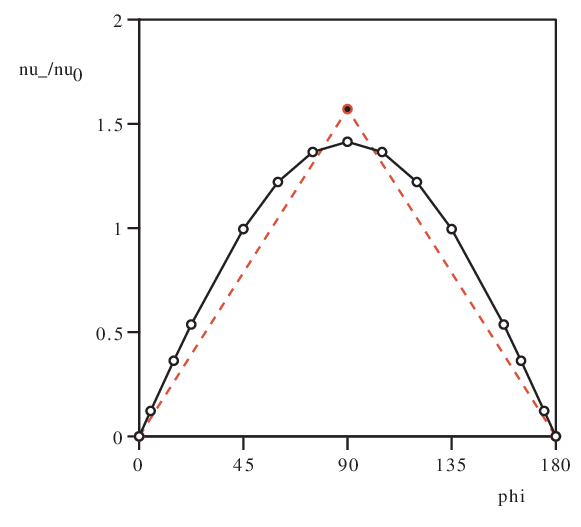}
    \vspace{-.5cm}
    \begin{quote}
Fig.\,3: The frequency distribution $\nu_-/\nu_0$ of the basic 
diatomic mode 
according to Eq.(28) with M/m = 100. The dashed line shows the 
distribution of the frequencies corrected for the group velocity. 
    \end{quote}
    \end{figure}

   From the frequency distribution of the diatomic oscillations 
(Eq.28), shown in Fig.\,3, follows the group velocity 
   $\mathrm{d}\omega/\mathrm{dk} 
= 2\pi\emph{a}\,\,d\nu/d\phi$\, at each point $\phi$. With $\nu = 
\nu_0f(\phi)$ and $\nu_0$ = $\sqrt{\alpha/4\pi^2\mathrm{M}}$ = 
c/2$\pi\emph{a}$ as in Eq.(10) we find

\begin{equation} \mathrm{c}_g = d\omega/dk = 
\emph{a}\sqrt{\alpha/\mathrm{M}}\cdot df(\phi)/d\phi\,,
\end{equation}
where M = m($\nu_\mu$). In order to determine the value of 
d$\omega/dk$ we have to know the value of 
$\sqrt{\alpha/\mathrm{M}}$. From Eq.(9) for $\alpha$ follows that 
$\alpha = \emph{a}\,c_{44}$ in the isotropic case with central forces. 
The group velocity is therefore

\begin{equation} \mathrm{c}_g = \sqrt{a^3c_{44}/\mathrm{M}}\cdot 
df(\phi)/d\phi\,.
\end{equation}

\noindent
In Eq.(29) we now set $\emph{a}\sqrt{\alpha/\mathrm{M}}$ = c,
as in Eq.(11), where c  is the velocity of light. It follows that

\begin{equation} \mathrm{c}_g = \mathrm{c}\cdot df(\phi)/d\phi\,,
\end{equation}

\noindent
as it was with the $\gamma$-branch. Equation (31) applies regardless 
whether we consider $\nu_+$ or $\nu_-$ in  
Eq.(28). That means that there are no separate transverse 
oscillations with their theoretically higher frequencies.

  The mass M of the heavy neutrino can be determined with lattice 
theory from Eq.(30), as we have shown in [12]. This involves the
inaccurately known compression modulus of the proton. We will, 
therefore, rather determine the mass  
of the muon neutrino with Eq.(33), which leads to m($\nu_\mu$) 
$\approx$  50\,milli-eV/c$^2$. It can be verified easily that
 m($\nu_\mu$) = 50\,milli-eV/c$^2$ makes sense.  
The energy in the rest mass of the $\pi^\pm$ mesons, which do not 
have spin, is\\

\hspace{3cm} m($\pi^\pm$)c$^2$ = 139.57\,MeV\,,\\

\noindent and we have N/4 = 0.7135$\cdot10^{\,9}$ muon neutrinos 
and the same number of antimuon neutrinos, each also with 
an energy of about 50\,milli-eV. It follows from Eq.(33) that the 
energy in the masses of all muon- and antimuon neutrinos in $\pi^\pm$   
is N/2\,$\cdot$\,m($\nu_\mu$)c$^2$  = 71.35\,MeV, that is 51\% of the 
energy in the rest mass of the $\pi^\pm$ mesons. In simple terms:
m($\pi^\pm$)/2 is approximately equal to the sum of the muon- and
antimuon neutrino masses in the $\pi^\pm$ lattice, or equal to 
N/2\,$\cdot$\,m($\nu_\mu$), if m($\nu_\mu$) = m($\bar{\nu}_\mu$), 
and if m($\nu_\mu$) is $\approx$ 50\,$\mbox{milli-eV}$/c$^2$. A very 
small part of m($\pi^\pm$)c$^2$ goes, as we will see, into the 
electron neutrino masses and into the electron, the rest of the energy 
in $\pi^\pm$ is in the lattice oscillations.
The energy in the rest mass of the $\pi^\pm$\,mesons is the sum 
of the oscillation energies plus the sum of the energy in the 
masses of the neutrinos, plus the energy in e$^\pm$.

\begin{itemize}\item
\emph{The $\pi^\pm$\,mesons are like cubic black bodies\\ 
filled with oscillating neutrinos}.
\end{itemize}

   For the sum of the energies of the frequencies of the oscillations 
in the diatomic lattice in the $\pi^\pm$ meson we use 
a modified form of Eq.(20). We replace the argument $\pi^0$  by 
$\pi^\pm$, and use the same N, there are N neutrinos in the $\pi^\pm$ 
lattice. But we assume that the neutrino oscillations are circular, that means
that there are 2N oscillations. We use the same $\nu_0$ = c/2$\pi$\emph{a}  
as for the $\gamma$-branch, but the limits of the integral are now \,-$\pi$/2 
and $\pi$/2, because in the diatomic case  
the increase of the frequencies ends at $\pi$/2, see Fig.\,3.
The modified equation for the energy of the standing waves 
in a diatomic lattice is given by Eq.(32). The energy of the circular 
neutrino oscillations in $\pi^\pm$ is then

\begin{eqnarray} \mathrm{E}_\nu(\pi^\pm) =  
\frac{\mathrm{2N\cdot h \nu_0}}{(\mathrm{e^{h\nu/kT}}\,\mathrm{-}\,1)}
\cdot\frac{1}{\pi}\int\limits_{-\pi/2}^{\pi/2}\phi\,d\phi  \cong   
1/2\cdot\mathrm{E}_\nu(\pi^0)  \nonumber\\
= \,67.82\,\mathrm{MeV} = 
0.486\,\mathrm{m}(\pi^\pm)\mathrm{c}^2(\emph{exp})\,.\end{eqnarray}
\noindent
The value of the integral in 
Eq.(32) for the diatomic frequencies $\nu = \nu_0\phi$ is 1/2 of 
the value $\pi^2$ of the same integral in the case of monatomic 
frequencies, because in the latter case the increase of the 
frequencies continues to $\phi$ = $\pi$, whereas in the diatomic case the 
increase of the frequencies ends at $\pi$/2, as can be seen on the plot of the 
diatomic frequencies in Fig.\,3. From Eq.(32) follows that  \emph{$\approx$
1/2 of the energy of $\pi^\pm$ is in the oscillation energy E$_\nu$$(\pi^\pm)$.}   

   In order to determine the sum of the masses of the neutrinos we
make use of E$_\nu(\pi^\pm)$ and obtain an approximate value of 
the sum of the masses of the neutrinos in $\pi^\pm$ from 
\begin{eqnarray} \mathrm{m}(\pi^\pm)\mathrm{c}^2 
\,-\, \mathrm{E}_\nu(\pi^\pm) = \sum_i\,[m(\nu_\mu) +
 m(\bar{\nu}_\mu) + m(\nu_e) + m(\bar{\nu}_e)]\mathrm{c}^2 \nonumber\\ 
=\mathrm{N}/2\cdot( m(\nu_\mu) + m(\nu_e))\mathrm{c}^2 = 
71.75\,\mathrm{MeV} = 1.0282\,\mathrm{m}(\pi^\pm)\mathrm{c}^2/2\,,
\end{eqnarray}
and find that \emph{$\approx$ 1/2 of the energy of $\pi^\pm$  is in 
the neutrino masses}. 

   The sum of the energy of the masses of all neutrinos in $\pi^\pm$, 
Eq.(33), plus the oscillation energy, Eq.(32), neglecting the energy in 
e$^\pm$ gives, after division by c$^2$,
the theoretical \emph{rest mass of the $\pi^\pm$\,mesons} without charge,
or of the $\pi^\pm$ lattice,\\
 
\hspace{3cm} m($\pi^\pm$) = N\,$\cdot$\,(m($\nu_\mu$) + m($\nu_e$))\,,    
\hspace {3.4cm}(33a)\\

\noindent 
assuming that m($\nu_\mu$) = m($\bar{\nu}_\mu$) and m($\nu_e$) = 
m($\bar{\nu}_e$). From Eq.(33a) follows that, since we 
use m($\pi^\pm)$  in the determination of the neutrino masses with
Eq.(33b), that (33a) is equal to the experimental rest mass m($\pi^\pm$) =
139.57\,MeV/c$^2$.

   If m($\nu_e$) $\ll$ m($\nu_\mu$) and m($\nu_\mu$)
 = m($\bar{\nu}_\mu$), as we will justify later with Eqs.(68,71), we arrive 
from Eq.(33) with $\Sigma_i$\,m($\nu_\mu$) = N/2\,$\cdot$\,m($\nu_\mu$) 
$\approx$ 71.75\\MeV/c$^2$, 
and with N/2 = 1.427$\cdot10^{\,9}$, at an approximate value for\\

\hspace{3cm} \emph{the mass of the muon neutrino}\\
 
\hspace{4cm} m($\nu_\mu$) $\approx$ 50\,milli-eV/$\mathrm{c}^2$.
\hspace{3.2cm} (33b)\\

\noindent
An accurate value of m($\nu_\mu$) will be given later, Eq.(70). The 
existence of the neutrino masses is the subject of Section 10.

Since nothing else but the charge e$^\pm$ contributes
to the rest mass of $\pi^\pm$ it appears that in a good 
approximation the oscillation energy in $\pi^\pm$ is equal to the
energy in the sum of the neutrino masses in $\pi^\pm$, that means that 
\begin{equation} \mathrm{E}_\nu(\pi^\pm) \cong 
\Sigma_i\,\mathrm{m(\nu_i)c^2} = 
\mathrm{N}/2\cdot(\mathrm{m}(\nu_\mu) + 
\mathrm{m}(\nu_e))\mathrm{c}^2
\cong 1/2\cdot\mathrm{m(\pi^\pm)c^2}\,.  \end{equation}
This applies only to the neutral neutrino lattice of the pion, the 
consequences of the charge of $\pi^\pm$ have not been considered.

   A cubic lattice and conservation of neutrino numbers during the reaction
$\gamma$  + p $\rightarrow \pi^+ + \pi^-$  + p  \emph{necessitates} 
that the $\pi^+$ and $\pi^-$  lattices contain just as many
neutrinos as antineutrinos. If the lattice is cubic it
must have a \emph{center neutrino} (Fig.\,4). Conservation of neutrino numbers
requires furthermore that the center neutrino of  $\pi^+$ is matched  
by an antineutrino in $\pi^-$. In the decay sequence of (say)
the $\pi^-$\,meson $\pi^- \rightarrow  \mu^- + \,\bar{\nu}_\mu$ and 
$\mu^- \rightarrow$ e$^-$ + $\bar{\nu}_e + \nu_\mu$ an electron
neutrino $\nu_e$ does not appear. But since (N\,-\,1)/4 electron 
neutrinos  $\nu_e$ must be in the cubic $\pi^-$ lattice it follows that 
(N\,-\,1)/4 electron neutrinos must go with the electron emitted 
in the $\mu^-$ decay.

\begin{figure}[h] 
    \hspace{2.2cm}
    \includegraphics{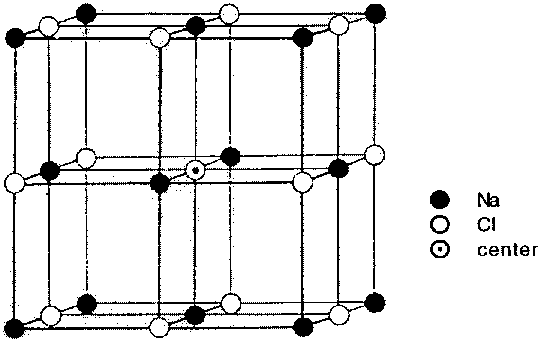}
    \begin{quote}
Fig.\,4: The center of a NaCl lattice. (After Born and Huang).
    \end{quote}
\end{figure}

    We must now be more specific about N, which is an odd 
number, \,27 in the case of Fig.\,4, because a cubic lattice  
has a center particle, just as the NaCl lattice.
In the $\pi^\pm$\,mesons are then (N\,-\,1)/4 muon 
neutrinos $\nu_\mu$ and the same number of 
antimuon neutrinos $\bar{\nu}_\mu$, as well as 
(N\,-\,1)/4 electron neutrinos $\nu_e$ and the same number  
of anti-electron neutrinos $\bar{\nu}_e$, \emph{plus} a center  
neutrino or antineutrino. We replace
N\,-\,1 by N$^\prime$. But N$^\prime$ differs from N by 
only one in $10^{\,9}$, so N$^\prime$ $\cong$ N, and we 
will in the following use N instead of N$^\prime$. Although the 
numerical difference between N and N$^\prime$ is negligible we 
cannot consider any integer number N because that would mean 
that there would be fractions of a neutrino. N/2\,$\cdot$\,$\nu_e$,
for example, is strictly speaking incorrect.
   
   The \emph{antiparticle} of the $\pi^+$\,meson is the particle in   
which all frequencies of the 
neutrino lattice oscillations have been replaced by frequencies with 
the opposite sign, all neutrinos replaced by their antiparticles and the 
positive charge replaced by the negative charge. If, as we will show, the 
antineutrinos have the same mass as the neutrinos it follows that the 
antiparticle of the $\pi^+$\,meson has the same mass as $\pi^+$ but 
opposite charge, i.e. is the $\pi^-$\,meson. As we will see, the 
explanation of the mass of the $\pi^\pm$\,mesons opens the door to 
the explanation of the mass of the muon and of the electron. 

   The decay $\pi^\pm$ $\rightarrow$ e$^\pm$$\nu_e$($\bar{\nu_e}$)
(1.230\,$\cdot$\,10$^{-4}$) shows that equal numbers of neutrinos 
and antineutrinos, as well as a single center neutrino, plus an 
electric charge e$^\pm$ are in the $\pi^\pm$\,meson. In the decay 
the neutrinos and antineutrinos annihilate, but for the center neutrino,
which is conserved as conservation of neutrino numbers requires. The 
charge e$^\pm$ is, of course, conserved too. Our explanation
of the mass of the $\pi^\pm$\,mesons with a cubic neutrino lattice plus
a charge e$^\pm$ leads also to the explanation of the absence
of spin of the $\pi^\pm$\,mesons, see Section 15.

   To summarize what we have learned about the $\pi^\pm$\,mesons. The
$\pi^\pm$\,me-sons consist of a cubic lattice of muon neutrinos, electron 
neutrinos and their antiparticles, plus a charge e$^\pm$. The neutrinos 
are held together by the weak nuclear force. The mass of the $\pi^\pm$
mesons so determined is, without considering the consequences of the charge 
for the mass of $\pi^\pm$, equal to the sum of the masses of the N neutrinos 
in the $\pi^\pm$ lattice, plus the mass in the energy of the lattice 
oscillations, or equal to N\,$\cdot$\,(m($\nu_\mu$) + m($\nu_e$)).

\medskip  
 
   Now we turn to the K\,mesons. The mass of K$^\pm$  is 
m(K$^\pm$) = 493.677 MeV/c$^2$ = 0.8843\,$\cdot$\,4m($\pi^\pm$)  = 
3.5372\,$\cdot$\,m($\pi^\pm$). The K$^\pm$ mesons do not have spin.  
The primary decay of the K$^\pm$\,mesons, 
K$^\pm \rightarrow \mu^\pm + \nu_\mu(\bar{\nu}_\mu)$ (63.55\%),  
leads to the same end products as the $\pi^\pm$\,meson decay 
$\pi^\pm \rightarrow \mu^\pm + \nu_\mu(\bar{\nu}_\mu)$ (99.98\%). 
From this and the decay of the $\mu^\pm$\,muons we learn that the 
K$^\pm$\,mesons must, at least partially, be made of the same four 
neutrino types as in the $\pi^\pm$\,mesons, namely of muon neutrinos, 
antimuon neutrinos, electron neutrinos and anti-electron neutrinos and 
their oscillation energies. The concept that a neutrino lattice, and a 
charge e$^\pm$, is in the K$^\pm$\,mesons is supported by 
the decay K$^\pm$ $\rightarrow$ e$^\pm$$\nu_e(\bar{\nu_e}$)
(1.581\,$\cdot$\,10$^{-5}$), 
in which the entire neutrino lattice dissolves, but for one neutrino, 
as it was in the corresponding decay of the $\pi^\pm$\,mesons.   
However the K$^\pm$\,mesons cannot be 
solely the second mode of the lattice oscillations of the 
$\pi^\pm$\,mesons, because the second mode of the 
$\pi^\pm$\,mesons has an energy of 
\begin{eqnarray}
{ \mathrm{E}((2.)\pi^\pm) = 4\mathrm{E}_{\nu}(\pi^\pm) + 
\mathrm{N}/2\cdot(\mathrm{m}(\nu_\mu) + 
\mathrm{m(\nu_e))\,c^2}\nonumber }\\
\cong 2\mathrm{m(\pi^\pm)c^2} + 1/2\cdot 
\mathrm{m(\pi^\pm)c^2} = 348.92\,\mathrm{MeV}\,,\end{eqnarray}

\noindent with E$_\nu$((2.)$\pi^\pm$) = 4E$_\nu$($\pi^\pm$) = 
2m$(\pi^\pm$)c$^2$ and $\Sigma_i$m($\nu_i$) =  N/2\,$\cdot$\,(m($\nu_\mu$) 
+ m($\nu_e$)) $\cong$ m($\pi^\pm$)/2 from Eqs.(33,34). The 348.9\,MeV are 
the energy in the second or (2.)\,\,mode of the $\pi^\pm$ mesons, which 
fails m(K$^\pm$)c$^2$ = 493.7\,MeV by a wide margin.

   The concept that the K$^\pm$\,mesons are alone a higher mode of the\\  
$\pi^\pm$\,mesons also contradicts our point that the particles consist of  
the particles into which they decay. The decays K$^\pm$ $\rightarrow$ 
$\pi^\pm \,+ \,\pi^0$ (20.66\%), as well as K$^+$ $\rightarrow$ $\pi^0$ + 
e$^+ + \nu_e$ (5.07\%), called K$^+_{e3}$, and 
K$^+$ $\rightarrow$ $\pi^0 + \mu^+ + \nu_\mu$ (3.53\%), called 
K$^+_{\mu3}$, make up 29.26\% of the K$^\pm$\,meson
decays. A $\pi^0$\,meson figures 
in each of these decays. If we add the energy in the rest mass of  
a $\pi^0$\,meson m($\pi^0$)c$^2$ = 134.97\,MeV to the 
348.9\,MeV in the second mode of the $\pi^\pm$\,mesons then 
we arrive at an energy of 483.9\,MeV, which is 98.0\% of 
m(K$^\pm$)c$^2$. Therefore we conclude that the 
K$^\pm$\,mesons consist of the second 
mode of the $\pi^\pm$\,mesons $\emph{plus}$ a $\pi^0$\,meson, 
or are the state (2.)$\pi^\pm$ + $\pi^0$. Then 
it is natural that $\pi^0$\,mesons from the $\pi^0$ component 
in the K$^\pm$\,mesons are among the decay products of the 
K$^\pm$\,mesons.
 
   In qualitative terms: The energy in the measured mass of the 
K$^\pm$\,mesons is about 3.5 times the energy in the mass of the pions. 
The second harmonic of the pion lattice oscillations contains 4 times 
the energy of the basic oscillation, Eq.(35). This adds up to two times 
the energy in the pion mass. To the 
2m($\pi^\pm)$c$^2$ caused by the second mode we must add  
the energy in the masses of the neutrinos in a $\pi^\pm$\,meson,
another 1/2 m($\pi^\pm$)c$^2$. Finally the energy in the mass 
of an additional $\pi^0$\,meson, $\approx$ m($\pi^\pm)$c$^2$, has to 
be added, as we suggested. So we arrive at m(K$^\pm$)c$^2$ 
$\cong$ (4$\times$1/2  + 1/2 + 1)m($\pi^\pm$)c$^2$ 
= 3.5\,$\cdot$\,m($\pi^\pm$)c$^2$. If m(K$^\pm$)c$^2$ = 
3.54\,$\cdot$\,m($\pi^\pm$)c$^2$, then 3.54/4 = 0.885 is practically
equal to the factor 0.884 before 4$\pi^\pm$ in the mass ratio 
m(K$^\pm$)/m($\pi^\pm$) in Table\,3.

\medskip

   The average factor 0.85 $\pm$ 0.025, which appears in Eq.(3)
for the ratios of the masses  
of the particles of the $\nu$-branch to the mass of the 
$\pi^\pm$\,mesons, is a consequence of the neutrino masses. 
They make it impossible that the ratios of the particle masses are  
outright integer multiples, because the particles consist of the 
energy in the neutrino oscillations, plus the energy in the neutrino 
masses, which are independent of the order of the lattice oscillations.

\bigskip    

   The K$^0$,$\overline{{\mathrm{K}}^0}$\,mesons have a rest 
mass m(K$^0$,$\overline{{\mathrm{K}}^0}$) = 
497.614\,MeV/c$^2$ = \\
1.00809\,$\cdot$\,m(K$^\pm$). They do not have spin.
We obtain the K$^0$\,meson if we superpose onto the 
second mode of the $\pi^\pm$\,mesons instead of 
a $\pi^0$\,meson a basic mode of the $\pi^\pm$\,mesons, 
with a charge opposite 
to the charge of the second mode of the $\pi^\pm$\,meson. 
The K$^0$ and $\overline{{\mathrm{K}}^0}$\,mesons, 
or the state (2.)$\pi^\pm$ + $\pi^\mp$, is made  
of neutrinos and antineutrinos only, without a photon component,
because the second mode of $\pi^\pm$ as well as the basic mode 
$\pi^\mp$ consist of neutrinos and antineutrinos only, neglecting 
the charge. That means that the lattice of the K$^0$
mesons consists of N neutrino \emph{dipoles}. 

   The K$^0$\,meson has a measured mean square charge radius
$\langle$r$^2$$\rangle$ = $-$\,0.077 $\pm$ 0.010\,fm$^2$, 
according to the Review of Particle Physics or to [39],
whereas the K$^\pm$ mesons have a charge radius $\langle$r$\rangle$
= 0.560 fm. This means that $\langle$r$\rangle$(K$^0$) = 
$\langle$r$\rangle$(K$^\pm$)/2.018. 
This can only be if there are \emph{two charges of opposite 
sign} within K$^0$, as this model implies. But the K$^0$
meson does not have a magnetic moment, because the K$^0$
meson does not have spin. Since the mass of a 
$\pi^\pm$\,meson is by 4.59\,MeV/c$^2$ larger than
m($\pi^0$), the mass of K$^0$ should be larger  
than m(K$^\pm$), if m(K$^0$) is m((2.)$\pi^\pm$ + $\pi^\mp$).
And indeed m(K$^0$)\,$-$\,m(K$^\pm$) = 
3.937\,MeV/c$^2$ according to [2]. Similar differences occur 
with m(D$^\pm$)\,$-$\,m(D$^0$) and 
m($\Xi_c^0$)\,$-$\,m($\Xi_c^+$). The decay 
K$^0_S \rightarrow \pi^++ \pi^-$ (69.20\%) creates directly the 
$\pi^+$ and $\pi^-$ mesons which are part of the (2.)$\pi^\pm$ + 
$\pi^\mp$ structure of K$^0$ we have suggested.
The decay K$^0_S \rightarrow \pi^0 + \pi^0$ (30.69\%) is similar 
to the 2$\gamma$ branch of electron positron annihilation. 
Both decays account for 99.89\% of the decays of K$^0_S$,\,\,see
Table\,4.

    \begin{table}\caption{The mass ratios and decays of K$^0$} 
    \begin{tabular}{llllrcl}\\

\hline\hline\\
 & m/m($\pi^\pm$) & multiples & decays
 & fraction & spin & lifetime  
 \\
 & & & & (\%) & & sec\\
[0.5ex]\hline
\\
K$^0_S$ & 3.56533 & 0.89133\,$\cdot$\,\,4$\pi^\pm$ & $\pi^+\,\pi^-$ & \,\,69.20
& 0 & 0.8953$\cdot10^{-10}$\\
 & & & $\pi^0\,\pi^0$  & \,\,30.69\\
\\
K$^0_L$ & 3.56533 & 0.89133\,$\cdot$\,\,4$\pi^\pm$ & $\pi^\pm$e$^\mp$$\nu_e$
 & \,\,40.55 & 0 & 5.116$\cdot10^{-8}$\\
 & & & $\pi^\pm$$\mu^\mp$$\nu_\mu$ & \,\,27.04\\
 & & & 3$\pi^0$ & \,\,19.52\\
 & & & $\pi^+$\,$\pi^-$\,$\pi^0$ & \,\,12.54\\
[0.3cm]\hline\hline

\vspace{0.1cm}
    \end{tabular}

    \end{table}

   The decay K$^0_L \rightarrow 3\pi^0$ (19.52\%) is similar 
to the 3$\gamma$ branch of electron positron annihilation. 
The two decays of K$^0_ L$ called K$^0_{\mu3}$ into 
$\pi^\pm\,\mu^\mp\,\nu_\mu$ (27.04\%) 
and K$^0_{e3}$ into $\pi^\pm$\,e$^\mp\,\nu_e$ 
(40.55\%), which together make up 67.59\% of the
K$^0_{L}$ decays, apparently originate from the decay of  
the second mode of the $\pi^\pm$\,mesons in the 
K$^0$ structure, either into $\mu^\mp$ + $\nu_\mu$  
or into e$^\mp$ + $\nu_e$. The same types of decay, 
apparently tied to the (2.)$\pi^\pm$ mode, accompany also the 
K$^\pm$  decay K$^\pm$ $\rightarrow \pi^\pm\pi^0$ (20.66\%)
in which, however, a $\pi^0$\,meson in  the K$^0_S$ decay 
replaces the $\pi^+$\,meson in the K$^\pm$  decay. Our rule 
that the particles consist of the particles into which they decay also 
holds for the K$^0$ and $\overline{{\mathrm{K}}^0}$\,mesons. The
explanation of the K$^0$,$\overline{{\mathrm{K}}^0}$\,mesons with the
state (2.)$\pi^\pm$ + $\pi^\mp$ confirms that the state (2.)$\pi^\pm$ +
$\pi^0$ was the correct choice for the explanation of the K$^\pm$\,mesons.
The state (2.)$\pi^\pm$ + $\pi^\mp$ is also crucial for the explanation 
of the absence of spin of the K$^0$,$\overline{{\mathrm{K}}^0}$\,mesons, 
as we will see in Section 15.   

\vspace{0.5cm}   
   The neutron, with mass m(n) = 939.565MeV/c$^2$ = 
0.95156\,$\cdot$\,2m(K$^\pm$)/c$^2$ \\= 0.9440\,$\cdot$\,2m(K$^0$)/c$^2$ 
and spin 1/2,  
is either the superposition of a K$^+$ and a K$^-$\,meson or of   
a K$^0$\,meson and a $\overline{{\mathrm{K}}^0}$\,meson. As
has been shown in [67], the spin rules out that the
neutron is the sum of a K$^+$ and a K$^-$\,meson. On the 
other hand, the neutron can be the superposition of a K$^0$   
and a $\overline{{\mathrm{K}}^0}$\,meson. This guarantees that 
the neutron is neutral, and that its lattice consists of neutrinos 
and antineutrinos only, without a photon component. The neutron lattice 
contains then at each of the N lattice points a 
$\nu_\mu,\bar{\nu}_\mu,\nu_e,\bar{\nu}_e$ \emph{neutrino quadrupole}, 
because in each K$^0$  and $\overline{{\mathrm{K}}^0}$\,meson
are neutrino pairs at the lattice points. There is also a single 
quadrupole of positive and negative charges e$^\pm$, because each K$^0$  
and $\overline{{\mathrm{K}}^0}$\,meson carries a pair of opposite 
charges e$^+$ and e$^-$. Opposite charges must be in 
the neutron, because it has a mean square charge radius
$\langle$r$^2\rangle$ = $-$ 0.1161\,fm$^2$ [2] and a magnetic moment. 
The lattice oscillations
in the neutron must be  coupled pairs of oscillations in 
order for the neutron to have spin 1/2, just as the $\Lambda$ 
baryon with spin 1/2 is a superposition of two $\eta$\,\,mesons. With 
m(K$^0$)(\emph{theor}) = m(K$^\pm$) + 4\,MeV/c$^2$ =
487.9\,MeV/c$^2$
from above it follows that m(n)(\emph{theor}) $\approx$ 
2m(K$^0)$(\emph{theor}) $\approx$ 975.8\,MeV/c$^2$ 
= 1.04\,m(n)(\emph{exp}).

\bigskip 

   The proton, with mass m(p) = 0.99862\,m(n) and spin 1/2, does 
not decay and does not tell which particles it is made of. 
However, we learn about the structure of the proton through the 
decay of the neutron n $\rightarrow$ p + e$^- + \bar{\nu}_e$ (100\%). 
A proton, an electron and one single anti-electron neutrino is emitted when  
the neutron decays and, according to [2], 1.29333 MeV are released. As it 
appears all N anti-electron neutrinos,
actually (N-1)/2\,$\cdot$\,$\bar{\nu}_e$ and (N-1)/2\,$\cdot$\,$\nu_e$ 
plus one $\bar{\nu}_e$, (N is odd according to Fig.\,4), are removed from the 
structure of the neutron in the neutron decay and converted into
the kinetic energy of the decay products. This type of 
process will be explained in Section 9. On the other hand,
it is certain that the proton has a neutrino lattice because the 
neutron has a neutrino lattice. One half of the mass of the proton
is equal to the sum of the masses of the neutrinos in the neutrino
lattice, the other half of the mass of the proton is in the energy of
the neutral neutrino oscillations. This is just as it is with the other 
particles of the $\nu$-branch, e.g. the $\pi^\pm$\,mesons. We neglect
the very small contribution of the mass of the charges e$^+$e$^-$e$^+$,
because m(p)/m(e) = 1836. The proton carries a net positive 
charge e$^+$e$^-$e$^+$, because the neutron carries an e$^+$e$^-$e$^+$e$^-$ 
quadrupole, of which one e$^-$ is lost in the $\beta$-decay.
The concept that the proton carries, just as the electron, a single 
but positive charge 
has been abandoned a long time ago, when it was said that the 
proton consists of three quarks carrying fractional electric charges. 
Each of the three charges e$^\pm$ in the proton has a magnetic moment,
all of them point in the same direction, because the spin
of the one e$^-$ must be opposite to the spin of the two e$^+$,
in order for the spin of the proton to be 1/2.
Each magnetic moment of the charges e$^\pm$ has a g-factor 
$\cong$\,\,2. All three charges in the proton must then have
a g-factor $\approx$\,\,6, whereas the measured g-factor of the 
magnetic moment of the proton is g(p) = 5.585 = 0.93\,$\cdot$\,6.

   It is obvious and fundamental that the proton is stable. In the 
literature the stability of the proton is blamed on conservation of the
baryon number, which does not provide insight into the reason for which 
the proton
does not decay. Actually one must wonder why the proton does not decay
in the process p $\rightarrow$ e$^+$ + $\nu_e$, just as the 
$\pi^+$\,meson can decay into $\pi^+$ $\rightarrow$ e$^+$ + $\nu_e$, which
is an experimental fact [2]. The lattice of the proton, which consists of
equal numbers of neutrinos and antineutrinos plus an extra electron neutrino,
could apparently dissolve conserving charge, the neutrino numbers and the 
extra electron
neutrino, just as in the case  $\pi^+$ $\rightarrow$ e$^+$\,$\nu_e$. When
the three charges e$^+$e$^-$e$^+$ in the proton are converted into e$^+$, 
conservation of charge is followed as well. So p $\rightarrow$ 
e$^+$\,$\nu_e$ should be possible. However, contrary to $\pi^+$ 
$\rightarrow$ e$^+$\,$\nu_e$ there is spin on the left side of
p $\rightarrow$ e$^+$\,$\nu_e$, because the proton has spin, whereas in
$\pi^+$ $\rightarrow$ e$^+$\,$\nu_e$ the $\pi^+$\,meson does not have spin.
The spin of the proton originates from 
the circular lattice oscillations in the proton, on the other hand, the 
linear lattice oscillations in $\pi^+$ do not cause spin.
The spin, or \emph{conservation of angular momentum rules out the decay}
 p $\rightarrow$ e$^+$\,$\nu_e$. Similarly, a decay analogous to $\pi^\pm$
$\rightarrow$ $\mu^+$ + $\nu_\mu$ (99.98\%) would not work, or   
decays as the decays of the K mesons. In our model of
the particles \emph{the proton is stable}. The Review of Particle Physics
lists about twenty possible proton decays, which are not necessarily possible,
because that list was made without knowledge of the structure of the proton.

  The D$^\pm$\,mesons with m(D$^\pm$) = 0.9954\,(m(p) + 
m($\bar{\mathrm{n}}$)) and spin s = 0
are the superposition of a proton and an anti-neutron of opposite 
spin or of their antiparticles. The superposition 
of a proton and a neutron with the same spin creates the deuteron 
with spin 1 and a mass m(d) = 0.9988\,(m(p) + m(n)). The deuteron 
has an oscillating neutrino lattice. The binding energy
of the deuteron must come out of the sum of the oscillation energies 
in the proton and neutron. The D$_s^\pm$\,mesons seem to be made of 
a body centered cubic lattice (Fig.\,5), as described in [40].
\begin{figure}[h] 
    \hspace{3cm}
    \includegraphics{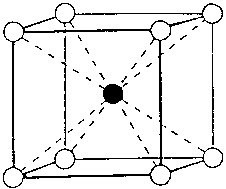}
    \begin{quote}
Fig.\,5: A body-centered cell. (After Born and Huang). 
\indent\qquad\qquad\,\,\,\,In the center of a D$^\pm_s$ cell is a 
$\tau$\,neutrino, in the corners are\\   
\indent $\nu_{\mu}$,\,$\bar{\nu}_\mu$,\,$\nu_e$,$\bar{\nu}_e$ 
neutrinos, as in Fig.\,2.\\ 
    \end{quote}\,
\end{figure}

\bigskip

   Summing up: The particles of the $\nu$-branch consist of lattices of
oscillating neutrinos and antineutrinos, 
and one or more positive and/or negative charges 
e$^\pm$. The characteristic feature of the $\nu$-branch particles is 
their cubic lattice consisting of $\nu_\mu,\bar{\nu}_\mu,\nu_e,\bar{\nu}_e$ 
neutrinos. The rest mass of the $\nu$-branch particles is the sum 
of the masses of the neutrinos and antineutrinos in the lattice, plus the 
mass in the energy of the  lattice oscillations, plus the mass in the 
charges e$^\pm$. The existence of the neutrino
lattice is a necessity if one wants to explain the spin, or the absence
of spin, of the $\nu$-branch particles. \emph{We do not use hypothetical 
particles} for the explanation of the $\nu$-branch particles.
 
\begin{center}
\section{The weak force in the interior of the particles}
\end{center}

\noindent
After we have explained the masses of the stable mesons and baryons 
with cubic lattices consisting of either photons or of  neutrinos,
we can now determine the strength of the weak and the strength 
of the strong nuclear forces. Both are 70 years old puzzles.
 We will use lattice theory to determine the strength of the weak 
force which holds the lattices of the elementary particles together. 
We will then show that the strong force between two elementary 
particles is nothing but the sum of the unsaturated weak forces 
emanating from the lattice points at the surface of the lattice. 

In order to determine the force in the interior of the cubic lattices 
with which we have explained the particle masses we will,
as we have done before in [23], use a classical paper by Born and 
Land\'{e}  [41], (B\&L), dealing with the potential and compressibility 
of regular ionic crystals. It is essential to realize that,
\bigskip

\begin{itemize}
\item
 \emph{for the existence of a cubic lattice it is necessary that the force 
between the lattice points has an attractive part and a repulsive part}.
\end{itemize}
 
\hspace{0.5cm}\emph{Otherwise the lattice would not be stable and collapse}.\\

For the ionic crystals considered by B\&L the Coulomb force between the ions
is the attractive force, whereas the repulsive force originates from the electron 
clouds surrounding the ions. When the electron clouds of the ions approach
each other during the lattice oscillations they repel each other. The magnitude 
of the repulsive force is not known per se and has to be determined from the 
properties of the crystal.

We follow exactly the procedure in B\&L in order to see whether their 
theory is also applicable to a cubic lattice made of neutrinos. In this case the
Coulomb force is, of course, irrelevant. As B\&L do, we say that the potential 
of a cell of the neutrino lattice has an attractive part $-$ a/$\delta$ and a 
repulsive part + b/$\delta^\mathrm{\,n}$ with the unknown exponent n. $\delta$ 
is the distance in the direction between two neutrinos of the same type, either
 muon neutrinos and anti-muon neutrinos or electron neutrinos and 
anti-electron neutrinos.

   The potential of a cell in an ionic  cubic  lattice is of the form
\begin{equation}
\phi = -\,\mathrm{a}/\delta + \mathrm{b}/\delta^\mathrm{\,n}\,, \end{equation}
Eq.(1) of B\&L. The constant b is eliminated with the equilibrium condition 
d$\phi$/d$\delta$ = 0. Consequently 
\begin{equation} \mathrm{b} = \frac{\mathrm{a}}{\mathrm{n}}
\,\delta_0^{\mathrm{\,n} -1}\,, \end{equation}
where $\delta_0$ is the lattice constant.
The unknown exponent n of $\delta$ in Eq.(36) was determined by B\&L with
the help of the compression modulus $\kappa$, which is known for ionic 
crystals. $\kappa$ is defined by 
\begin{equation} \kappa = -\,1/V\cdot dV/dp\,, \end{equation}
with the volume V.

   The compression modulus  of an ionic lattice is given by  
\begin{equation} \kappa = 9\,\delta_0^4/\mathrm{a}\,(\mathrm{n} -1)\,, 
\end{equation}
Eq.(4) in B\&L.
The interaction constant a of the Coulomb force in cubic ionic
crystals resulting from the contributions of all ions of a lattice on a 
single ion is given by Eq.(5) of B\&L, it is 
\begin{equation}\mathrm{a} = 13.94\,\mathrm{e}^2 = 
3.2161\cdot10^{-18}\,\mathrm{erg\cdot cm}\,,\end{equation}
where e is the elementary electric charge. This equation is fundamental for
the theory of ionic lattices and is based on an earlier 
paper by Madelung [42]. Consequently we find that
\begin{equation} (\mathrm{n} -1) = 10.33\,\mathrm{r}_0^4/\mathrm{e}^2\kappa\,, 
\end{equation}
where r$_0$ = $\delta_0$/2 is the distance between a pair of neighboring
Na and Cl ions. For the alkali-halogenids, such as NaCl or KCl, B\&L found
that n $\approx$ 9. If n = 9 is used in Eq.(39) to determine theoretically the 
compression modulus $\kappa$, then the theoretical values of $\kappa$ 
agree, in a first approximation, with the experimental values of $\kappa$,
thus confirming the validity of the ansatz for the potential in Eq.(36).

   We now apply Eq.(41) to the neutrino lattice of the elementary particles in
order to determine the potential in the interior of the particles. We must 
use for r$_0$  the distance between two neighboring neutrinos in the lattice, 
which is equal to the range of the weak nuclear force. The range of the weak
force is 1$\cdot10^{-16}$ cm, as in Eq.(8), and so we have
\begin{equation} \mathrm{r}_0 = 1\cdot 10^{-16}\,\mathrm{cm}\,. \end{equation}
We have used this value of r$_0$ throughout our 
explanation of the masses of the mesons and baryons, though r$_0$ 
was previously designated by the symbol \emph{a}. We must, furthermore, 
replace e$^2$ by the interaction constant g$_w^2$ of the weak force which
holds the nuclear lattice together. According to p.\,25 of Perkins [22] 
\begin{equation} \mathrm{g}_w^2 = 4\pi\hbar\,\mathrm{c}\cdot 1.02\cdot10^{-5}\,
(\mathrm{M}_W/\mathrm{M}_p)^2\,,  \end{equation}
where M$_W$ is the mass of the W boson, M$_W$ = 80.399\,GeV/c$^2$,
and M$_p$ is the mass of the proton, M$_p$ = 0.938\,272\,GeV/c$^2$.
That means that 
\begin{equation} \mathrm{g}_w^2 = 2.9758\cdot 10^{-17}
\,\mathrm{erg\cdot cm}. \end{equation}

   We must also use the compression modulus $\kappa$ of the nucleon.
The value of the compression modulus of the nucleon  has been determined
theoretically by Bhaduri et al. [43], following earlier theoretical and experimental investigations of the compression moduli of nuclei. Bhaduri et al. found that the
compression modulus K$_\mathrm{A}$(1) of the nucleon ranges from 900 to 
1200\,MeV, or is 940\,MeV  or  900\,MeV for particular sets of parameters. 
We determine $\kappa$ of the nucleon with
\begin{equation} \kappa = 9/\rho_\#\mathrm{K}_\mathrm{nm}\,,
\end{equation}
from Eq.(1) in [43], with the number density $\rho_\#$ being the number 
density per fm$^3$. Bhaduri et al. write that ``the compression modulus
K$_\mathrm{nm}$ of nuclear matter is calculated by considering the 
nucleons as point particles", which they are not. Other assumptions are also 
sometimes made such as infinite nuclear matter, periodic boundary conditions, 
etc. Recent theoretical studies of the compressibility of ``nuclear matter" 
[44,45,46]
place the compressibility K$_\mathrm{{nm}}$ at values from between 
250 to 270 MeV, not much different from what is was twenty years earlier
in [43]. Considering the uncertainty of K$_\mathrm{nm}$ it seems to be 
justified to set, in the case of the nucleon, K$_\mathrm{nm}$ = 
K$_\mathrm{A}$, where K$_\mathrm{A}$ is defined as the compression 
modulus for a finite system with A nucleons. It then follows from Eq.(45)
with the radius of the nucleon R$_0$ = 0.88\,$\cdot$\,10$^{-13}$\,cm, 
and with 1\,MeV = 1.6022\,$\cdot$\,10$^{-6}$\,erg, that the compression 
modulus of the nucleon is  
\begin{equation} \kappa_n = 1.603\cdot10^{-35}\,\,\mathrm{cm^2/dyn}\,,
\end{equation}
if we use for K$_\mathrm{A}$(1) the value 1000\,MeV. We will keep in mind 
that $\kappa_n$ is not very accurate, because K$_\mathrm{A}$(1) is not very
accurate.

   If we insert (42), (44), and (46) into n $-$ 1 = 10.33\,r$_0^4$/e$^2\kappa$
(Eq.41) we find an equation for the exponent  n of the term 
b/$\delta^\mathrm{n}$  in the repulsive part of the potential in a nuclear 
lattice, 

\begin{equation} \mathrm{n} = 1 + 2.164\cdot10^{-12} = 1 + \epsilon\,,
\end{equation}
with r$_0^4$ = O(10$^{-64}$), g$_w^2$ = O(10$^{-17}$) and $\kappa$ = 
O(10$^{-35}$).
 
\bigskip

   With Eq.(37) that means that
 
\bigskip
\emph{the potential $\phi$ in the interior of an elementary 
particle is given by}
\begin{equation} \phi = - \,\frac{\mathrm{a}}{\delta} + \frac{\mathrm{b}}
{\delta^{\,1 + \epsilon}} = \frac{\mathrm{a}}{\delta}\,[\frac{(\delta_0/\delta)\,^\epsilon}{\mathrm{n}} - 1\,]\,,
\end{equation}
which we can reduce with n $-$ 1 = $\epsilon$, neglecting a term multiplied 
by $\epsilon\,^2$ = O(10$^{-24}$), using also a = 13.94\,e$^2$ (Eq.40) 
and a$^x$  $\cong$ 1 + x\,ln\,a + ..., to
\begin{equation} \phi \cong -\,\frac{\mathrm{a}\,\epsilon}{\delta}\,
[1 - \mathrm{ln}(\delta_0/\delta)] \cong -\,\frac{13.94\,\mathrm{g}_w^2\,\epsilon}
{\delta} \, [1 - \mathrm{ln}(\delta_0/\delta)\,]\,, \end{equation}
setting e$^2$ $\cong$ g$_w^2$.
In equilibrium the value of $\phi$ in the nuclear lattice is about 
g$_w^2\cdot\epsilon$/e$^2$ $\approx$ 2.7$\cdot$10$^{-10}$ times 
smaller than the corresponding electrostatic potential in an ionic lattice. A 
graph of the potential in Eq.(49) versus $\delta$ is shown in Fig.\,6.

\begin{figure}[h]
    \vspace{0.5cm}
    \hspace{2.5cm}
    \includegraphics{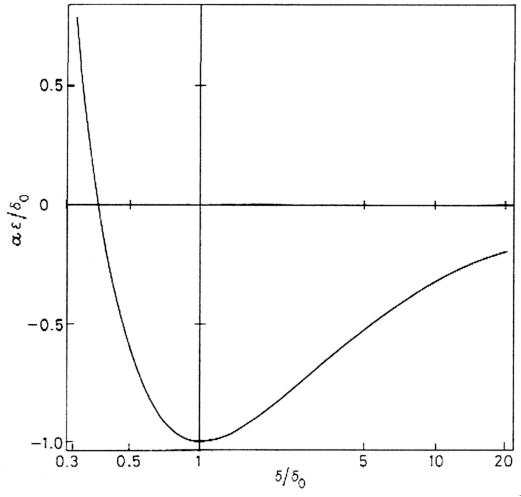}
    \vspace{-0.2cm}
    \begin{quote}
Fig.6: The potential $\phi$ of the weak force as a function of $\delta$.\\
\indent\hspace{1.1cm} After [23].
    \end{quote}
\end{figure}

\bigskip

 The minimum of the curve marks the equilibrium. From Eq.(49) follows with
F$_\mathrm{w}$ = d$\phi$/d$\delta$  and $\delta$ = 2r that

\bigskip 

\emph{the weak force in the interior of the nuclear lattice is approximately}  

\begin{equation} \mathrm{F}_\mathrm{w} \cong \,\frac {3.485\cdot \mathrm{g}_w^2\,\epsilon}
{\mathrm{r}^2}\cdot \mathrm{ln}(\frac{\delta}{\delta_0})\,. 
\qquad(\mathrm{dyn})\end{equation}
For all distances $\delta$\,$>$$\delta_0$ the force F$_\mathrm{w}$ is attractive, 
for all distances $\delta$\,$<$$\delta_0$ the force is repulsive. The small value of 
$\epsilon$ $\approx$ 10$^{-12}$ means that small displacements 
$\delta/\delta_0$\,$<$1 of the neutrinos from their equilibrium position, 
which carry the neutrinos into the domain of their neighboring neutrino, 
cause a very strong repulsive force between both neutrinos. 

   We have thus determined the potential of the weak force in the interior of the lattice in the elementary particles with lattice theory. Let us consider how this was done. 

\bigskip
\noindent
\emph{Following exactly the procedure used by B\&L to determine 
the potential in the interior of an ionic crystal},

\bigskip
\noindent
\emph{we have determined the 
potential  in the interior of the lattice in an elementary particle}

\medskip
\noindent
by using the parameters of the nuclear lattice. As in an
ionic lattice the potential in a nuclear lattice has an attractive and a 
repulsive part, as is necessary for the stability of the lattice.

\begin{center}
\section{The strong force between two elementary particles}
\end{center}

   Crucial for the understanding of the existence of a strong force between 
the sides of two cubic lattices is the observation that

\bigskip
\begin{itemize}
\item
 \emph{the sides of two halves of a cubic lattice cleaved in vacuum 
exert a strong, attractive force on each other}.
\end{itemize}

 It is an automatic consequence of lattice theory that\\

\emph{the weak force, which holds the lattice together, is accompanied by a}\\ 
\indent\emph{strong force which emanates from the sides of the 
lattice}.\\
 
 This seems to contradict the simple 
observation that two salt crystals stacked upon each other can be 
separated without any effort. This is only so because the surface of 
a cubic crystal cleaved in air oxidizes immediately, and then the sides  
do not attract each other any longer. The origin of the force emanating 
from the sides of a cubic ionic lattice is the Coulomb force between
ions of opposite polarity, i.e.\,\,the force which
holds the lattice together. The attractive force emanating from
the side of a crystal cleaved in vacuum has a macroscopic value.
The force between the sides of two cubic lattices was first studied 
by Born and Stern [47] (B\&S).

   If U$_{12}$ is the potential between two sides of a crystal with the 
equal surfaces A, or the work that is necessary to separate the two sides 
of a cleaved crystal, then the capillary constant $\sigma$ is given by
Eq.(2) of B\&S 
\begin{equation} \sigma = -\,\mathrm{U}_{12}/\mathrm{A}\,. \end{equation}
The capillary constant is, in the following, taken at zero degree absolute 
and against vacuum for the square outside area A of a cubic crystal. 
$\sigma$ has been explained by B\&S, but their formula cannot be 
used here because they use the value  n = 9 of the alkali-halegonids.
We will instead use Eq.(463) from Born [48] for the capillary constant 
$\sigma_{100}$ of the (100) front surface of an ionic cubic crystal
\begin{equation} \sigma_{100} = -\,\frac{\mathrm{e}^2}{\mathrm{r}_0^3}\,
\frac{s_0(1)}{2}\cdot[\,1 - \frac{1}{\mathrm{n}}\,\frac{s_0(\mathrm{n})}
{s_0(1)}\cdot\frac{S_0(1)}{S_0(\mathrm{n})}\,]\,. \end{equation}
The sums s$_0$(n) and S$_0$(n) originate from the contributions of the 
different lattice points to the repulsive part of the potential. The sign of 
the second term on the right hand side in Eq.(52) comes from the repulsive 
part of the potential in Eq.(36). For the capillary constant in a nuclear 
lattice we set e$^2$ =  g$_w^2$, n = 1 + $\epsilon$, (Eq.47), and  
s$_0$(1) = $-$\,0.0650 from [48] p.743. We find that 
s$_0$(n) $\cong$ s$_0$(1) since n = 1 + $\epsilon$ and 
$\epsilon \cong$ 10$^{-12}$.  Similarly we have 
S$_0$(n) $\cong$ S$_0$(1). Then we have
\begin{equation} \sigma_{100} \cong \frac{0.065}{2}\,\frac{\mathrm{g}_w^2}
{\mathrm{r}_0^3}\, \epsilon\,. \qquad(\mathrm{dyn/cm}) \end{equation}

The work required to separate one half of a nuclear lattice from the other  
half is according to Eq.(51) given by 
\begin{equation} \mathrm{U}_{12} \cong -\,\frac{0.065}{2}\,\, 
\frac{\mathrm{g}_w^2\,\epsilon}{\mathrm{r}_0^3}\cdot\mathrm{A}\,.
\end{equation}    
\noindent
We determine the area A with the number of the lattice points in the 
cubic nuclear lattice 

\begin{equation} \mathrm{N}  = 2.854\cdot10^{\,9}\,, \end{equation}
\noindent
from Eq.(17). It follows that A = ($\sqrt[3]{\mathrm{N}}\,\cdot\,$r$_0)^2$. 
And it follows that  the strong attractive force between the sides
of two nuclear lattices  is

\medskip

\begin{equation} \mathrm{F_s} = \frac{\mathrm{dU}_{12}}{\mathrm{dr}} =
-\,\frac{\mathrm{d}\sigma}{\mathrm{dr}}\cdot\mathrm{A} = 
\frac{3\cdot0.065\,\mathrm{g}_w^2\,\epsilon}
{2\,\mathrm{r}^4}\cdot\mathrm{A}\,. \end{equation}

\medskip

\noindent
\emph{The force emanating from the front surface of a cubic nuclear lattice,\\
\noindent 
the strong force, is}

\begin{equation} \mathrm{F_s} = \frac{0.0975\,\,\mathrm{g}_w^2\,\,\epsilon}
{\mathrm{r}^4}\cdot(\sqrt[3]{\mathrm{N}}\cdot\mathrm{r}_0)^2\,. 
 \qquad (\mathrm{dyn}) \end{equation}

\medskip
\noindent
The strong force depends, first of all, on the \emph{weak interaction constant}
g$_w^2$, that means on the force between neighboring lattice points, 
and secondly on the number of lattice points on the side of 
the lattice, ($\sqrt[3]{\mathrm{N}})^2\,$ = 2.012\,$\cdot\,10^{\,6}$.
The strong force decreases rapidly with increasing r because it is inversely 
proportional to the \emph{fourth} power 
of the distance between the particles. In our model the strong force depends,
other than on constants and r$^{-4}$, on the number N$^{2/3}$ of the lattice 
points at the side of the lattice. That means that the force which 
emanates from the sides of the $\pi$\,mesons is the same as the force
which emanantes from the sides of the proton, because both have, in this
model, the same number of lattice points. We also note that the strong force, 
(Eq.57), is independent of a charge. The entire force which goes out from 
the surface of the lattice is six times as much as given by Eq.(57).

   The ratio of the strong force F$_\mathrm{s}$ emanating from the side 
of a cubic nuclear lattice to the weak force F$_\mathrm{w}$ in the interior 
of the lattice, Eq.(50), is
\begin{equation}\mathrm{F_s} = 
\frac{0.056\cdot10^{\,6}}{\mathrm{(r/r_0)^2\,ln}(\mathrm{r/r_0})}
\cdot \mathrm{F}_w\,.
\end{equation}
 
   The ratio F$_\mathrm{s}$/F$_\mathrm{w}$ is not constant, but depends on the 
ratio r/r$_0$. For r$\,>\,$r$_0$ = 10$^{-16}$\,cm the strong force decreases 
with increasing r, 
for r $\rightarrow$ r$_0$ we have F$_s$ $\rightarrow \infty$, and for 
r$\,<\,$r$_0$ the strong force is
repulsive, as it must be when one lattice enters another lattice. The constant 
factor in the ratio of the strong and weak forces originates from the number of 
lattice points at a side of the lattice ($\sqrt[3]{\mathrm{N}})^2$ = 
2.012\,$\cdot\,10^{\,6}$.

\bigskip 

   To summarize: According to lattice theory 
\begin{itemize}
\item
\emph{the existence of the 
strong nuclear force between two elementary particles is an automatic
consequence of our explanation of the masses of the elementary
particles with cubic nuclear lattices,} 
\end{itemize}
held together by the weak nuclear force.  The lattices we have used for 
the explanation of 
the masses of the particles consist of photons or neutrinos. That means: 
\hspace{1.2cm}\emph{We do not use hypothetical particles.}  

  We have found long sought after answers to the questions what is the weak 
nuclear force and the strong nuclear force, and why is the strong force so much
stronger than the weak force\,?  The strong force between two nuclear particles 
is nothing but the sum of the large number of unsaturated weak forces at the 
surface of a nuclear lattice. In order to understand the origin of the strong 
nuclear force 
one has to understand the structure of the elementary particles, which 
we have explained with nuclear lattices.  We have also understood 
the strength of the weak force which holds the nuclear lattice together, 
and thereby the cause of the strong force between two nuclear lattices.

\section{The rest mass of the muon}

Surprisingly one can also explain the mass of the $\mu^\pm$ muons, 
originally called the $\mu^\pm$\,mesons, 
with our explanation of the $\pi^\pm$\,mesons. The existence of the 
muons has been a puzzle since their discovery about 75 years ago. 
The muons belong to the lepton family. 
The leptons are distinguished from the mesons and baryons by the 
absence of strong interaction with the mesons and baryons. The charged 
leptons make up 1/2 of the number of the charged stable elementary 
particles. The standard model of the particles does not deal with the 
leptons. Barut [49] has given a simple and quite 
accurate empirical formula relating the masses of the electron, 
muon and $\tau$\,lepton, which formula has been 
extended by Gsponer and Hurni [50] to the quark masses.

    The origin of most of the muons, which have spin 1/2, is the decay
of the $\pi^\pm$\,mesons $\pi^\pm$ $\rightarrow \mu^\pm
+ \nu_\mu(\bar{\nu}_\mu$) (99.98770\%),
or the most frequent decay (63.55\%) of the K$^\pm$\,mesons  
K$^\pm \rightarrow \mu^\pm + \nu_\mu(\bar{\nu}_\mu)$. 
The rest mass of the muons is\\
 
\hspace{2.5cm} m($\mu^\pm$) = 105.658\,367 $\pm$ 
4$\cdot$$10^{-6}$\,MeV/c$^2$,\\

\noindent 
according to the Review of Particle Physics. The mass of the muons  
is usually compared to the mass of the electron and is very often said to be
\begin{equation} 
\mathrm{m}(\mu^\pm)(emp) = \mathrm{m(e}^\pm)\cdot(1 + 3/2\alpha_f) =
 206.554\,\mathrm{m(e}^\pm) =  0.99896\,\mathrm{m}(\mu^\pm)(exp)\,,
\end{equation}
(with the fine structure constant $\alpha_f$ = 1/137.036). 
The experimental value of m($\mu^\pm$) is 206.768\,m(e$^\pm$).  
The formula (59) for m($\mu^\pm$) was given by Barut [51], following 
an earlier suggestion by Nambu [10] that the mass of the $\pi$\,meson
is $\approx 2/\alpha_f\,\cdot\,$m(e$^\pm$) and that m($\mu^\pm$) 
$\approx$  3/2$\alpha_f$\,$\cdot$\,m(e$^\pm$). The muons are 
``stable", their lifetime  
$\tau(\mu^\pm) = 2.19703\cdot10^{-6} \pm 2.2\cdot10^{-11}$\,sec is
about a hundred times longer than the lifetime of the $\pi^\pm$\,mesons,  
that means longer than the lifetime of any other elementary particle, but 
for the electrons, protons and neutrons.

   Comparing the mass of the muons to the mass of the 
$\pi^\pm$\,mesons, from which the muons emerge
we find, with m($\pi^\pm$) = 139.570\,18\,MeV/c$^2$, that
\begin{equation} 
\bullet\hspace{1.0cm}\mathrm{m}(\mu^\pm)/\mathrm{m}(\pi^\pm)(exp) = 
0.757027 = 1.00937\cdot3/4\,.
\end{equation}

\noindent
The mass of the muons is in a good approximation 3/4 
of the mass of the $\pi^\pm$\,mesons. We have also m($\pi^\pm)\,-$ 
m($\mu^\pm$) = 33.9118\,MeV/c$^2$ =\\
0.24297m($\pi^\pm$)  
or approximately 1/4\,$\cdot$\,m$(\pi^\pm$). On the other hand, the 
mass of the muon is about 207 times larger than the mass of the 
electron, the contribution of m(e$^\pm$) to m($\mu^\pm$) will therefore 
be neglected in the following. We assume, as we have done before   
and as appears to be natural, that the particles, including the muons, 
$\emph{consist of the particles into}$ \emph{which they decay}.

  The muons decay via $\mu^\pm$ $\rightarrow$ e$^\pm$ +
$\bar{\nu_\mu}$($\nu_\mu)$ + $\nu_e$($\bar{\nu_e})$ ($\approx$ 100\%). 
If the particles consist of the particles into which they decay, then the, say, 
$\mu^-$ muons consist of $\nu_\mu$\,\,and $\bar{\nu}_e$  neutrinos, and the 
charge e$^-$. That raises the question, what happened to the 
$\bar{\nu}_\mu$ neutrinos, as well as to the $\nu_e$ neutrinos, which were in 
the $\nu_\mu$,\,$\bar{\nu}_\mu$,\,$\nu_e$,\,$\bar{\nu}_e$ neutrino lattice of  
the $\pi^\pm$\,meson, from which the $\mu^\pm$ muons emerged\,? The 
$\nu_\mu$ or the $\bar{\nu}_\mu$  neutrinos have been lost in the decay of 
$\pi^\pm$, $\pi^\pm$ $\rightarrow \mu^\pm + \nu_\mu (\bar{\nu}_\mu$) 
(99.98770\%), and therefore  only either $\nu_\mu$ or
$\bar{\nu}_\mu$ neutrinos remain in the $\mu^\pm$ lattice, not both of them.  
That seems to mean that only
three neutrino types, namely $\nu_\mu$,\,$\nu_e$,\,$\bar{\nu}_e$
or $\bar{\nu}_\mu$,\,$\nu_e$,\,$\bar{\nu}_e$ neutrinos  
are in the muons, in other words 3/4\,$\cdot$\,N neutrinos, as on Fig.(7), 
not N neutrinos as they are needed for a cubic 
lattice. We will see that the fourth lattice point on the sides of a cubic 
lattice is filled by neutrinos from the charge e$^\pm$.

  The muons, with a mass m($\mu^\pm$) $\cong$ 3/4\,$\cdot$\,m($\pi^\pm$), 
seem to be related to the $\pi^\pm$\,mesons,
rather than to the electron with which the muons have
been compared traditionally, although m($\mu^\pm$) is separated 
from m(e$^\pm$) by a factor $\cong$ 207. The muons are a fragment of
the $\pi^\pm$\,meson decay, not a massive form of the electron.
The decay of the muons is 
described in the literature with the help of the W$^\pm$ bosons.
Since m(W$^\pm$) $\cong$ 760.9\,m($\mu^\pm$), the participation 
of W$^\pm$ in the decay of the muons violates conservation of energy.
On the other hand, the decay of the muons we propose conserves energy. 

   From Eq.(33b) followed that the mass of a muon neutrino   
should be about 50\,milli-eV/c$^2$. Provided that 
the mass of an electron neutrino m($\nu_e)$ is small as compared  
to m($\nu_\mu$), as will be shown by Eq.(72), we find, with 
N\,\,=\,\,2.854$\cdot10^{\,9}$, that:
\bigskip

(a) The difference of the rest masses of the muons and the $\pi^\pm$ 
mesons is, not considering the consequences of the charge e$^\pm$, 
nearly equal to the sum of the masses of all \emph{muon neutrinos}, 
respectively anti-muon neutrinos, which are in the $\pi^\pm$\,mesons.
\bigskip

\noindent
    m($\pi^\pm)\,\,-\,\,$m$(\mu^\pm$) = 33.912\,MeV/$\mathrm{c}^2$ 
\quad versus \quad $\mathrm{N}/4\cdot\mathrm{m}(\nu_\mu)$ 
$\approx$ 35.68\,$\mathrm{MeV}/\mathrm{c}^2$\,,
approximating in the following N\,-\,1 by N.
 
\bigskip
    
(b) The energy in the oscillations of all 
$\nu_\mu,\bar{\nu}_\mu,\nu_e,\bar{\nu}_e$
neutrinos in the $\pi^\pm$\,mesons
is nearly the same as the energy in the oscillations of 
all $\bar{\nu}_\mu,\nu_e,\bar{\nu}_e$, respectively 
$\nu_\mu,\bar{\nu}_e,\nu_e$, neutrinos in the muons. The oscillation 
energy is the rest mass of a particle minus the sum of the 
masses of all neutrinos in the particle as in Eq.(33). With 
m($\nu_\mu$) = m($\bar{\nu}_\mu$) and m($\nu_e$) = 
m($\bar{\nu}_e)$ from Eqs.(68,71) we have
\begin{equation} \mathrm{E}_{\nu}(\pi^\pm) =
 \mathrm{m}(\pi^\pm)\mathrm{c}^2 - 
\mathrm{N}/2\cdot[\mathrm{m}(\nu_\mu) + 
\mathrm{m}(\nu_e)]\mathrm{c}^2 = 68.22\,\mathrm{MeV}
 \end{equation}
\quad versus 
\begin{equation} \mathrm{E}_{\nu}(\mu^\pm) =
 \mathrm{m}(\mu^\pm)\mathrm{c}^2 - 
\mathrm{N/4}\cdot \mathrm{m}(\nu_\mu)\mathrm{c}^2 - 
\mathrm{N}/2\cdot\mathrm{m}(\nu_e)\mathrm{c}^2 = 
69.98\,\mathrm{MeV}\,.\end{equation}
\noindent
Equation (62) means that either all N/4 muon neutrinos or  
all N/4 anti-muon neutrinos have been removed from the 
$\pi^\pm$  lattice in its decay.  If, e.g.,  any $\nu_\mu$ neutrinos   
were to remain in $\mu^+$  after the decay of the 
$\pi^+$\,meson they ought to appear in the decay of $\mu^+$, 
but they do not.

   We attribute the 1.768\,MeV difference 
between the left and right side of (a) to the second order effects
which cause the deviations of the masses of the particles from
the integer multiple rule. There is also the difference that the left 
side of (a) deals with two charged particles, whereas the right side 
deals with neutral particles. (b) seems to say that the
oscillation energy of all neutrinos in the $\pi^\pm$ lattice is conserved
in the $\pi^\pm$ decay, which seems to be necessary because the 
oscillation frequencies in $\pi^\pm$ and $\mu^\pm$ must follow Eq.(14).
If indeed 
\begin{equation} \mathrm{E}_\nu(\pi^\pm) = \mathrm{E}_\nu(\mu^\pm)
\end{equation} 
then it follows from the difference of Eqs.(61) and (62) that
\begin{equation}
 \mathrm{m}(\pi^\pm)\,-\,\mathrm{m}(\mu^\pm) = 
\mathrm{N}/4\cdot\mathrm{m}(\nu_\mu) =
\mathrm{N}/4\cdot \mathrm{m}(\bar{\nu}_\mu)\,. \end{equation}
This equation applies only to the neutral neutrino lattices of the pions 
and muons. The energy in N/4 muon neutrinos is 35.675\,MeV, if 
m($\nu_\mu$)c$^2$ = m($\bar{\nu}_\mu$)c$^2$ = 50\,milli-eV, as in 
Eq.(33b). That means that the difference in the energy of the rest masses of 
$\pi^\pm$ and $\mu^\pm$, m($\pi^\pm$)c$^2$ - m($\mu^\pm$)c$^2$
= 33.912\,MeV, originates from the energy in N/4 muon neutrinos. A small part  
of this energy, $\Delta$ = 1.763\,MeV, is retained by the muon lattice.

   The charges e$^\pm$ in $\mu^\pm$ consist of N/4 charge elements 
Q$_k$\,\,and N/4 electron neutrinos or anti-electron neutrinos, as we
will see in Section 11. The N/4 $\nu_e(\bar{\nu}_e)$ neutrinos from e$^\pm$ 
pick up 1/4 of the oscillation energy E$_\nu(\pi^\pm$) of the $\pi^\pm$\,mesons, 
which becomes available when the muon neutrinos $\nu_\mu$ or antimuon 
neutrinos $\bar{\nu}_\mu$ leave the $\pi^\pm$ lattice in the $\pi^\pm$ decay. The
$\nu_e(\bar{\nu}_e)$ neutrinos coming from e$^\pm$ can pick up oscillation energy, 
because they move in a free charge e$^\pm$ with frequencies proportional to
$\nu_0\alpha_f$, whereas in the $\pi^\pm$ or $\mu^\pm$ lattices the frequencies 
are proportional to $\nu_0$. The neutrinos coming with e$^\pm$ into $\mu^\pm$ 
make it possible that $\mathrm{E}_\nu(\pi^\pm) = \mathrm{E}_\nu(\mu^\pm)$.  
After the $\pi^\pm$ decay
the remaining muon neutrinos in $\mu^\pm$ retain their original oscillation 
energy E$_\nu(\pi^\pm$)/4, the remaining electron neutrinos and 
anti-electron neutrinos in $\mu^\pm$ retain their original oscillation 
energies E$_\nu(\pi^\pm$)/4 as well. The oscillation energy 
E$_\nu(\pi^\pm$)/4 of the $\pi^\pm$ lattice so far not accounted for 
is picked up by the neutrinos from e$^\pm$, brought into $\mu^\pm$ by  
e$^\pm$. There are then in total N neutrinos in the $\mu^\pm$ lattice. 
Without a recipient for the oscillation energy E$_\nu(\pi^\pm$)/4 
picked up by the neutrinos from e$^\pm$ a stable new particle can apparently 
not be formed in the $\pi^\pm$ decay, that means there is no $\mu^0$\,particle.

   All parts of the muon lattice were already in the $\pi^\pm$ lattice. The 
lattice of $\pi^\pm$ consists, considering also the charge e$^\pm$, of 
N/4 $\nu_\mu$ and N/4 $\bar{\nu}_\mu$, N/4 $\nu_e$ and N/4 $\bar{\nu}_e$ 
neutrinos, plus N/4 $\nu_e(\bar{\nu}_e)$ from e$^\pm$,
and also of N/4 charge elements Q$_k$, because e$^\pm$ consists of 
N/4 $\nu_e$($\bar{\nu}_e$) neutrinos and N/4 Q$_k$ charge elements, as 
we will see in Section 11. The lattice of $\mu^\pm$ consists, considering 
also the charge e$^\pm$, of N/4 $\nu_\mu$($\bar{\nu}_\mu$) and 
N/4 $\nu_e$($\bar{\nu}_e$) and N/4 $\bar{\nu}_e$($\nu_e$) neutrinos,
as well as of N/4 $\nu_e(\bar{\nu}_e)$ neutrinos from e$^\pm$, plus 
N/4 Q$_k$ charge elements, which sit in the centers of the cell sides.
The only difference between the $\pi^\pm$ and 
the $\mu^\pm$ lattice is the absence of N/4 $\nu_\mu$($\bar{\nu}_\mu$) muon
neutrinos, which have been lost in the decay $\pi^\pm$ $\rightarrow$ $\mu^\pm$
+ $\nu_\mu$($\bar{\nu}_\mu$). The actual presence of the neutrinos in the 
$\mu^\pm$ lattice is shown in the decay 
of the muons, $\mu^\pm$ $\rightarrow$ e$^\pm$ 
+ $\bar{\nu}_\mu$($\nu_\mu$) + $\nu_e$($\bar{\nu}_e$). This all follows our 
rule that the particles consist of the particles into which they decay.

   We should note that in the $\pi^\pm$ decays only \emph{one single} 
muon neutrino or single antimuon neutrino is emitted, not N/4 
of them, but that in the 
$\pi^\pm$ decay 33.912\,MeV are released. Since, according to (b) the 
oscillation energy of the neutrinos in the $\pi^\pm$\,mesons is conserved 
in the $\pi^\pm$  decay, the 33.912\,MeV released in the $\pi^\pm$ 
decay can come from
\emph{no other source} then from the energy in the masses of the muon 
neutrinos or the anti-muon neutrinos in the $\pi^\pm$\,mesons. The 
average kinetic energy  of the muon neutrinos in the $\pi^\pm$ lattice 
is about 33.9\,MeV/(N/4) = 47.5\,milli-eV, it is therefore not possible
that a single neutrino in the $\pi^\pm$ lattice possesses an energy
of 33.9\,MeV. The 33.9\,MeV can come only from the energy in the 
sum of the muon neutrino or anti-muon neutrino masses in  the 
$\pi^\pm$\,mesons. However, what happens then to the neutrino 
numbers\,? 

   Either conservation of neutrino numbers is violated or the 
decay energy comes from equal numbers of muon neutrinos and 
anti-muon neutrinos. Equal numbers N/8 muon neutrinos 
and  N/8 antimuon neutrinos would then be in
$\mu^\pm$, instead of straight N/4 $\nu_\mu(\bar{\nu}_\mu)$ neutrinos.
This would not make a difference in either the 
oscillation energy or in the sum of the masses of the neutrinos or in 
the spin of the muons. A situation similar to the $\pi^\pm$ 
decay occurs in the $\mu^\pm$ decay. The 105.147\,MeV released in
the $\mu^\pm$ decay comes, in our model, mainly from the energy in 
the masses of either N/4\,$\cdot$\,$\nu_\mu$  or 
N/4\,$\cdot$\,$\bar{\nu}_\mu$
neutrinos and their oscillations, because the masses of the 
$\nu_e,\bar{\nu}_e$ neutrinos, which are also in $\mu^\pm$, are so
small. Conservation of neutrino numbers in the $\mu^\pm$ decay 
requires that N/8 muon neutrinos and N/8 anti-muon neutrinos are in the 
$\mu^\pm$ lattice. Since m($\nu_\mu$) = m($\bar{\nu}_\mu$) 
we will, however, for the sake of simplicity, write 
N/4\,$\cdot$\,m($\nu_\mu$)
or N/4\,$\cdot$\,m($\bar{\nu}_\mu$) for 
N/8\,$\cdot[$m($\nu_\mu$) + m($\bar{\nu}_\mu$)].   

   Inserting m($\pi^\pm$)\,$\mathrm{-}$\,m($\mu^\pm$) = 
N/4\,$\cdot$\,m($\nu_\mu$)
from Eq.(64) into Eq.(62) we arrive at an equation for the 
theoretical value of the rest mass of the muons. It is
\begin{equation} \mathrm{m}(\mu^\pm)\mathrm{c}^2(theor) = 
1/2\cdot[\,\mathrm{E}_\nu(\pi^\pm) + 
\mathrm{m}(\pi^\pm)\mathrm{c}^2 +
\mathrm{N}\mathrm{m}(\nu_e)\mathrm{c}^2/2\,] = 
103.95\,\mathrm{MeV}\,,\end{equation}
which is 0.9838\,m($\mu^\pm$)c$^2$(\emph{exp}) and  expresses
m($\mu^\pm$) through the well-known mass of $\pi^\pm$,
the calculated oscillation energy of $\pi^\pm$, and a small 
contribution (0.4\%) of the electron neutrino and anti-electron 
neutrino masses. Eq.(65) shows that 
our explanation of the mass of the muons comes close (1.6\%) to 
the experimental value m($\mu^\pm$) = 105.658\,MeV/c$^2$. With
E$_\nu(\pi^\pm$) = E$_\nu(\mu^\pm$), and with m($\pi^\pm)$ which 
follows from Eq.(33), we find a different form of Eq.(65) which is, 
in the case of $\mu^+$, 
\begin{equation}\mathrm{m}(\mu^+)(theor) =
 \mathrm{E}_\nu(\mu^\pm)/\mathrm{c}^2 + 
\mathrm{N}\mathrm{m}(\bar{\nu}_\mu)/4 + 
\mathrm{N}\mathrm{m}(\nu_e)/4 + \mathrm{N}\mathrm{m}(\bar{\nu}_e)/4 \,\,.
 \end{equation}

\noindent
Or Eq.(66) means, with E$_\nu$$(\mu^\pm$)/c$^2$ = 
N/2\,$\cdot$\,(m($\nu_\mu)$ + m$(\nu_e$)) and m($\nu_\mu$) = m($\bar{\nu}_\mu$),
m($\nu_e$) = m($\bar{\nu}_e$), and without considering the charge, that

\newpage

 \emph{the rest mass of the muons should be equal to} \\

$\bullet$ \hspace{1.0cm} m($\mu^\pm$)(\emph{theor}) = 
3/4\,$\cdot$\,Nm($\nu_\mu$) + Nm($\nu_e$) = 
107.88\,MeV/c$^2$\,, (66a)\\

\noindent 
which is 1.021\,$\cdot$\,m($\mu^\pm$)(\emph{exp}).\\

Eq.(66) tells that the rest mass of the muons is the sum of the masses
of the \emph{muon neutrinos}, respectively antimuon neutrinos, and of the 
masses of the electron neutrinos and anti-electron neutrinos which are in the 
muon lattice, plus the oscillation energy of these neutrinos, neglecting the 
mass of e$^\pm$, which is 0.510998\,MeV/c$^2$. 
The ratio  m($\mu^\pm$)/m($\pi^\pm$) is 3/4, as it must be, if we 
divide Eq.(66a) by m($\pi^\pm$) from Eq.(33a), and if we neglect the small 
masses of the electron neutrinos and anti-electron neutrinos.\\

   \emph{The  muons cannot be point particles} because  
they have a neutrino lattice.  The commonly held belief that the 
muons are point particles  is based on the results of scattering
experiments. But at a true point the density of a ``point particle" would
be infinite, which poses a problem. It is odd that the muons,
which emerge from the $\pi^\pm$\,mesons and have a mass which is nearly 
3/4\,$\cdot$\,m($\pi^\pm$), should have a mass which is concentrated in a 
point, whereas it is accepted and measured that the $\pi^\pm$\,mesons have
a body of finite size. Since, on the other hand, neutrinos do not interact, in  
a very good approximation, with electrons or positrons it will not be possible  
to determine the size of the muon lattice through conventional 
scattering experiments. The muons \emph{appear} to be 
point particles because only the charge of the $\mu^\pm$ muons 
participates in the usual scattering processes, and  
electrons or positrons scatter like point particles.

\vspace{0.5cm}

\indent 
\emph{99.5\% of the energy in the rest mass of the muon} \\
\indent 
\emph{consists of neutrinos and their oscillation energy},

\vspace{0.5cm}

\noindent
both of which do not interact with an incoming electron.

   Our model of the muon seems to make it possible to provide a qualitative 
explanation for the fact that the mean lifetime of the muon $\tau(\mu^\pm)$ =
2.197034$\cdot$10$^{-6}$ sec is 84 times longer than the mean lifetime 
of the pion $\tau(\pi^\pm)$ = 2.6033$\cdot$10$^{-8}$ sec. If the corners of
a square cell-side of the cubic  muon lattice do, indeed, contain a 
\emph{single} muon (or anti-muon) neutrino and 
two electron and/or anti-electron neutrinos, then the force that is exerted from 
the single, say, muon neutrino in a corner of the cell-side, on any electron (or 
anti-electron) neutrino in the other corners of the 
cell-side, should be stronger in a muon cell-side, 
than in a pion cell-side. In the cell-sides of the pion, the force exerted on, say,
an electron neutrino by a muon neutrino in a corner of a cell-side, is counteracted
by the force coming from the anti-muon neutrino in the opposite corner of the 
square cell-side. The bond in a cell-side of the pion lattice should therefore be 
weaker than the bond in a cell-side of the muon lattice, and that means that the 
pion lattice should be less stable than the muon lattice.       

   Finally we must address the question for what reason do the 
muons or leptons not interact strongly with the mesons 
and baryons\,? We have shown in  Section 8 that a strong force emanates 
from the sides of a cubic lattice caused by the unsaturated weak 
forces of about $10^{\,6}$  lattice points at the surface of the lattice of 
the mesons and baryons. This follows from the study of Born and Stern [47] 
which dealt with the forces between two parts of a cubic lattice cleaved 
in vacuum. The strong force between two particles is an automatic 
consequence of the weak internal force which holds the particles together.
If the $\mu^\pm$ muons have a charged lattice consisting of N/8 muon neutrinos 
and N/8 antimuon neutrinos and, say, of N/2 electron neutrinos and N/4 
anti-electron neutrinos, their lattice surface is not the same as the surface 
of the cubic $\nu_\mu,\,\bar{\nu}_\mu,\,\nu_e,\,\bar{\nu}_e$ lattice of the 
mesons and baryons. Therefore  the muon lattice does  
not bind with the cubic lattice of the mesons and baryons.

   To summarize what we have learned about the muons.
Eq.(66) says that the energy in m($\mu^\pm$)c$^2$ is the sum of 
the oscillation energies plus the energy in the sum of the masses of 
the neutrinos and antineutrinos in m($\mu^\pm$), neglecting the energy  
in e$^\pm$. The three neutrino types in the lattice of the muons  
shown in Fig.\,7 are the remains of the cubic neutrino lattice in the 
$\pi^\pm$\,mesons. Since  N/8\,$\cdot$\,$\nu_\mu$ and
N/8\,$\cdot$\,$\bar{\nu}_\mu$ neutrinos have been
removed from the $\pi^\pm$ lattice in the $\pi^\pm$ decay,
and since m(N/8\,$\cdot$\,$\nu_\mu$ + N/8\,$\cdot$\,$\bar{\nu}_\mu$) 
 = N/4\,$\cdot$\,m($\nu_\mu$)\,\,$\cong$\,\,m($\pi^\pm$)/4, 
the rest mass of the muons must be
$\cong$ 3/4\,$\cdot$\,m($\pi^\pm$), as the experiments find. The
absence of a neutrino in the center of the lattice of Fig.\,7 is crucial
for the explanation of the spin $\hbar$/2 of the muon in Section 15, or in 
[73].

\begin{figure}[h] 
    \vspace{0.5cm}
    \hspace{0.2cm}
    \includegraphics{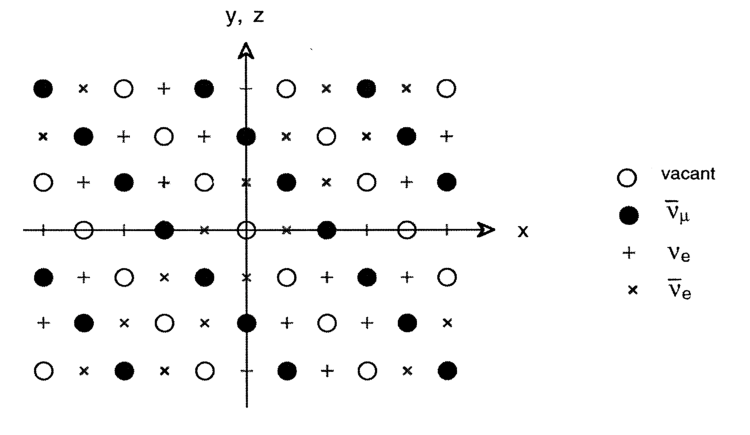}
    \vspace{-0.3cm}
    \begin{quote}
Fig.\,7: A section through the central part of the neutrino\\   
\indent \hspace{1.1cm} lattice of the $\mu^+$ muon without the charge.
    \end{quote}
\end{figure}

   The muons are not point particles. Just as the 
gravitational force of the mass of the Earth is described 
as coming from a \emph{point} in the center of the Earth, for all points 
outside of the Earth, so can the force originating from the mass and 
charge of the cubic muon lattice be described as coming from a 
\emph{point} in the center of the lattice, for all points outside of the
lattice. The \emph{points} in question are theoretical abstractions,
they do not have any extension. In spite of the central forces which
emerge from them, neither the Earth nor the muon are point particles. 

   The mass of the $\tau^\pm$ lepton, also referred to as the tauon,
whose mass is m($\tau^\pm$) =
1776.82\,MeV/c$^2$  = 16.8166\,m($\mu^\pm$) or
1.8937\,m(p), follows from the most frequent  
leptonic decay of the D$^\pm_s$\,mesons, D$^\pm_s 
\rightarrow \tau^\pm + \nu_\tau(\bar{\nu}_\tau)$ (5.54\%). It can be 
shown readily that the oscillation energies 
of the lattices in D$^\pm_s$ and in $\tau^\pm$ are the same. 
From that follows that the energy in the mass
of the $\tau^\pm$ lepton is the sum of the oscillation energy in the 
$\tau$  lattice plus the energy in the sum of the masses of all 
neutrinos and antineutrinos in the $\tau$ lattice, just as with the muon. 
We will skip the details. 

\section{The neutrino masses}

   Now we come to the neutrino masses. The neutrinos move with the velocity 
of light but have a mass, just as photons have a mass, because E = mc$^2$. 
Within the rest masses of the $\nu$-branch particles the neutrinos cannot 
progress with the velocity of light, but they can form standing waves in the 
particles. We assume that the energy in the standing waves is equal to mc$^2$ 
of the neutrinos in the lattices. As mentioned before, the 
experiments show that the electron neutrinos, muon neutrinos and tau
neutrinos are different. It is undisputed that all of them have no charge 
and the same spin. However, if the three neutrino types have different 
masses then the electron neutrinos, muon neutrinos and tau
neutrinos are different, as they are supposed to be.  There is no certain 
knowledge what the neutrino masses are. Numerous values for m($\nu_e$) and
m($\nu_\mu$) have been proposed and upper limits for them have
been established experimentally which have, with time, decreased 
steadily. The Review of Particle Physics gives 
for the mass of the electron neutrino the value $<$\,2\,eV/c$^2$. Neither  
the Superkamiokande [37] nor the Sudbury [38] experiments determine 
a neutrino mass, however, both experiments make it very likely that the 
neutrinos have masses. We will now determine the neutrino masses 
from the composition of the $\pi^\pm$\,mesons and from the 
$\beta$-decay of the neutron.
 
   If the same principle that applies to the decay of the 
$\pi^\pm$\,mesons, namely that in the decay the oscillation energy
of the decaying particle is conserved (Eq.63), and that an entire 
neutrino type supplies the energy released in the decay (Eq.64),  
also applies to the decay of the neutron
n $\rightarrow$ p + e$^-\,+\,\bar{\nu}_e$ (100\%), then the mass of  
the electron neutrino can be determined from the known difference 
$\Delta$ =  m(n)\,$-$\,m(p) = 1.293\,332\,MeV/c$^2$ [2]. 
The standard description of the $\beta$-decay
of the neutron uses the W$^\pm$ bosons. Since m(W$^\pm$) 
$\cong$ 85\,m(n) the presence of W$^\pm$ in the decay of the
neutron violates conservation of energy. The decay of the neutron 
considered in the following does, on the other hand, conserve energy.
 
   Nearly one half of $\Delta$ comes from the energy lost by the emission 
of the electron, whose mass is $\cong$ 0.510\,9989\,MeV/$\mathrm{c}^2$. 
N electron neutrinos are in the neutrino quadrupoles in the neutron. 
As we have seen on p.\,32  
the decay sequence of the $\pi^\pm$\,mesons requires that the 
electron carries with it N/4 electron neutrinos. The composition 
of the electron will be discussed in further detail in Section\,11.  
After the neutron has lost N/4 electron neutrinos to the electron 
emitted in the $\beta$-decay, another 3/4\,$\cdot$\,N
electron neutrinos in the neutron apparently provide
the energy ($\Delta$ $-$ m(e$^-$))c$^2$ = 0.782\,333\,MeV
released in the decay of the neutron. However, conservation of 
neutrino numbers makes it impossible that the energy $\Delta^\prime$
= ($\Delta$ - m(e$^-$))c$^2$ comes from only one type of neutrinos.
In order to conserve the neutrino numbers in the $\beta$-decay of the 
neutron the energy $\Delta^\prime$ of the 
3/4\,$\cdot$\,N $\nu_e(\bar{\nu}_e)$ neutrinos must be given by the equation\\

\hspace{2cm} N/2\,$\cdot$\,m($\bar{\nu}_e)$c$^2$ + 
N/4\,$\cdot$\,m($\nu_e)$c$^2$  = 0.782\,333\,MeV.

\medskip

\noindent Only N/4 electron neutrinos are in this equation, because another 
N/4 electron neutrinos have left the neutron with the electron emitted 
in the $\beta$-decay. In order to determine m($\nu_e$) and 
m($\bar{\nu_e}$) we need a second equation, which comes from the 
decay of the anti-neutron, 
$\bar{\mathrm{n}} \rightarrow\bar{\mathrm{p}}$ + e$^+$ + $\nu_e$.
This leads to\\

\hspace{2cm} N/2\,$\cdot$\,m($\nu_e$)c$^2$ + 
N/4\,$\cdot$\,m($\bar{\nu}_e)$c$^2$  = 0.782\,333\,MeV.\\

\noindent From 0.782\,333\,MeV follows after division by 3/4\,$\cdot$\,N, 
with N = 2.854\,$\cdot$\,10$^{\,9}$,  that
\begin {equation} \mathrm{m}(\bar{\nu}_e) = 
0.365\,\mbox{milli-eV}/\mathrm{c}^2\,,\end{equation}

\noindent and that  
\begin{equation} \mathrm{m}(\nu_e) = \mathrm{m}(\bar{\nu}_e)\,. 
\end{equation}
We note that it follows from Eq.(67) that 
\begin{equation}\mathrm{N}/4\cdot \mathrm{m}(\nu_e) =
 \mathrm{N}/4\cdot \mathrm{m}(\bar{\nu}_e) =
 0.51\,\mathrm{m(e}^\pm)\,. \end{equation}
This equation is, as we will see, fundamental for the explanation of  
the mass of the electron.
 
   Inserting Eq.(67) into Eq.(33) for the sum of the masses of all 
neutrinos in $\pi^\pm$ we find that 
\begin{equation} \mathrm{m}(\nu_\mu) = 
49.91\,\mbox{milli-eV}/\mathrm{c}^2\,.
\end{equation}
Since the same considerations apply for either the $\pi^+$ or 
the $\pi^-$ meson it follows that  
\begin{equation}\mathrm{m}(\nu_\mu) = \mathrm{m}(\bar{\nu}_\mu)\,.
\end{equation}
Experimental values for the masses of the different neutrino types are
not available. However, it appears that for the $\nu_\mu \leftrightarrow 
\nu_e$ oscillation the value
for $\Delta$m$^2$ = m$^2_2$ $\mathrm{-}$ m$^2_1$  =
3.2$\times10^{-3}$\,eV$^2$ given on p.1565 of [37] can be used to
 determine m$_2$ = m($\nu_\mu$) if m$_1$ = m$(\nu_e$) is much smaller 
than m$_2$. We have then m($\nu_\mu$) $\approx$ 56.56\,milli-eV/c$^2$,
which is compatible with the value of m($\nu_\mu$) given in Eq.(70).

   From Eqs.(67,70) follows that, within the particles,\\

\begin{equation}
\bullet\hspace{1.5cm}\mathrm{m}(\nu_e) = 1/136.74\cdot\mathrm{m}(\nu_\mu)
 \cong \alpha_f\mathrm{m}(\nu_\mu)\,.
\end{equation}\\
\noindent
1/136.74 is 1.00217 times the fine structure constant $\alpha_f$ = 
e$^2/\hbar$c =\\ 1/137.036. It does not seem likely that Eq.(72) is just 
another `coincidence'. The probability for this being a `coincidence' 
is zero, considering the infinite pool of numbers on which the 
ratio m($\nu_e$)/m($\nu_\mu)$ could settle.

\bigskip

   The mass of the $\tau$\,neutrino $\nu_\tau$ can be 
determined from the decay D$_s^\pm$  $\rightarrow$ 
$\tau^\pm$ + $\nu_\tau\,(\bar{\nu}_\tau)$,
and the subsequent decay $\tau^\pm \rightarrow \pi^\pm + \bar{\nu}_\tau
(\nu_\tau)$, stated in [2]. The appearance of $\nu_\tau$
in the decay of D$^\pm_s$ 
and the presence of $\nu_\mu,\bar{\nu}_\mu,\nu_e,\bar{\nu}_e$ neutrinos 
in the $\pi^\pm$ decay product of the $\tau^\pm$\,leptons means that 
$\nu_\tau,\bar{\nu}_\tau,\nu_\mu,\bar{\nu}_\mu,\nu_e,\bar{\nu}_e$ 
neutrinos must be  in the D$^\pm_s$ lattice. 
The additional $\nu_\tau$ and $\bar{\nu}_\tau$ neutrinos can be
accomodated in D$^\pm_s$ by a body-centered cubic lattice,
in which there is in the center of each cubic cell one particle different
from the particles in the eight cell corners (Fig.\,5). In a body-centered 
cubic lattice are N/8  cell centers. If the particles in the cell 
centers are tau neutrinos, then N/16 tau neutrinos $\nu_\tau$ and
N/16 anti-tau neutrinos $\bar{\nu}_\tau$  must be present, 
because of conservation of neutrino numbers. From m(D$^\pm_s$) = 
1968.47\,MeV/c$^2$ and m($\tau^\pm$) = 
1776.82\,MeV/c$^2$  follows that 

\begin{equation} \mathrm{m}(\mathrm{D}^\pm_s) - \mathrm{m}(\tau^\pm) = 
191.15\,\mathrm{MeV/c^2} = \mathrm{N}/8
\cdot\mathrm{m}(\nu_\tau)\,.\end{equation}
\noindent
The theoretical mass of the $\tau$\,neutrinos is therefore
\begin{equation} \mathrm{m}(\nu_\tau) = \mathrm{m}(\bar{\nu}_\tau) = 
0.537\,\mathrm{eV/c}^2\,.
\end{equation}
\noindent
From the neutrino masses given by Eq.(74) and Eq.(70) follows that 
\begin{equation}
\mathrm{m}(\nu_\tau) = 10.76\,\mathrm{m}(\nu_\mu) = 
1.048\,(\alpha_w/\alpha_f)\mathrm{m}(\nu_\mu)\,,
\end{equation}
where $\alpha_w$  is the weak coupling constant $\alpha_w$  = 
$g_w^2/4\pi\hbar$c, (Eq.43), 
and $\alpha_f$  is the fine  structure constant. We keep in mind that 
g$_w^2$ in $\alpha_w$  is not nearly as accurately known  
as e$^2$ in $\alpha_f$ = e$^2$/$\hbar$c. With Eq.(72) we find that 
\begin{equation}
\mathrm{m}(\nu_\tau) = 1.048\cdot(\alpha_w/\alpha_f^2)\cdot
\mathrm{m}(\nu_e) = 1474\,\mathrm{m}(\nu_e)\,. \end{equation}

   According to Eqs.(67,70,74) the sum of the masses of the electron
neutrino, muon neutrino and tau neutrino is 0.586\,eV/c$^2$, primarily
because of the mass of the tau neutrino. According to the Review of 
Particle Physics (2004,\,p.439) it follows from astrophysical data that 
the sum of the neutrino masses $\sum_{i}$\,m${(\nu{_i})}$ $\leq$ 
0.7\,eV/c$^2$. We arrive at essentially the same result.

\smallskip

   To summarize what we have learned about the masses of the leptons: 
After we have found in Section 9 an explanation for the mass of the
muon and $\tau$ lepton, we have also determined the masses of the 
electron-, muon-, and tau neutrinos and antineutrinos. In other words,
we have found the masses of all leptons, exempting the electron, 
which will be dealt with in the next Section.

\section{Neutrinos in the electron}

The electron or positron differ from the other particles we have considered 
so far as it appears that their charge e$^\pm$ cannot be separated from 
their mass m(e$^\pm$), 
whereas in the other charged particles the mass of the charge is,
in a first approximation, unimportant for the mass of the particles. Even 
in the rest mass of the muons the mass of the electron contributes only 
five thousandth of the muon mass. On the other hand, the electron or
positron are fundamental for the stability of the charged 
particles, whose lifetime is sometimes orders of magnitude larger than the 
lifetime of their neutral counterparts. For example the lifetime of the 
$\pi^\pm$\,mesons is eight orders of magnitude larger than the lifetime of 
$\pi^0$, the lifetime of the proton is infinite, whereas the neutron decays 
in about 900 seconds and, as a startling example, the lifetime of 
$\Sigma^\pm$ is O($10^{-10}$) seconds, whereas the lifetime of 
$\Sigma^0$ is O($10^{-20}$) seconds. There is something particular 
to the interaction of the elementary electric charge with the particle masses.  

   J.J. Thomson [52] discovered the small corpuscle, which soon 
became known as the electron, more than 110 years ago. An enormous 
amount of theoretical work
has been done to explain the existence of the electron. Some of the most
distinguished physicists have participated in this effort. Lorentz [53], 
Poincar\'{e} [54], Ehrenfest [55], Einstein [56], Pauli [57], and others 
showed that it is fairly certain that the electron cannot be explained as  
a purely electromagnetic particle following Maxwell's equations. In 
particular it was not clear 
how the charge of the electron could be held together in its  
small volume, because the internal parts of the charge repel each other. 
Poincar\'{e} [58] did not leave it at showing that such an electron
could not be stable, but suggested a solution for the problem by 
introducing what has become known as the Poincar\'{e} stresses, whose 
origin however remained unexplained. These studies were concerned with the
static properties of the electron, its mass m(e) and its 
charge e\,; the positron, the spin and the neutrinos were not known at that
time. In order to explain the electron with its existing mass and 
charge it appears to be necessary to add to Maxwell's equations a 
non-electromagnetic mass and a non-electromagnetic force which could 
hold the charge of the electron together. We shall see what this mass and 
force is. 

   The discovery of the spin of the electron by Uhlenbeck and Goudsmit in 
1925 [59] increased the difficulties of the problem in so far as it now
had also to be explained how the angular momentum $\hbar$/2 and the 
magnetic moment $\mu_e$ come about.
The spin of a point-like electron seemed to be explained by Dirac's  
equation [60], however it turned out later [61] that Dirac type equations can  
be constructed for any value of the spin. Afterwards Schr\"{o}dinger [62] 
tried to explain the spin and the magnetic moment of the electron with 
the so-called Zitterbewegung. Dirac [63] suggested a model of an electron
without spin, consisting of a charged, hollow sphere held together by surface 
tension. The first higher mode of oscillation appeared to be the muon, to 
quote ``one can look upon the muon as an electron excited by radial
oscillations". Many other models 
of the electron were proposed. On p.74 of his book ``The Enigmatic 
Electron" Mac\,Gregor [64] lists more than thirty such attempts.

   At the end none of these models has been successful because 
the problem developed a seemingly insurmountable difficulty
when it was shown, through electron scattering experiments, 
that the charge radius of the electron must be smaller
than $10^{-16}$\,cm [65], in other words that the electron appears 
to be a point particle, at least by three orders of magnitude smaller  
than the classical electron radius r$_e$  =  e$^2$/m$_e$c$^2$  = 
2.8179$\cdot10^{-13}$\,cm. This, of 
course, makes it very difficult to explain how a particle can have
a finite angular momentum when its radius goes to zero, and how a 
charge e$^\pm$ can be confined in an infinitesimally small volume. 
If the charge e$^\pm$ 
would be in a volume with a radius of O($10^{-16}$)\,cm the
Coulomb self-energy would be orders of magnitude larger than the 
rest mass of the electron, which is not realistic. The choice is between 
a massless point charge and a finite size particle with a non-interacting 
mass to which the charge e$^\pm$ is attached. It seems fair  
to say that at present, more than 110 years after the discovery of the 
electron, we do not have an accepted theoretical explanation of the 
electron.

   We propose in the following, as in [66], that the non-electromagnetic 
mass which seems to be necessary in order to explain the electron 
consists of neutrinos. If it is true that the electron cannot be explained 
with Maxwell's equations, what else could be in the electron but something
neutral\,? What neutral thing with a mass smaller than the electron do we 
know but the neutrinos\,? Neutrinos in the electron are
actually a necessary consequence of our 
model of the mass of the $\pi^\pm$\,mesons and of the decay 
sequence of $\pi^\pm$. And we propose that the non-electromagnetic 
force required to hold the charge and the neutrinos in the electron 
together is the weak nuclear force which, as we have suggested, holds 
together the masses of the mesons and baryons and the mass of 
the muons as well. Since the range of the weak nuclear force is on 
the order of $10^{-16}$\,cm the neutrinos in the electron must be 
arranged in a lattice, with the weak force extending from each lattice  
point only to the nearest neighbors.  An O($10^{-13}$)\,cm size of  
the neutrino lattice in the electron does not at all contradict the result 
of the scattering experiments that the radius of the electron should 
be O(10$^{-16}$)\,cm, just as the  explanation of the mass of 
the muons with our model does not contradict the 
apparent point particle characteristics of the muon. Neutrinos 
are in a very good approximation non-interacting, and therefore are 
not noticed in scattering experiments with electrons.

\medskip 

   The rest mass of the electron is\\ 

\hspace{2cm} m(e) = 0.510\,998\,91 $\pm$ 1.3$\cdot10^{-8}$\,
MeV/c$^2$\,,\\ 

\noindent and the electrostatic charge of the
electron is e = 4.803\,204\,27\,$\cdot\,10^{-10}$\,esu, as stated 
in the Review  of Particle Physics. \emph{The objective of a 
theory of the electron must, first of all, be the explanation of 
m(e$^\pm$) and e$^\pm$}, but also of the spin of (e$^\pm$) and 
of the magnetic moment $\mu_e$. We will first  
explain the rest mass of the electron, making use of what we
have learned  about the explanation of the mass of the 
muons in Section 9. The muons are leptons,
just as the electrons, that means 
that they interact  with other particles exclusively through 
the electric force. The muons have a mass which is 206.768 
times larger than the mass of the electron, but they have the same 
charge as the electron or positron and the same spin. 
Scattering experiments tell that the muons are point particles 
with a size $<$\,$10^{-16}$\,cm, just as the electron. In other 
words, the muons have the same characteristics as the electrons
and positrons but for a mass which is about 200 times larger. 
Consequently the muon is often referred to as a ``heavy"  electron. 
If a non-electromagnetic  mass is required to explain
the mass of the electron, then a non-electromagnetic mass 200 times
as large as in the electron is required to explain the mass of the muons.
These non-electromagnetic masses  \emph{must be non-interacting}, 
otherwise scattering experiments could not find the size of either the 
electron or the muon at 10$^{-16}$\,cm. 
 
   We have explained the mass of the muons in Section 9. According 
to our model the muons consist of an oscillating lattice of muon 
neutrinos, or\,\,antimuon\,\,neutrinos,
electron neutrinos or anti-electron neutrinos, and a charge 
e$^\pm$. Neutrinos are the only non-interacting matter we know of. 
In the muon are, according to our model,
(N\,-\,1)/4 $\cong$ N/4 muon neutrinos $\nu_\mu$  
(respectively antimuon neutrinos $\bar{\nu}_\mu$), 
N/4 electron neutrinos $\nu_e$ and the same 
number of anti-electron neutrinos $\bar{\nu}_e$,  one 
electric charge e$^\pm$ and the energy of the lattice oscillations. 
The letter N stands for the number of all neutrinos and antineutrinos in
the cubic lattice of the $\pi^\pm$ mesons, N = 2.854$\cdot10^{\,9}$, 
Eq.(15). It is a necessary consequence of the
$\nu_\mu$,\,$\bar{\nu}_\mu$,\,$\nu_e$,\,$\bar{\nu}_e$ neutrino lattice 
of $\pi^\pm$ and the decay sequence $\pi^-$ $\rightarrow$  
$\mu^-$ + $\bar{\nu}_\mu$ and   
$\mu^-$  $\rightarrow$ e$^-$ + $\bar{\nu}_e$ + $\nu_\mu$,
that \emph{there must be N/4 electron neutrinos $\nu_e$ in 
the emitted electron}, as stated on p.\,32, quote: ``Since (N - 1)/4 
electron neutrinos must be in the $\pi^-$ lattice it follows that
(N - 1)/4 electron neutrinos $\nu_e$ must go with the electron 
emitted in the $\mu^-$ decay". Since N - 1 differs from N by 
one in 10$^{\,9}$ we replace N - 1 by N. 

   The explanation of the $\pi^\pm$\,mesons led to the 
explanation of the muons and now leads  
to the explanation of the mass of e$^\pm$. For the 
mass of the electron neutrinos or anti-electron neutrinos  
we found in Eq.(67) that m($\nu_e$) = m($\bar{\nu}_e$) = 
0.365\,milli-eV/c$^2$. The energy in the sum of the
masses of the\\ (N - 1)/4 $\cong$ N/4 electron neutrinos or
anti-electron neutrinos in the lattice of the electron or positron is then 
\begin{equation} \sum_i{\,\mathrm{m(\nu_e)c^2}} =
 \mathrm{N}/4\cdot\mathrm{m}(\nu_e)\mathrm{c}^2 
= 0.260\,43\,\mathrm{MeV}
 = 0.5096\,\mathrm{m(e^-)}\mathrm{c}^2\,. \end{equation}
\noindent
In other words: 

\begin{itemize}
\item
 \emph{1/2 of the rest mass of the electron is approximately equal\\ 
to the sum of the masses of the neutrinos in the electron}.
\end{itemize}

 In modern parlance this is the ``bare" 
part of the electron. The bare part is not observable.  
The other half of the rest mass of the electron must originate from 
the charge e$^-$ carried by the electron.  

   From pair production 
$\gamma$ + M   $\rightarrow$  e$^-$ + e$^+$ + M, (M being any
nucleus), and from conservation of neutrino numbers follows 
necessarily that there must also be a neutrino lattice 
 composed of N/4 anti-electron neutrinos, which make 
up the lattice of the positrons, which lattice has, since m($\nu_e$) = 
m($\bar{\nu}_e$), the same mass as the neutrino lattice of the 
electron, as it must be for the antiparticle of the electron. 
Conservation of charge depends on the conservation of neutrino 
numbers. If the electron consists to one-half of electron neutrinos then 
it cannot decay, because that would violate conservation of neutrino 
numbers. In our model the electron is stable. Stability is an 
essential part of a realistic model of the electron.      
   
   Fourier analysis dictates that a continuum of high frequencies must be 
in the electrons or positrons created by pair production in a timespan of 
$10^{-23}$ seconds. We will now determine the  oscillation energy 
E$_\nu$(e$^\pm$) in the  interior of the electron. 
Since we want to explain the \emph{rest mass} of the electron we can 
only consider the frequencies of non-progressive waves, either standing 
waves or circular waves. The sum of the energies of linear lattice 
oscillations is, in the case of the $\pi^\pm$\,mesons,  given by 

\begin{equation} \mathrm{E}_\nu(\pi^\pm) =  
\frac{\mathrm{Nh} \nu_0}{(\mathrm{e^{h\nu/kT}}\,\mathrm{-}\,1)}
\cdot\frac{1}{2\pi}\int\limits_{-\pi}^{\pi}\phi\,d\phi\,. \end{equation}
\noindent
This equation was used to determine the oscillation energy in  
the $\pi^\pm$ mesons, Eq.(32). This type of equation 
was introduced by Born and v.\,Karman [14] in order to explain the 
internal energy of cubic crystals.  If we apply Eq.(78) to the  
electron, which has N/4 oscillating electron neutrinos $\nu_e$, we 
arrive at  E$_\nu$(e$^\pm)$  = 1/4\,$\cdot$\,E$_\nu(\pi^\pm$). 
This is mistaken because E$_\nu(\pi^\pm$) $\approx$ 
m($\pi^\pm$)c$^2$/2 and m($\pi^\pm$) = 273\,m(e$^\pm$),
so E$_\nu$(e$^\pm$) would be 273/8\,$\cdot$\,m(e$^\pm$)c$^2$ = 
34.1\,m(e$^\pm$)c$^2$. 
Eq.(78) must be modified in order to be suitable for the monatomic 
oscillations in a free electron. It turns out that we must use
\begin{equation} \mathrm{E}_\nu(\mathrm{e}^\pm) =  
\frac{\mathrm{Nh}\nu_0\cdot\alpha_f}{(\mathrm{e^{h\nu/kT}}\,
\mathrm{-}\,1)}\cdot\frac{1}{2\pi}\int\limits_{-\pi}^{\pi}\phi\,d\phi\,,
\end{equation}
\noindent
where $\alpha_f$ is the fine structure constant. As is well-known the 
fine structure constant $\alpha_f$ characterizes the strength of the 
electromagnetic forces. The appearance of $\alpha_f$ in Eq.(79)
means that the nature of the oscillations in the electron is different 
from the oscillations in the $\pi^0$ or $\pi^\pm$ lattices. With 
$\alpha_f$  = e$^2/\hbar$c 
and $\nu_0$ = c/2$\pi$\emph{a} we have
\begin{equation} \mathrm{h}\nu_0\alpha_f = \mathrm{e}^2/\emph{a}\,, 
\end{equation}
which shows that the oscillations in the electron are \emph{electric 
oscillations}. The appearance of e$^2$  in Eq.(80) guarantees that
the oscillation energy of the electron and positron are the same. 

   There must be N/2 oscillations of the elements of the charge 
in e$^\pm$, because we deal with non-progressive circular waves, which
are the superposition of two waves. That means that  N in Eq.(79) must  
be replaced by N/2. As we will see later the spin requires 
that the oscillations are circular. From Eqs.(78,79) then follows that
\begin{equation} \mathrm{E_\nu(e}^\pm) =
 \alpha_f/2\cdot\mathrm{E}_\nu(\pi^\pm)\,. \end{equation}
\noindent
E$_\nu(\pi^\pm)$  is the oscillation energy in the $\pi^\pm$\,mesons 
which  can be calculated with Eq.(78). According to Eq.(32) it is
\begin{equation} \mathrm{E}_\nu(\pi^\pm)(theor) = 67.82 \,\mathrm{MeV} =
0.486\,\mathrm{m}(\pi^\pm)\mathrm{c}^2 \approx 
\mathrm{m}(\pi^\pm)\mathrm{c}^2/2\,.
 \end{equation}
With E$_\nu(\pi^\pm$) $\approx$ m($\pi^\pm$)c$^2$/2 = 
139.57/2\,MeV and $\alpha_f$  = 1/137.036  follows from Eq.(81) 
that the oscillation energy of the electron or positron is 
\begin{equation} \mathrm{E_\nu(e}^\pm) = \frac{\alpha_f}{2}\cdot
\frac{\mathrm{m}(\pi^\pm)\mathrm{c}^2}{2} = 
0.254\,623\,\mathrm{MeV} = 
0.996\,570\,\mathrm{m}(\mathrm{e}^\pm)\mathrm{c}^2/2\,. 
\end{equation}
\noindent
Another `coincidence'\,? If we replace in Eq.(83) the experimental value 
for m($\pi^\pm$) by the good empirical approximation 
m($\pi^\pm$) $\cong$ m(e$^\pm$)(2/$\alpha_f$ - 1), Eq.(103), then 
it follows likewise that  
\begin{equation}\mathrm{E}_\nu(\mathrm{e}^\pm) 
 \cong 1/2\cdot \mathrm{m(e^\pm)c}^2 \,.\end{equation} 
In other words\,:

\medskip

\hspace{2cm} \emph{1/2 of the energy in the rest mass of a free}\\ 
 \hspace*{2cm} \emph{electron is made up by the oscillation energy in e$^\pm$}.

\bigskip

\noindent This equation corresponds to Eq.(34) for the oscillation energy in 
the $\pi^\pm$\,me-sons. The other half of the energy in e$^\pm$ is in the
energy of the sum of the neutrino masses, Eq.(77).

   In Eq.(83) we have determined the value of the oscillation energy in 
e$^\pm$  from the product of the very accurately known fine structure 
constant and the very accurately 
known rest mass of the $\pi^\pm$\,mesons. This establishes 
a firm value of the oscillation energy of e$^\pm$.  We can confirm 
Eq.(83) without  using E$_\nu(\pi^\pm)$ with the formula for the 
oscillation energy in the form of Eq.(88) with N/2 = 1.427$\cdot10^{\,9}$, 
e = 4.803$\cdot10^{-10}$\,esu,
 $\emph{a}$ = 1$\cdot10^{-16}$\,cm, f(T) = 1.305$\cdot10^{13}$, 
and with the integral being $\pi^2$, we obtain E$_\nu$(e$^\pm$) = 
0.968\,m(e$^\pm$)c$^2$/2. This calculation involves more 
parameters than in Eq.(83) and is consequently less accurate than Eq.(83).

   In a good approximation the oscillation energy of e$^\pm$ in 
Eq.(83) is equal to the energy in the sum of the masses
of the electron neutrinos or anti-electron neutrinos in the e$^\pm$ lattice 
in Eq.(77). That means that, with\\ N\,-\,1 $\cong$ N,

\hspace{2.1cm} E$_\nu$(e$^-$) = $\Sigma_i$\,m($\nu_e$)c$^2$ = 
N/4\,$\cdot$\,m($\nu_e$)c$^2$. \hspace{2.6cm} (84a)\\

\noindent 
We find from the sum of the neutrino masses and the oscillation energy that 
 \begin{equation} \mathrm{m(e}^-) 
= \mathrm{N/2}\cdot\mathrm{m}(\nu_e) \,. 
\end{equation}
\noindent
The mass of the positron m(e$^+$) is, likewise, equal to N/2 
times the mass of the anti-electron neutrino m($\bar{\nu_e}$). From 
Eq.(85) follows with m($\nu_e)$ = m($\bar{\nu}_e$) =
0.365\,milli-eV/c$^2$, Eq.(67), that in our model\\

$\bullet$\hspace{1.1cm}\emph{the rest mass of a free electron or positron is} 
    
\begin{eqnarray} \mathrm{m(e^\pm)c^2}(theor) =
\mathrm{N/2}\cdot{\mathrm{m}(\nu_e)\mathrm{c}^2}
= \mathrm{N/2}\cdot{\mathrm{m}(\bar{\nu}_e)\mathrm{c}^2}\nonumber\\ 
 = 0.5208\,\mathrm{MeV} = 1.019\,\mathrm{m(e}^\pm)\mathrm{c}^2(exp)\,.
 \end{eqnarray}\\
The theoretical rest mass of the electron in our model agrees, within the 
accuracy of the parameters N and m($\nu_e)$,  with the measured rest 
mass of the electron. If, for a comparison, we add the numerical
value of Eq.(77) to the numerical value of Eq.(83), we have 
m(e$^\pm$)c$^2$(\emph{theor}) = 0.5151\,MeV = \newline
1.0080\,m(e$^\pm$)c$^2$(\emph{exp}). 
This is also compatible with m(e$^\pm$)c$^2$(exp).

   From Eq.(81) follows with E$_\nu(\pi^\pm)$ $\cong$ 
m($\pi^\pm$)c$^2$/2  from Eq.(82) that

\vspace{0.5cm} 
\centerline{m(e$^\pm$)c$^2$ $\cong$  2E$_\nu$(e$^\pm) =
 \alpha_f$E$_\nu(\pi^\pm)$ $\cong$ $\alpha_f$m$(\pi^\pm)$c$^2$/2  \,,}

\vspace{0.4cm}
\noindent
or that
\begin{equation} \mathrm{m(e^\pm)}\cdot2/\alpha_f  =
 274.072\,\mathrm{m(e^\pm)} 
= 1.0034\,\mathrm{m(\pi^\pm)}\,. \end{equation}
The ratio m($\pi^\pm$)/m(e$^\pm$) $\cong$ 2/$\alpha_f$ = 274.072, which
follows from Eq.(87), comes close to m($\pi^\pm$)/m(e$^\pm$)(\emph{exp}) = 
273.13. It is a necessary condition for the validity of our model of 
the electron that we come up with a correct value of 
m($\pi^\pm$)/m(e$^\pm$) which, of course, depends on a valid
explanation of m(e$^\pm$). 

   We have thus shown that the \emph{rest mass of the electron or 
positron can be explained} by the sum of the masses of the 
electron neutrinos or anti-electron neutrinos in a cubic  lattice, 
with N/4 electron neutrinos or N/4 anti-electron 
neutrinos, plus the mass in the sum of the energy of 
N/2 standing electric oscillations in the lattice, Eq.(83). A section 
through the lattice is shown in Fig.\,8. 

\begin{figure}[h]
\vspace{0.5cm}
\hspace{2.2cm}
\includegraphics{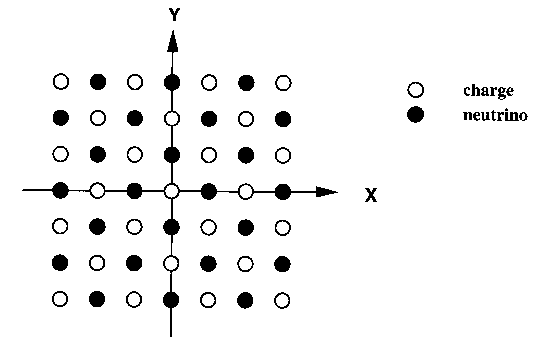}
\vspace{-0.2cm}
\begin{quote}
Fig.\,8: Horizontal or vertical section through the central part \\ 
\indent\hspace{1.1cm} of the electron lattice.
\end{quote}
\end{figure}

   From our model follows, 
since it deals with a cubic neutrino lattice, that \emph{the 
electron is not a point particle}. However, since neutrinos are 
non-interacting their presence will not be detected in the usual
electron scattering experiments. The charge radius of the 
electron determined by electron scattering experiments is 
$<$\,$10^{-16}$\,cm [65] and seems to contradict the 
model of the electron proposed here, whose size is on the order  
of 10$^{-13}$\,cm as shown in Appendix B. However, the
experimental charge radius does not apply to the circumstances 
considered here. One has to find the scattering formula for finite 
size charged cubic  lattices  and analyze the experimental data 
with such a scattering formula, in order to see whether our model 
is in contradiction to the experiments. As with the muon, the forces 
which originate from the distribution of mass and charge in the
electron are likely to come from the \emph{center of mass} and
\emph{center of charge}, which are true points. That, however, does 
not mean that the electron and the muon are point particles.

\bigskip   
 
   In order to confirm the \emph{validity} of our preceding explanation  
of the mass of the electron we must show that the sum of the charges 
in the electric oscillations in the interior of the electron is equal 
to the charge of the electron.   
We recall that Fourier analysis requires that, after pair production,
there must be a continuum of frequencies in the electron and positron.
With h$\nu_0\alpha_f$ = e$^2$/\emph{a} from Eq.(80) follows 
from Eq.(79) that the oscillation energy of the simple cubic lattice
in e$^\pm$ is the sum of N/2 circular electric oscillations
\begin{equation} \mathrm{E}_\nu(\mathrm{e}^\pm) =  
\frac{\mathrm{N}}{2} \cdot \frac{\mathrm{e^2}}{\emph{a}}\cdot 
\frac{1}{f(T)} \cdot \frac{1}{2\pi}
\int\limits_{-\pi}^{\pi}\phi\,\mathrm{d}\phi\,, \end{equation}
with f(T) = (e$^{h\nu/kT} \mathrm{-}$ 1) = 1.305$\cdot10^{13}$
from Eq.(21). Inserting the values for N, e, f(T) and \emph{a} we find
that E$_\nu$(e$^\pm$) = 0.968\,m(e$^\pm$)c$^2$/2 $\cong$ 
m(e$^\pm$)c$^2$/2, similar to Eq.(83). The discrepancy
between m(e$^\pm$)c$^2$/2 and E$_\nu$(e$^\pm$) so calculated
must originate from the uncertainty of the parameters N, f(T) and \emph{a}
in Eq.(88).

   In order to determine the charge e in the electric oscillations we 
replace the integral divided by 2$\pi$ in Eq.(88), which has the value
$\pi$/2 and in which $\phi$ is continous,
by the sum over the elements $\Delta\phi$, that means by 
$\Sigma_k\,\phi_k\Delta\,\phi$, where k is an integer 
number with the maximal value k$_m$ = (N/4)$^{1/3}$.  The number 
$k_m$  is the number of the charge elements between the center and 
the end of the lattice on the $\phi$ axis. $\phi_k$ is equal to 
k$\pi$/k$_m$ and we have

\begin{displaymath}  \Sigma_k\,\phi_k\,\Delta\phi = 
\sum_{k=1}^{k_m}\,\frac{k\pi}{k_m}\cdot\frac{1}{k_m}
= \frac{ k_m(k_m + 1)\pi}{2\,k_m^2} \cong \frac{\pi}{2}\,.
\end{displaymath}
\noindent The energy in the \emph{individual 
electric  oscillation} with index k is  
\begin{equation} \Delta\mathrm{E}_\nu(k) = 
\mathrm{const}\cdot\mathrm{e}^2\cdot\phi_k\,\Delta\phi = 
\mathrm{const}\cdot\mathrm{e}^2\cdot k\pi/k_m^2\,, \end{equation}
and increases linearly with k. The charge in the electron is not only 
distributed over N/4 charge elements, but each charge element 
has a different quantized energy. 

   Suppose that the energy of the electric oscillations is correctly 
described by the self-energy of an electric charge Q 
\begin{equation}\mathrm{U} = 1/2\,\cdot\,\mathrm{Q}^2/\mathrm{r}\,.
\end{equation}
The self-energy of the charge e$^\pm$ is normally used to 
determine the mass of the electron from its charge, here we use Eq.(90) 
the other way around, we determine the charge from the energy  
in the oscillations. 

   The charge of the electron is contained in the electric oscillations. 
That means that the charge e$^\pm$ is not concentrated in a
point, but is distributed over N/4 = O($10^9)$ charge elements
Q$_k$, each charge element consisting of two perpendicular 
oscillations. \emph{The charge elements are distributed in a cubic lattice} 
and the resulting electric field is cubic, not spherical. In the absence of a 
central force, which originates at the center of the particle and affects all  
parts of the particle, the configuration of the particle is not spherical but 
cubic, just as it was with the shape of the $\pi^\pm$\,mesons. For 
distances large as compared 
to the sidelength of the cube, (which is O($10^{-13}$)\,cm), say at the 
first Bohr radius which is on the order of $10^{-8}$\,cm, the deviation 
of the cubic field from the spherical field will have been reduced by
about $10^{-10}$.
 
   The charge in all electric oscillations in the electron is  
\begin{equation} \mathrm{Q} = \sum_{k}\,\mathrm{Q}_\mathrm{k}\,. 
\end{equation}

\noindent Setting the radius r in the formula for the self-energy (Eq.90) equal  
to 2\,\emph{a}, one charge element is separated from the nearest other 
by 2\emph{a},  we find, with Eqs.(88,89,90), that the charge in the 
individual electric oscillations is 

\begin{equation} \mathrm{Q_k} = 
\pm\,\sqrt{2\pi\,N\,e^2/f(T)k_m^2}\,\cdot\,\sqrt{k} = \pm\,\mathrm{e}/k_m
\cdot\sqrt{2\pi\,N\,/f(T)}\cdot\sqrt{k}\,,
\end{equation}  
\noindent
with k$_m$ = 1/2\,$\cdot$\,(N/4)$^{1/3}$ = 447.  The number $k_m$ 
is the number of the charge elements between the center and the end of
the lattice on the $\phi$ axis. 

   In order to find Q from Eq.(91) we need the sum of the $\sqrt{k}$ in 
Eq.(92), which has been computed to be

\begin{equation}  \sum_{k=1}^{k_m}\,\sqrt{k} = 6310.8\,. \end{equation}

It follows, after we have doubled the sum over $\sqrt{k}$, 
because for each index k there is a second oscillation on the negative 
axis of $\phi$, that

\begin{equation}
\mathrm{Q} = \Sigma_k\,\mathrm{Q_k} = \pm\,1.0467\cdot\mathrm{e}\,\,=
\pm\,5.027\cdot10^{-10}\,\,\mathrm{esu}\,.\end{equation}

\noindent Because of the plus-minus sign the absolute value of the sum 
of the charges is the same in the electron or positron, as it must be.   
The elementary electric charge is e$^\pm$ = 
$\pm$\,4.803\,$\cdot\,10^{-10}$\,esu.  
Within the uncertainty of the parameters the theoretical charge 
of the electron agrees with its experimental value, as well as,
with the opposite sign, with the charge of the positron. That means that
we have confirmed that it follows from our model of the 
electron that the electron has, within a 5\% error, the correct electric 
charge.

   Each element of the charge distribution is surrounded in the horizontal
plane by four electron neutrinos as in Fig.\,8, and in vertical direction by   
an electron neutrino above and also below the element. The electron 
neutrinos hold the charge elements in place.  We must assume that 
the charge elements are bound to the neutrinos by the weak nuclear 
force. The weak nuclear force plays here a role similar 
to its role in holding, for example,  the $\pi^\pm$ or $\mu^\pm$ lattices 
together. It is not possible, in the absence of a definitive explanation 
of the neutrinos, to give an explanation for the electro-weak 
interaction between the electric oscillations and the neutrinos. 
However, the presence of the range
\emph{a} of the weak nuclear force in e$^2$/\emph{a} is a sign  
that the weak force is involved in the electric oscillations. The 
attraction of the charge elements by the neutrinos
overcomes the Coulomb repulsion of the charge elements.
The weak nuclear force is the missing non-electromagnetic force   
or the Poincar\'{e} stress which holds the electric charge 
together. The same considerations apply 
for the positive charge of the positron, only that then the 
electric oscillations are all of the positive sign and that they are 
bound to anti-electron neutrinos.

   As far as the charge in the $\pi^\pm$\,mesons is concerned, the
N/4 electron neutrinos from, say, e$^-$ react with N/4 anti-electron
neutrinos, or possibly with N/4 antimuon neutrinos, in the neutral 
neutrino lattice of the pions. The neutrinos coming with e$^\pm$ into 
the pion lattice can only interact with antineutrinos because of the Pauli 
principle. They then create N/4 neutrino dipoles. In the formation of the
dipoles energy is lost, which sums up as the binding energy of e$^\pm$
to the neutral neutrino lattice of the pions.

   The N/4 charge elements of e$^\pm$ are apart by the distance 2\emph{a},
just as Cl ions in Fig.(4) are apart by twice the lattice constant. The
volume filled by the charge elements coming with the electron or positron 
into the neutrino lattice of $\pi^\pm$ is equal to 
N/4\,$\cdot$\,(2\emph{a})$^2$\,$\cdot$\,\emph{a} = N\emph{a}$^3$. That means 
that the charge e$^\pm$ added to the neutrino lattice of $\pi^\pm$ fills the 
entire volume of the neutrino lattice, in other words the volume of the 
$\pi^\pm$\,mesons.

   Finally we learn that  Eq.(88) precludes the possibility that the 
charge of the electron sits only on its surface. The number N in 
Eq.(88) would then be on the order 
of $10^{\,6}$, whereas N must be on the order of $10^{\,9}$ so that
E$_\nu$(e$^\pm$) can be m(e$^\pm$)c$^2$/2, as is necessary. 
In other words, the charge of the electron must be distributed 
throughout the interior of the electron, as we postulated.

\medskip  
 
   Summing up: The rest mass of the electron or positron originates 
from the sum of the masses of N/4 electron neutrinos or 
anti-electron neutrinos in cubic lattices plus the mass in the energy of 
the electric oscillations in their neutrino lattices. The neutrinos, as 
well as the electric oscillations, make
up 1/2 of the rest mass of e$^\pm$ each.
The electric oscillations are bound to the neutrinos by the weak 
nuclear force. The sum of the charge elements of the electric oscillations 
accounts for the charge of the electron, respectively positron. The 
electron or the positron are not point particles. The electron is stable.
  
   One hundred years of sophisticated theoretical work have made it 
abundantly clear that the electron is not a purely electromagnetic particle
described by Maxwell's equations. 
There must be something else in the electron but electric charge, 
otherwise the electron could not be stable. It is equally clear from the 
most advanced scattering experiments that the  ``something else"  in the 
electron must be non-interacting, otherwise it could not  be that we find that 
the charge radius of the electron must be smaller than $10^{-16}$\,cm. 
The only non-interacting matter we know of with certainty are the neutrinos. 
So it seems to be natural to ask whether neutrinos are not part of the 
electron. Actually we did not introduce the neutrinos in the electron in 
an axiomatic way, but rather as a 
consequence of our explanation of the stable mesons, baryons
and muons. It follows necessarily from this model that after the decay 
of, say, the $\mu^-$\,meson there must be electron neutrinos in the 
emitted electron, and that they make up one half of the rest mass of the 
electron. The other half of the energy in the electron originates from the 
energy of the electric oscillations. With a cubic lattice of 
anti-electron neutrinos we also arrive with the same considerations as 
above at the correct  mass and charge of the positron.

\section{The magnetic moment of the electron}

    If one half of the mass of the electron does indeed  consist of 
neutrinos, then the magnetic moment of the electron can be explained 
immediately. The magnetic moment of the electron is
known with extraordinary accuracy, $\mu_e$  = 
1.001\,159\,652\,186\,$\mu_B$, according to the Review of Particle 
Physics, $\mu_B$ is Bohr's magneton. The decimals after 
1.00\,$\mu_B$  are caused by the anomalous magnetic moment which 
we will not consider. As is well-known the magnetic dipole moment of 
a particle with spin is, in Gaussian units, given by
\begin{equation} \vec{\mu} = g\,\frac{e\hbar}{2mc}\,\vec{s}\,,
 \end{equation}
where g is the dimensionless Land\'{e} factor, m the rest mass of  
the particle that carries the charge e, and $\vec{s}$  the spin vector. 
The g-factor has been introduced in order to bring the magnetic 
moment of the electron into agreement with the experimental facts. 
As Uhlenbeck and Goudsmit [59] postulated, and as has been 
confirmed experimentally, 
the g-factor of the electron is 2. With the spin s = 1/2  and g = 2  
the magnetic dipole moment of the electron is  
\begin{equation} \mu_e = e\hbar/2m(e^\pm)c\,,
 \end{equation}
or one Bohr magneton $\mu_B$, in agreement with the experiments, 
neglecting the anomalous moment. For a structureless point particle
Dirac [60] or quantum electrodynamics has explained why g = 2 for 
the electron. However, we consider here an electron with structure 
and a finite size, which is at rest. 
When it is at rest the electron has still its magnetic moment. Dirac's 
theory does therefore not apply here. In order to arrive at an 
explanation of the magnetic moment of the electron it will be necessary
to consider the structure of the electron.

  The only part of Eq.(96) that can be changed in order to explain the 
g-factor of an electron with structure is the ratio e/m, which deals with 
the spatial distribution of charge and mass. If part of the mass of the 
electron is non-electromagnetic and the non-electromagnetic part of 
the mass does
not contribute to the magnetic moment of the electron, which to all 
that we know is true for neutrinos, then the ratio e/m in Eq.(95) is not
e/m(e$^\pm$). The charge e 
certainly remains unchanged, but  e/m depends on what fraction of the 
mass of the electron is of electromagnetic origin and what fraction of 
the mass is non-electromagnetic. Only the current, not the mass of a 
current loop, determines the magnetic moment of a loop. From the  
very accurately known values of $\alpha_f$, m($\pi^\pm$)c$^2$ 
and m(e$^\pm)$c$^2$ and from Eq.(83)
for the energy in the electric oscillations in the electron 
$\mathrm{E}_\nu$(e$^\pm$) =  
0.996570\,m(e$^\pm$)c$^2$/2 follows
that  very nearly one half of the mass of the electron is of electric 
origin, the other half of m(e$^\pm$) is made of neutrinos, Eq.(77),
and neutrinos do not contribute to the magnetic moment. That means     
that in the electron the mass that carries the charge e is
m(e$^\pm$)/2. The magnetic moment of the electron in our model is then 
\begin{equation}
\vec{\mu}_e = g \frac{e\hbar}{2m(e^\pm)
/2\cdot c}\vec{s}\,, \end{equation}
and with s = 1/2 we have
 
\hspace{4.2cm} $\mu_e$ = ge$\hbar$/2m(e$^\pm$)c . \hspace{4cm}(97a)

\vspace{0.5cm}

The g-factor in Eq.(97a) must be equal to one and is
unnecessary. What we have found is akin to what Perkins [22, p.320]
states as follows: ``The magnetic moment of a charged particle depends
on the ratio e/m and thus, classically, for a rotating structure, on the 
spatial distributions of charge and mass. \emph{If the two distributions 
are the same, a value g = 1 is obtained on classical arguments}"(emphasis 
added).

   In other words, if exactly one half of the mass of the 
electron consists of neutrinos, then it follows automatically from our 
model of the electron that it has the correct magnetic moment 
$\mu_e$ = $e\hbar/2m(e^\pm)c$, without the artificial g-factor. 
This result is proof that indeed one half of the mass of the electron
consists of neutrinos. For the explanation of the magnetic moment  
of the muon see Appendix C.  

\section{The ratios m($\mu^\pm$)/m(e$^\pm$), m($\pi^\pm$)/m(e$^\pm$) \\ 
and m(p)/m(e)}

An obvious about 100 years old question is, why is the mass of the proton,
which originally seemed to be the carrier of the positive elementary charge,
about 1800 times larger than the mass of the electron\,? Later the questions 
arose why is the mass of the muon 207 times larger, and the mass of the pion 273
times larger than the mass of the electron\,? Both questions found empirical 
answers. More than 60 years ago Nambu [10] found that m($\mu^\pm$)/m(e$^\pm$)
is about 3/2$\alpha_f$, and m($\pi^\pm$)/m(e$^\pm$) is about 2/$\alpha_f$.
Twenty-five years later Barut [51] published more accurate values of these 
ratios. It seems fair to split the credit for these equations to both of them. 
We will call m($\mu^\pm$)/m(e$^\pm$) = 3/2$\alpha_f$ + 1 Barut’s formula,
and m($\pi^\pm$)/m(e$^\pm$) = 2/$\alpha_f$ - 1 will be Nambu’s formula. 
Any valid explanation of the masses of either the muon or the pion must be 
able to explain Barut’s and Nambu’s equations. It must also be explained how 
it is possible that these mass ratios depend solely on the fine structure 
constant, which is of electrodynamic origin.
   
   In order to determine m($\mu^\pm$)/m(e$^\pm$) we first
modify Eq.(69) by setting  N/4\,$\cdot$\,m($\nu_e$)
= 0.5\,m(e$^\pm$), not at 0.51\,m(e$^\pm$). In other words we say that 
1/2 of the mass of the electron is made of neutrinos. If the other half of
the mass of the electron originates from the charge of the electron, 
as we have shown in Section\,11, then the mass of the electron
is twice the sum of the masses of the neutrinos
in the electron (Eq.85) and we have 
\begin{equation}  \mathrm{m(e}^\pm) = \mathrm{N}/2\cdot
\mathrm{m}(\nu_e) \quad\mathrm{or}\quad 
\mathrm{N}/2\cdot\mathrm{m}(\bar{\nu}_e)\,.
\end{equation}
We also set E$_\nu(\pi^\pm$) = 0.5\,m($\pi^\pm$)c$^2$, not 
at 0.486\,m($\pi^\pm$)c$^2$
as in Eq.(32). With E$_\nu(\pi^\pm$) = E$_\nu(\mu^\pm$) from Eq.(63)
follows from the formula for the oscillation energy of the pion, Eq.(61), 
and with m($\pi^\pm$)c$^2$ = 2\,E$_\nu(\pi^\pm$), that 
\begin{equation} \mathrm{E}_\nu(\mu^\pm) = 
\mathrm{N}/2\cdot[\mathrm{m}(\nu_\mu) + 
\mathrm{m(\nu_e)]c}^2\,. \end{equation} 
From Eq.(66a) for the mass of  $\mu^+$  and with m($\nu_e)$ = 
m($\bar{\nu_e}$) and m($\nu_\mu$) = m($\bar{\nu_\mu}$) we find,
considering only the neutrino masses, not the charge in $\mu^\pm$, that 
\begin{equation} \mathrm{m}(\mu^\pm) = 3/4\cdot\mathrm{N}
\mathrm{m}(\nu_\mu) + \mathrm{N}\mathrm{m}(\nu_e)\,.
\end{equation}
With m(e$^\pm$) = N/2\,$\cdot$\,m($\nu_e$) from
Eq.(98) we then have
\begin{equation} \frac{\mathrm{m}(\mu^\pm)}{\mathrm{m(e}^\pm)}(theor) = 
\frac{3}{2}\cdot\frac{\mathrm{m}(\nu_\mu)}{\mathrm{m}(\nu_e)} + 2\,.
\end{equation}
The ratio m($\mu^\pm$)/m(e$^\pm$) is independent of N. It turns out,
with m($\nu_e$) = $\alpha_f$\,$\cdot$\,m($\nu_\mu$) from Eq.(72), that
\begin{equation} \frac{\mathrm{m}(\mu^\pm)}{\mathrm{m(e}^\pm)}(theor)
\cong \frac{3}{2}\cdot \frac{1}{\alpha_f} + 2 = 207.55
= 1.0038\cdot\frac{\mathrm{m}(\mu^\pm)}{\mathrm{m(e}^\pm)}(exp)\,,
\end{equation}
\noindent
whereas m($\mu^\pm$)/m(e$^\pm$)(\emph{exp}) =  206.768. We attribute the 
difference between the experimental value of m($\mu^\pm$)/m(e$^\pm$) 
and the theoretical value of \\ m($\mu^\pm$)/m(e$^\pm$) to the binding energy
of the charge to the neutral neutrino lattice. The consequences of the 
charge for the mass of the muon and pion have been treated in [74].    
In order to arrive from Eq.(101) at the ratio of 
m($\mu^\pm$)/m(e$^\pm$) in Eq.(102)
\begin{itemize}
\item
\hspace{1cm}\emph{it is necessary that m($\nu_e$) = 
$\alpha_f$\,$\cdot$\,m($\nu_\mu$)}, 
\end{itemize}
as we established in Eq.(72). It is critical for the validity of our model
of the particles that Eq.(102) comes close to the actual value of 
m($\mu^\pm$)/m(e$^\pm$). Eq.(102) does not depend on N, 
nor on m($\nu_\mu$) or m($\nu_e$).

   The mass of the muon is, according to Eq.(102), 207 times larger than 
the mass of the electron.  Let us compare the \emph{rest mass of the muon},
which was explained in Section 9 with an oscillating lattice of 
muon neutrinos, electron neutrinos and anti-electron neutrinos and the 
charge e$^\pm$, to \emph{the rest mass of the electron}.  
The electron has the 
most simple neutrino lattice, consisting of only one neutrino type,
either electron neutrinos or anti-electron neutrinos, and it has the 
smallest sum of the masses of the neutrinos in a particle. The 
heavy weight of the  muon m($\mu^\pm$) = 206.768\,m(e$^\pm$) is 
a consequence of the heavy weight of the
N/4 muon neutrinos or N/4 antimuon neutrinos in
the muon lattice. The mass of either a muon neutrino or an antimuon 
neutrino is 137 times the mass of an electron or anti-electron neutrino
according to Eq.(72). That makes the electron neutrinos and 
anti-electron neutrinos in $\mu^\pm$, as well as the mass of the 
charge e$^\pm$, in a first approximation negligible for m($\mu^\pm$). 
It then follows from 
Eq.(101) that m($\mu^\pm$)(\emph{theor}) $\cong$ 
3/2\,$\cdot$\,(m($\nu_\mu$)/m($\nu_e$))\,$\cdot$\,m(e$^\pm$) = 
3/2$\alpha_f\,\cdot$\,m(e$^\pm$) = 205.554\,m(e$^\pm$) = 
0.99413\,m($\mu^\pm$)(\emph{exp}), 
which proves that the heavy mass of the muons is caused by 
the heavy $\nu_\mu,$ \,or \,$\bar{\nu}_\mu$ neutrinos.  

   Equation (102) is nearly the same as Barut's [51] empirical formula\\ 
(Eq.59) according to which the muon/electron mass ratio is 
\begin{displaymath}\mathrm{m}(\mu^\pm)
/\mathrm{m(e}^\pm)(\emph{emp}) = 
3/2\alpha_f + 1 = 206.554 = 0.99896\,\mathrm{m}(\mu^\pm)/
\mathrm{m(e}^\pm)(\emph{exp})\,. \end{displaymath}
A much better approximation to the experimental mass ratio is obtained 
when the +\,1 in Barut's formula is replaced by +\,1.25. The thus calculated 
m($\mu^\pm$)/m(e$^\pm$) = 206.804
differs then from the measured m($\mu^\pm$)/m(e$^\pm$) =
206.7683 by the factor 1.000\,17. 
 
\bigskip
   Similarly we obtain for the $\pi^\pm$\,mesons from Eq.(33a) the ratio 
\begin{equation} \frac{\mathrm{m}(\pi^\pm)}{\mathrm{m(e^\pm)}}(theor)
= 2\,[\frac{\mathrm{m}(\nu_\mu)}{\mathrm{m}(\nu_e)} +1] 
 \cong \frac{2}{\alpha_f} + 2 = 276.07 = 
1.0108\,\frac{\mathrm{m}(\pi^\pm)}{\mathrm{m(e^\pm)}}(exp)\,,
\end{equation}
with m($\pi^\pm$)/m(e$^\pm$)(\emph{exp}) =  273.1321. We have,
however, only considered the ratio of the masses of the neutrinos in 
$\pi^\pm$  and e$^\pm$, not the consequences of the charge for 
m($\pi^\pm$), which have been treated in [74]. For a comparison Nambu's 
(improved) empirical formula for the ratio m($\pi^\pm$)/m(e$^\pm$) is
\begin{equation} \frac{\mathrm{m}(\pi^\pm)}
{\mathrm{m}(\mathrm{e}^\pm)}(emp) 
= \frac{2}{\alpha_f}\,-\,1 = 273.07 = 
0.99978\,\frac{\mathrm{m}(\pi^\pm)}{\mathrm{m}(\mathrm{e}^\pm)}(exp)\,.
 \end{equation}
In simple terms: The measured ratio m($\pi^\pm$)/m(e$^\pm$) is 
273.132, and 273.132  = 1.9931/$\alpha_f$ $\cong$ 2/$\alpha_f$. That 
means that m($\pi^\pm$)/m(e$^\pm$) is practically equal to the
ratio of the mass of two muon neutrinos to the mass of one electron
neutrino, or equal to 2m($\nu_\mu$)/m($\nu_e$) = 2/$\alpha_f$. This has   
to be so because, neglecting the mass of the electron neutrinos and 
anti-electron neutrinos, the mass of 
m($\pi^\pm$) $\cong$ N\,m($\nu_\mu$) (Eq.33a) and m(e$^\pm$) =
 N/2\,$\cdot$\,m($\nu_e$) (Eq.85).

\bigskip

From the ratio of Eq.(102) for the muon and Eq.(103) for the pion follows that
\begin{equation} \frac{\mathrm{m}(\mu^\pm)}{\mathrm{m}(\mathrm{\pi}^\pm)}
(theor) = 1.002\,43 \cdot 3/4 \cong 3/4\,, \end{equation}
which is 0.9931 times the experimental ratio  
m($\mu^\pm$)/m($\pi^\pm$) = 1.00937$\cdot$3/4, Eq.(60).   

\bigskip

   In order to determine m(n)/m(e$^\pm$) we start with K$^0$ = 
(2.)$\pi^\pm$ + $\pi^\mp$ and E((2.)$\pi^\pm$) = 4E$_\nu(\pi^\pm$) + 
N/2\,$\cdot$\,[m($\nu_\mu)$ + m$(\nu_e)$]c$^2$, Eq.(35),
and with m($\pi^\pm$) = N\,$\cdot$\,[m($\nu_\mu$) + m$(\nu_e$)], Eq.(33a).
Then m(K$^0$) =  7N/2\,$\cdot$\,[m($\nu_\mu)$ + m($\nu_e$)],
and with m(n) $\cong$ m(K$^0$ +  $\overline{{\mathrm{K}}^0}$) =
2m(K$^0$) follows that
\begin{equation} \frac{\mathrm{m(n)}}{\mathrm{m(e^\pm)}}(theor) = 
14\,[\frac{\mathrm{m}(\nu_\mu)}{\mathrm{m}(\nu_e)} + 1] = 
14/\alpha_f + 14 = 1932.5 = 
1.051\frac{\mathrm{m(n})}{\mathrm{m(e^\pm)}}(exp)\,,
\end{equation}
with m(n)/m(e)(\emph{exp}) = 1838.68. But we have only
considered the mass of the neutrino lattice in the neutron, not the 
consequences of the quadrupole of two positive and two negative charges
e$^\pm$ in the neutron. The empirical value of m(n)/m(e$^\pm$)
is equal to 14/$\alpha_f$ $-$ 0.9977\,$\cdot$\,80.

\bigskip  

The ratio of the mass of the proton to the mass of the electron, for 
which an explanation has been looked for since about a hundred years, 
is m(p)/m(e)(\emph{exp}) = 1836.15. From our theoretical explanation
of the neutron follows that 
\begin{equation} \frac{\mathrm{m(p)}}{\mathrm{m(e)}}(theor) \cong
 14\,[\frac{\mathrm{m}(\nu_\mu)}{\mathrm{m}(\nu_e)} + 1] - 5/2 
= \frac{14}{\alpha_f} + 23/2 = 1930.00\,, \end{equation}
because the energy lost in the $\beta$-decay of the neutron is 
1.29333\,MeV or 1.01239\,$\cdot$\,5/2\,$\cdot$\,m(e$^\pm)\,$c$^2$. 
1930.0  is 1.051 times the experimental mass ratio 1836.15. We  
have, again, considered only the mass of the neutrino lattice in the 
proton, not the consequences of the three  
charges in the proton. An empirical formula for m(p)/m(e) is
\begin{equation}
\mathrm{m(p)}/\mathrm{m(e)}(emp) = 14\,[1/\alpha_f - 6]  =
14/\alpha_f - 84 =  0.9901\,\mathrm{m(p)}/\mathrm{m(e)}(exp)\,. 
\end{equation}

   The persistent appearance of the fine structure constant $\alpha_f$
in the leading term of the ratio of the masses of the particles to the 
mass of the electron is a consequence of the preponderance of the 
mass of the muon neutrinos in the lattices of the particles, following 
the relation m($\nu_e$) = $\alpha_f\,\cdot$\,m($\nu_\mu$).   
\medskip

\bigskip
   Our theoretical calculations of  m($\pi^\pm$)/m(e$^\pm$) and of
m($\mu^\pm$)/m(e$^\pm$) agree, within the percent range, with their
experimental values. This can only be if our explanation of m($\pi^\pm$),
m($\mu^\pm$) 
and m(e$^\pm$) are correct in the same approximation, and if the ratio 
m($\nu_e$) = $\alpha_f$m($\nu_\mu$) (Eq.72) is valid. In other words, 
our theoretical values of 
m($\pi^\pm$)/m(e$^\pm$) and of m($\mu^\pm$)/m(e$^\pm$) are

\begin{itemize}
\item
\emph{proof of the validity of our explanation of m($\pi^\pm$), 
m($\mu^\pm$) and of m(e$^\pm$),\\ as well as of the validity of the relation
m($\nu_e$) = $\alpha_f$m($\nu_\mu$)}.  
\end{itemize}

\section{The spin of the $\gamma$-branch particles}

  It appears to be crucial for the validity of a model of the elementary 
particles that the model can also explain the spin of the particles 
without additional assumptions. The spin or the intrinsic angular 
momentum is, after the mass and charge, the third most important 
property of the elementary particles. The standard model does 
not explain the spin, the spin is imposed on the quarks. 
As is well-known the spin of the electron 
was discovered by Uhlenbeck and Goudsmit [59] 90 years ago. 
Later on it was established that the baryons have spin as 
well, but not the mesons. We have proposed an explanation of the spin 
of the particles in [67].  For current efforts to 
understand the spin of the nucleon see Jaffe [68] and of the 
spin structure of the $\Lambda$ baryon see G\"ockeler et al.\,[69]. 
Rivas has described his own model of the spin and other spin models
in his book [70]. The explanation of the spin requires an unambiguous 
answer, the spin must be 0 or 1/2 or integer multiples thereof, nothing 
else. At present we do not have an accepted explanation of the spin. 
   
   For the explanation of the spin of the particles it seems  
to be necessary to have 
an explanation of the structure of the particles. The spin of  
a particle is, of course, the sum of the angular momentum 
vectors of the waves in the particle, plus the sum of the spin 
vectors of its  neutrinos and antineutrinos, plus the spin of 
the charges which the particle carries. It is
striking that the particles which consist of 
a single mode do not have spin, as the $\pi^0, 
\pi^\pm$ and $\eta$ mesons do, see Tables\,1 and 3. 
It is also striking that particles whose mass is approximately 
twice the mass of a smaller particle have spin 1/2, as is the 
case with the $\Lambda$ baryon, 
m($\Lambda$) $\approx$ 2m($\eta$),  and with the nucleon 
m(n) $\approx$ 2m(K$^\pm$) $\approx$ 2m(K$^0$). The $\Xi^0_c$ 
baryon, which is a doublet of one mode, has also spin 1/2. Composite 
particles which consist of a doublet of one mode plus one or two other
single modes have spin 1/2, as the  $\Sigma^0$, $\Xi^0$\, 
and $\Lambda_c^+$, $\Sigma_c^0$, $\Omega_c^0$ baryons do. 
The only particle which seems to be the triplet of a single mode,
the $\Omega^-$ baryon with m($\Omega^-$) $\approx$ 3m($\eta$),
has spin 3/2. It appears that the relation between
the spin and the modes of the particles is straightforward.

   The $\pi^0$\,meson does not have spin, s($\pi^0$) = 0. 
In our model of the $\gamma$-branch particles the $\pi^0$ 
and $\eta$ mesons consist of N = 2.854\,$\cdot\,10^{\,9}$ standing 
electromagnetic waves, each with its  own frequency.
Each of the electromagnetic waves in the $\pi^0$ and 
$\eta$  mesons may have spin s = 1, 
because circularly polarized electromagnetic waves have an angular 
momentum, as was first suggested by Poynting [71] and verified by, 
among others, Allen [72]. The creation of the $\pi^0$\,meson in the 
reaction $\gamma$ \,+\, p $\rightarrow \pi^0$ + p and conservation 
of angular momentum in this process dictates that the sum of the 
angular momentum 
vectors of the N electromagnetic waves in the $\pi^0$\,meson must 
be zero, $\sum_ij_i$ = 0. The waves in the $\pi^0$ meson can be 
either linear or circular. Linearly polarized electromagnetic waves do not  
have an angular momentum. That this is actually so was proven by 
Allen [72]. If the $\pi^0$ meson consists of linearly polarized 
electromagnetic waves, the $\pi^0$ meson does not have spin per se.

   On the other hand, if the waves in the interior of the $\pi^0$ or
$\eta$ meson are circular, the waves have an angular momentum, or 
spin s = 1. That could mean that the spin of the entire
particle could be s = N, whereas it must be zero. This discrepancy 
disappears when we recall that the waves in the particles must be 
standing waves. That means that we must superpose on a circular 
electromagnetic wave with the angular momentum j$_i$ an 
electromagnetic wave with the same frequency and same amplitude 
traveling in opposite direction. That introduces an angular momentum 
vector of the opposite direction -\,j$_i$. In a standing electromagnetic
wave the angular momentum of both waves \emph{cancel}. For later 
considerations it is important to note that this applies also for the wave
at the center of the lattice. Since the angular momentum vectors of the 
circular waves in the $\pi^0$ and $\eta$ mesons cancel, or 
since the sum of the spin vectors $s_i$ of the N 
circular electromagnetic  waves is zero, the intrinsic angular 
momentum of the $\pi^0$ and $\eta$ mesons is zero, or
\begin{equation} j(\pi^0,\eta) =  \sum_{i} \,j_i = 0 
\quad (0\,\le\,i\le \mathrm{N})\,.\end{equation}
In this model the $\pi^0$ and $\eta$ mesons do not
have an intrinsic angular momentum or spin,
whether the waves are linear or circular.

\medskip

   The $\Lambda$ baryon with spin s = 1/2 is the next on the list of 
the $\gamma$-branch particles. The $\Lambda$ baryon is the 
superposition of two $\eta$ mesons, based on the fact that 
m($\Lambda$) = 1.019\,$\cdot$\,2\,m($\eta$). An $\eta$ meson 
consists of N standing electromagnetic waves, and has zero spin,
Eq.(109). Suppose the waves in the two $\eta$ mesons
are linear, perpendicular to each other, and shifted 
in phase by $\pi$/2. At each of the N lattice points the superposition 
of the two waves creates a circular wave. The circular waves have the 
frequency $\omega_i$ and an angular momentum j$_i$. We will see 
that  all but one of the angular momentum vectors j$_i$ cancel. 
According to Eq.(14) the frequency of the waves in the lattice is
given by $\nu$ = $\nu_0\,\phi$. Consequently the angular 
momentum vectors j$_i$ change sign with $\phi$. At each lattice 
point there is, in the opposite quadrant of the lattice, a circular wave 
with frequency -\,$\omega_i$ and with the angular momentum -\,j$_i$, 
because j = mr$^2\,\omega$ = 2$\pi\,\nu_0$ mr$^2\,\phi$. Consequently  
the angular momentum vectors of all N electromagnetic waves in the 
$\Lambda$  lattice \emph{cancel}, but for the angular momentum of the 
wave at the \emph{center of the lattice}, which is not countered by a 
wave of opposite angular momentum.  

It remains to be shown what the angular momentum of the center 
wave is. The total energy of a traveling wave is obviously
the sum of the potential and the kinetic energy 

\begin {equation} \mathrm{E}_{pot} + \mathrm{E}_{kin} = 
\mathrm{E}_{tot} = \hbar\omega\,.\end {equation}
\noindent
In a traveling wave the kinetic energy is equal to the 
potential energy. From this follows
\begin{equation} \mathrm{E}_{tot} = 2\mathrm{E}_{kin} = 
2\frac{\Theta\omega^2}{2}\, = \hbar\omega, 
\end{equation}
\noindent
with the moment of inertia $\Theta$. The angular momentum is then
 
\begin{equation} j = \Theta\omega = \hbar\,.\end{equation}
\noindent 
But we consider now the superposition of two perpendicular waves.  
The energy is then the sum of the energy of both individual waves, 
and we have with Eq.(111) 

\begin{equation} 4\mathrm{E}_{kin} = 4\Theta\omega^2/2 = 
\mathrm{E}_{tot} = \hbar\omega\,,\end{equation}
\noindent
from which follows that the circular wave, which is the consequence 
of the superposition of two linear waves, has an angular momentum

\begin{equation} j = \Theta\omega = \hbar/2\,. \end{equation}
\noindent
The superposition of two linear, monochromatic waves, of equal
amplitudes and frequencies $\omega$ and $\mathrm{-}\,\omega$,
produces an angular momentum  j = $\hbar$/2, or spin s = 1/2. Since 
the center wave of the lattice has the angular momentum $\hbar/2$, 
and all the other angular momentum vectors cancel, the spin of the 
$\Lambda$ baryon is $\hbar/2$ or s = 1/2, if the waves are linear. 

  We consider now the possibility that the $\Lambda$ baryon consists 
of the superposition of 2N circular waves. Therefore we add 
at one lattice point to one monochromatic circular wave  
with frequency $\omega$ a second circular wave with
$\mathrm{-}\,\omega$,  of the same absolute value as $\omega$, but 
shifted in phase by $\pi$,  having the same amplitude, as we have 
done in [67]. Negative frequencies are permitted solutions of the 
equations for the lattice oscillations, Eq.(7). In other words we consider 
the circular waves

\begin{equation} x(t) = exp[i\omega t] + exp[-\,i(\omega t + 
\pi)]\,,\end{equation}

\begin{equation} y(t) = exp[i(\omega t + \pi/2)] + exp[-\,i(\omega t + 
  3\pi/2)]\,\,.\end{equation}
\noindent
This can also be written as 
\begin{equation} x(t) = exp[i\omega t] - exp[-i\omega t]\,,\end{equation}

\begin{equation} y(t) = i\cdot(exp[i\omega t] + exp[-i\omega t] 
)\,.\end{equation}
If we replace \emph{i} in the Eqs. above by $-$\,\emph{i} we have  
a circular wave turning in opposite direction. 

    The angular momentum vectors of all circular waves in the lattice  
of the $\Lambda$ baryon \emph{cancel},   
except for the wave at the center of the crystal. Each 
wave with frequency $\omega$ at $\phi$\,$>$\,0 has at its mirror 
position $\phi$\,$<$\,0 a wave with the frequency $-\,\omega$, 
which has a negative angular momentum, since j = 
mr$^2\omega$  and $\omega = \omega_0\phi$. 
Consequently the angular momentum vectors of both waves cancel. 
The center of the lattice oscillates too, as all the other lattice points. 
As the other circular waves in the lattice, the 
circular wave at the center has the angular momentum $\hbar$/2 
according to Eq.(114). The angular momentum of the wave at the
center of the lattice is the only angular momentum which is not 
canceled by a wave of opposite circulation. The wave at the 
\emph{center of the lattice} determines the spin of the particle.     
 
    The net angular momentum of the circular waves in the 
lattice reduces to the angular momentum of the center wave and 
is $\hbar$/2.  Since the waves in the $\Lambda$ baryon are 
the only possible contribution to an internal angular momentum, the 
intrinsic angular momentum of the $\Lambda$ baryon is $\hbar$/2 or
\begin{equation} j(\Lambda) = \sum_i\,j(\omega_i) = j(\omega_0) =  
\hbar/2\,.\end{equation} 
In this model the $\Lambda$ baryon has a net angular momentum 
$\hbar$/2, regardless whether the waves are linear or 
circular. We have thus explained that the $\Lambda$
and likewise the $\Xi_c^0$ baryon satisfy the necessary condition 
that j = $\hbar$/2 or s = 1/2.  Spin 1/2 is caused by the 
composition of the particles, it is not a contributor to the mass 
of a particle.

   The other particles of the $\gamma$-branch, the 
$\Sigma^0$, $\Xi^0$, $\Lambda^+_c$, $\Sigma^0_c$ and 
$\Omega^0_c$ baryons are composites of a baryon with spin 1/2 
plus one or two $\pi$\,mesons
 which do not have spin. Consequently the spin of these particles is 
1/2. The spin of all particles of the $\gamma$-branch, exempting the 
spin of the $\Omega^-$\,baryon, has thus been explained.
 For an explanation of  s($\Sigma^{\pm,0}$)  = 1/2, of s($\Xi^{-,0}$) 
 = 1/2 and of s($\Xi_c^{0,+})$  = 1/2, regardless
whether the particles are charged or neutral, we refer to [67]. In these 
cases the charge does not seem to be added as 
an electron or positron to the neutral baryon lattices, but rather through 
either $\pi^-$ or $\pi^+$\,mesons, which do not have spin. The presence 
of the $\pi^\pm$\,mesons in the charged versions of $\Sigma^0$, $\Xi^0$
and $\Xi^0_c$ is documented by the appearance of  $\pi^\pm$\,mesons  
in the decay products of $\Sigma^{\pm}$, $\Xi^{-}$  and $\Xi_c^{+}$, 
whereas in the decays of $\Sigma^0$ and $\Xi^0$  
charged mesons do not appear.

\section{The spin of the $\nu$-branch particles}

The characteristic particles of the neutrino-branch are the 
$\pi^\pm$\,mesons which do not have spin, s($\pi^\pm$) = 0. 
At first glance it seems to be 
odd that the $\pi^\pm$\,mesons do not have spin, because
 it seems that the $\pi^\pm$\,mesons should have spin 1/2 from 
the spin of the charges e$^\pm$ in $\pi^\pm$, but 
s($\pi^\pm)$ = 0. What happens to
the spin of e$^\pm$ in $\pi^\pm$\,? The solution of
this puzzle is in the composition of the $\pi^\pm$\,mesons which 
are, in our model of the particles, made of a lattice, Fig.2, of 
neutrinos and antineutrinos, each having spin 1/2, 
the lattice oscillations, plus a charge e$^\pm$.

   There is a fundamental difference between 
the $\pi^0$ and $\pi^\pm$\,mesons. All lattice points 
in $\pi^\pm$ have spin per se, because neutrinos have spin,  
whereas in the $\pi^0$\,meson the lattice points do not have 
spin because they are standing electromagnetic waves. 
Longitudinal oscillations in the neutrino lattice of the 
$\pi^\pm$\,mesons do not cause an angular momentum, 
$\sum_i\,j(\nu_i)$ = 0, because for longitudinal oscillations
$\vec{r}\,\times\,\vec{p}$  =  0. In the 
cubic lattice of N = O($10^9$) neutrinos and 
antineutrinos of the $\pi^\pm$\,mesons the spin of nearly all 
neutrinos and antineutrinos \emph{must cancel}, because conservation 
of angular momentum during the creation of the $\pi^\pm$\,mesons    
requires that the total angular momentum of the lattice is 
either 0 or $\hbar$/2. 
In fact the spin vectors of all neutrinos cancel, but for the neutrino 
or antineutrino in the 
\emph{center of the lattice}. In order for this to be so
the spin vector of any particular neutrino in the lattice 
has to be opposite to the spin vector of the neutrino at its mirror 
position. As is well-known only left-handed neutrinos and right-handed 
antineutrinos exist. From $\nu$ = $\nu_0\phi$ (Eq.14) follows 
that the direction of motion of the neutrinos in e.g.\,\,the upper right 
quadrant ($\phi$\,$>$\,0) is opposite to the direction of motion in the 
lower left quadrant ($\phi$\,$<$\,0). Consequently the spin vectors of all 
neutrinos or antineutrinos in opposite quadrants are opposite and cancel.
The only angular momentum remaining from the spin of the neutrinos of 
the lattice is the angular momentum of the neutrino or antineutrino at the
\emph{center of the lattice}, which does not have a mirror particle. 
\emph{The electrically neutral neutrino lattice of the 
$\pi^\pm$\,mesons}, consisting of N/2 neutrinos 
and N/2 antineutrinos and the center neutrino or antineutrino,
each with spin j($n_i$) = $\hbar$/2, \emph{has therefore an intrinsic angular 
momentum}  j = $\sum_i\,j(n_i)$ = \emph{j(n$_0$}) = $\hbar$/2. 

   But electrons or positrons added to the neutral neutrino lattice 
of the $\pi^\pm$\,mesons have spin 1/2. If the spin of the 
electron or positron added 
to the neutrino lattice is opposite to the spin of the neutrino 
or antineutrino in the center of the lattice, then the net spin of the
$\pi^+$ or $\pi^-$ mesons is zero, or
 \begin{equation} j(\pi^\pm) = \sum_i\,j(\nu_i) +\sum_i\,j(n_i) +
 j(\mathrm{e}^\pm) = j(n_0) + j(\mathrm{e}^\pm) = 0
\,\quad(0\leq i \leq \mathrm{N})\,.\end{equation}
It is important for the understanding of the structure of 
the $\pi^\pm$\,mesons to realize that s($\pi^\pm$) can only be 0, 
if the $\pi^\pm$\,mesons consist of a \emph{neutrino lattice} 
to which an electron or positron is added, whose spin is opposite to the 
net spin of the neutrino lattice. \emph{Spin 1/2 of the electron or 
positron can only be canceled by something that has also spin 1/2, and in
$\pi^\pm$ the only conventional choice for that is a single neutrino}.

\vspace{0.5cm}

   \emph{The absence of spin, the rest mass and the decay of $\pi^\pm$ 
require that the $\pi^\pm$\,mesons are made of a cubic neutrino lattice 
and a charge} e$^\pm$.

\bigskip

   The K$^{\pm}$\,mesons do not have spin, s(K$^\pm$) = 0. 
With the spin of the K$^\pm$\,mesons we 
encounter the same oddity we have just observed with the spin of  
the $\pi^\pm$\,mesons, namely we have a particle which carries a  
charge e$^\pm$ with spin 1/2, 
and nevertheless the particle does not have spin. The explanation 
of s(K$^\pm$) = 0 follows the same lines as the explanation of 
the spin of the $\pi^\pm$\,mesons. In our model the
K$^\pm$\,mesons are described by the state (2.)$\pi^\pm$ + $\pi^0$,
that means by the second mode of the lattice oscillations 
of the $\pi^\pm$\,mesons plus a $\pi^0$\,meson. The second mode
of the longitudinal oscillations of a neutral neutrino lattice does  
not have a net intrinsic angular momentum $\sum_i\,j(\nu_i)$ = 0. But
the spin of the neutrinos contributes an angular momentum $\hbar$/2, 
which originates from the neutrino or antineutrino in the center of the 
lattice, just as it is with the neutrino lattice in the 
$\pi^\pm$\,mesons, so $\sum_i\,{ j(n_i)} = j(n_0) = \hbar/2$. Adding 
a charge e$^\pm$ with a spin opposite to the net intrinsic  
angular momentum of the neutrino lattice creates the charged 
(2.)$\pi^\pm$ mode which has zero spin
\begin{equation}j((2.)\pi^\pm) = \sum_i\,{j(n_i)} + j(\mathrm{e}^\pm) 
= j(n_0) + j(\mathrm{e}^\pm) = 0\,.\end{equation}
 As discussed in Section 6 it is 
necessary to add a $\pi^0$\,meson to the second mode of the 
$\pi^\pm$\,mesons in 
order to obtain the correct mass and the correct decays of the 
K$^\pm$\,mesons. Since the $\pi^0$\,meson does not 
have spin the addition of the $\pi^0$\,meson does not add to the 
intrinsic angular momentum of the K$^\pm$\,mesons. So 
 s(K$^\pm$) = 0, as it must be.

\medskip

  The explanation of s = 0 of the K$^0$ and 
$\overline{{\mathrm{K}}^0}$\,mesons described by the
state (2.)$\pi^\pm$ + $\pi^\mp$ is different, because there is 
now no charge whose spin could cancel the spin of the 
neutrino lattice.  The longitudinal oscillations of the second mode
of the neutrino oscillations of (2.)$\pi^\pm$  in K$^0$ as well 
as of the basic $\pi^\mp$ mode do not create an angular momentum, 
$\sum_i\,j(\nu_i)$ = 0. The sum of the spin vectors of the neutrinos   
in K$^0$ and $\overline{{\mathrm{K}}^0}$ is determined by the 
neutrinos in the second mode of the $\pi^\pm$\,mesons,
or the (2.)$\pi^\pm$ state, and the basic  $\pi^\mp$ mode, 
each have (N - 1)/2 neutrinos and 
(N - 1)/2 antineutrinos plus a center neutrino or antineutrino, so 
the number of all neutrinos and antineutrinos in the sum of both 
states, the K$^0$,$\overline{{\mathrm{K}}^0}$\,mesons, is 2N. 
It follows that two neutrinos are at each 
lattice point of the K$^0$ or $\overline{{\mathrm{K}}^0}$\,mesons.  
We assume that Pauli's exclusion principle applies to neutrinos 
as well. Consequently each neutrino at each lattice point must 
share its location with an antineutrino. In other words, the neutrinos 
and antineutrinos form \emph{dipoles}. These dipoles do not have spin. 

   That means that  
the spins of all neutrinos and antineutrinos in the K$^0$\,meson 
cancel, or that $\sum_i\,j(2n_i)$ = 0. It also means that the center 
of the lattice does, in this case, not contribute to the spin of the 
lattice, because at the center of K$^0$ is also a neutrino dipole 
without spin. The sum of the spin vectors of the two opposite charges 
in either the K$^0$  or the  $\overline{{\mathrm{K}}^0}$\,mesons, 
or in the (2.)$\pi^\pm$ + $\pi^\mp$ state, is also 
zero. Since neither the lattice oscillations nor the spin of the 
neutrinos and antineutrinos 
nor the two opposite charges contribute an angular momentum 
\begin{equation} j(\mathrm{K}^0) = \sum_i\,j(\nu_i) + 
\sum_i\,j(2n_i) + j(\mathrm{e}^+ + \mathrm{e}^-) = 0\,. \end{equation}
The intrinsic angular momentum of our model of 
the K$^0$ and $\overline{{\mathrm{K}}^0}$\,mesons
is zero, or s(K$^0$,$\overline{{\mathrm{K}}^0}$) = 0, as it 
must be. In simple terms, since the structure of e.g. K$^0$ is 
(2.)$\pi^+$ + $\pi^-$, the spin of K$^0$ is the sum of the spin of  
(2.)$\pi^+$ and of $\pi^-$, both of which do not have spin. 
It does not seem possible to arrive at
s(K$^0$,$\overline{{\mathrm{K}}^0}$) = 0 if both particles 
do not contain the N \emph{pairs} of neutrinos and antineutrinos 
required by the (2.)$\pi^\pm$ + $\pi^\mp$ state which we have 
suggested in Section 6.

\bigskip  

   The neutron has spin s = 1/2. One must 
wonder how it comes about that a particle, which seems to be the 
superposition of two particles without spin, ends up with spin 1/2. 
The neutron,
which has a mass $\approx$ 2m(K$^\pm$) or 2m(K$^0$), is either  
the superposition of a K$^+$ and a K$^-$ meson or of a K$^0$ and a 
$\overline{\mathrm{K}^0}$ meson. The intrinsic angular momentum of 
the superposition of K$^+$ and K$^-$ is either 0 or $\hbar$, 
which means that the neutron cannot be the superposition of K$^+$ 
and K$^-$. For a proof of this statement we refer to [67].

   On the other hand, the neutron can be the superposition of a 
K$^0$  and a $\overline{\mathrm{K}^0}$\,meson. A significant 
change in the lattice occurs when a K$^0$ and a  
$\overline{\mathrm{K}^0}$ meson are superposed. 
Each K$^0$\,meson contains N neutrinos 
and N antineutrinos, as we discussed in context with the spin of 
K$^0$. The number of all neutrinos and antineutrinos in superposed 
K$^0$ and $\overline{\mathrm{K}^0}$ lattices, i.e. in the neutron, 
is consequently 4N. Each of the N lattice 
points of the neutron contains four neutrinos, a muon 
neutrino and an anti-muon neutrino as well as an 
electron neutrino and an anti-electron neutrino. The 
$\nu_\mu,\bar{\nu}_\mu,\nu_e,\bar{\nu}_e$ \emph{quadrupoles}  
oscillate just like individual neutrinos do, because we learned from 
Eq.(7) that the ratios of the oscillation frequencies are independent 
of the mass as well as of the interaction constant between the lattice 
points. In the neutrino \emph{quadrupoles} the spin of the neutrinos 
and antineutrinos cancels, $\sum_i\,j(4n_i)$ = 0. The superposition of 
two  neutrino lattice oscillations of frequency $\omega_i$  contribute  
an angular momentum at all lattice points, 
which all cancel, but for the center oscillation, so 
$\sum_i\,j(\omega_i) = j(\omega_0) = \hbar/2$. The spin 
and charge of the four charges e$^+$e$^-$e$^+$e$^-$ 
hidden in the sum of the K$^0$ and $\overline{\mathrm{K}^0}$ 
mesons cancel too, j(4e$^\pm$) = 0. It follows  
that the intrinsic angular momentum of a neutron created by the 
superposition of a K$^0$ and a $\overline{\mathrm{K}^0}$ meson 
comes from the neutrino lattice oscillations only and is 
\begin{equation} j(\mathrm{n}) = \sum_i\,j(\omega_i) + \sum_i\,j(4n_i) +
 j(4\mathrm{e}^\pm) =  \sum_i\,j(\omega_i) = j(\omega_0) = \hbar/2\,.  
\end{equation}
In simple terms, in this model of the neutron the spin originates from 
the superposition of two neutrino lattice 
oscillations with the frequencies $\omega$ and $\mathrm{-}\,\omega$
at all lattice points. From those only the angular momentum
$\hbar$/2 of the oscillation at the center of the lattice remains.

\medskip

   The spin of the proton is 1/2 and is unambiguously defined by the  
decay of the neutron n $\rightarrow$ p + e$^-$ + $\bar{\nu}_e$. We 
have suggested in Section 10 that 3/4$\cdot$N anti-electron neutrinos 
of the neutrino lattice of the neutron are removed in
 the $\beta$-decay of the neutron and that the other N/4
 anti-electron neutrinos leave with the emitted electron. The intrinsic 
angular momentum of the proton originates then from the spin of the 
central $\nu_\mu\bar{\nu}_\mu\nu_e$ triplet, from the spin of
 the e$^+$e$^-$e$^+$ triplet which is part of the remains of the neutron,
 and from the angular momentum of the center of the lattice oscillations 
with the superposition of two oscillations. The spin of the 
central ($\nu_\mu\bar{\nu}_\mu\nu_e$)$_0$ triplet is canceled by 
the spin of the e$^+$e$^-$e$^+$ triplet. The intrinsic 
angular momentum of the proton is

\begin{equation}
 j(\mathrm{p}) = j(\nu_\mu\bar{\nu}_\mu\nu_e)_0\,  +  
j{(\mathrm{e}^+\mathrm{e}^-\mathrm{e}^+})
+ j(\omega_0) = j(\omega_0) = \hbar/2\,.  \end{equation}
                         
   The other mesons of the neutrino branch, the D$^{\pm,0}$ and 
D$_s^\pm$ mesons, both having zero spin, are 
superpositions of a proton and an anti-neutron of opposite spin, or of 
their antiparticles, or of a superposition of a  neutron and an anti-neutron
of opposite spin in D$^0$.  The spin of D$^\pm$ and D$^0$ does 
therefore not pose a new problem.

\bigskip 

   The muons have spin, s($\mu^\pm$) = 1/2.
For an explanation of the spin of $\mu^\pm$ we refer to [73]. 
Since N/4 muon neutrinos or N/4 anti-muon  
neutrinos have been removed from the $\pi^\pm$ lattice in the
$\pi^\pm$ decay, it follows that a neutrino vacancy is at the center of 
the $\mu^\pm$ lattice (Fig.\,7). Without a neutrino in the center of
the lattice the spin vectors of all neutrinos cancel. In the absence of 
a center 
neutrino the angular momemtum vectors of the lattice oscillations
cancel as well. So the neutrino lattice of the muon does not have spin.
However, the muons consist of the neutrino  
lattice plus a charge e$^\pm$, whose spin is 1/2. 
The spin of $\mu^\pm$  originates therefore from 
the spin of the charge e$^\pm$ carried
by the muons, and is s($\mu^\pm$) = 1/2. 

   Both the $\pi^\pm$\,mesons and the $\mu^\pm$\,muons carry a charge. 
The $\pi^\pm$\,mesons do not have spin, whereas the $\mu^\pm$\,muons have 
spin 1/2. The presence or absence of a neutrino at the center of the 
lattice makes the difference. The spin of the charge in $\pi^\pm$ is 
canceled by the spin of the central neutrino, whereas the spin of the 
charge in $\mu^\pm$ remains, because there is no central neutrino to
cancel the spin of the charge. 

\bigskip   

   An explanation of the spin of the mesons and baryons can only be 
valid if the same explanation also applies to the antiparticles of these 
particles, whose spin is the $\emph{same}$ as that of the ordinary 
particles. The antiparticles of the $\gamma$-branch consist of 
electromagnetic waves whose frequencies differ from the 
frequencies of the ordinary particles only by their sign. The
angular momentum of the superposition of two circular oscillations 
with $\mathrm{-}\,\omega$ and $\omega$  has the same angular 
momentum as the superposition of two circular oscillations 
with frequencies of opposite  sign, as in $\Lambda$. Consequently 
the spin of the antiparticles of the $\gamma$-branch is the same as 
the spin of the ordinary 
particles of the $\gamma$-branch. The same considerations apply 
to the neutrino lattice oscillations which cause the spin of the 
neutron and proton, the only particles 
of the $\nu$-branch with spin. In our model of the particles 
the spin of the neutron and the anti-neutron is the same.

   Let us summarize: The spin of the particles of the $\nu$-branch 
is the result of the sum of the angular momentum vectors of the 
lattice oscillations plus the sum of the spin vectors 
of the neutrinos in the particles, to which the spin vector 
of the charge or charges a particle carries is added. The 
contribution of all or all but one of the O($10^{\,9}$) oscillations 
and O($10^{\,9})$  neutrinos to the intrinsic
angular momentum of the particles must cancel, otherwise the 
spin cannot be either 0 or 1/2. It requires the symmetry of a cubic 
lattice for this to happen. The center of the lattices alone determines 
the intrinsic angular momentum of the oscillations and neutrinos in the 
lattice. Adding to that the spin vector of one (or more)  
charges e$^\pm$ with spin 1/2 and we arrive at the total intrinsic 
angular momentum of a particle. The most illuminating case are the 
$\pi^\pm$\,mesons which do not have spin although they carry 
the charge e$^\pm$. Actually the neutrino lattice 
of the $\pi^\pm$\,mesons has the net-spin 1/2 from its central neutrino, 
but this spin vector is canceled by the spin of the charge e$^\pm$, 
so s($\pi^\pm$) = 0.

\medskip 

The explanation of the spin of the particles
follows from our explanation of the mass of the particles. We  
explain the spin of the particles through the structure of the particles. 
We did not introduce any new assumption. Cubic lattices are crucial for 
the explanation of the spin of the particles. We have thus confirmed the
validity of our model of the masses of the stable mesons and baryons.

\medskip    

   From the foregoing we arrive also at an understanding of the reason  
for the astonishing fact that \emph{the intrinsic angular momentum or spin  
of the particles is independent of the mass of the particles},
as exemplified by the spin $\hbar$/2 of the electron being the
same as the spin $\hbar/2$ of the proton, notwithstanding the fact that
the mass of the proton is 1836 times larger than the mass of the 
electron. However, in our model, the spin of the particles  
is determined solely by the angular momentum $\hbar$/2 at the
center of the lattice, the other angular momentum vectors in the 
particles cancel. The spin does not depend on the number of the
lattice points in a cubic lattice. Hence the mass of the particles in the 
other $10^{\,9}$ lattice points is inconsequential for the intrinsic angular
momentum of the particles. In this model of the particles the spin 
is independent of the mass of the particles. 
    
\section*{Conclusions}

   This investigation of the elementary particles makes the following points:

\newpage

\indent The measured rest masses of the stable mesons and baryons are, in a\\  
\indent very good approximation, integer multiples of the mass of the $\pi^0$\\  
\indent or $\pi^\pm$ mesons.

\begin{itemize}
\item
\emph{The $\pi^0$\,meson is like a cubic black body filled with\\  
standing electromagnetic waves}.
\end{itemize}

\begin{itemize}
\item
\emph{The $\pi^\pm$\,mesons are like cubic black bodies\\ 
filled with oscillating neutrinos}.
\end{itemize}

The measured rest mass of the muons is, in a good approximation, \\
\indent 3/4 of the mass of the pion 
\begin{itemize}
\item 
\hspace{1.5cm}m($\mu^\pm$)/m($\pi^\pm$)(\emph{exp}) = 
0.757027 = 1.00937\,$\cdot$\,3/4\,. 
\end{itemize}

The rest mass of the muon should be equal to \\
 
\indent$\bullet$\hspace{1cm}m($\mu^\pm$)(\emph{theor}) = 
3/4\,$\cdot$\,Nm($\nu_\mu$) + Nm($\nu_e$) = 
107.88\,MeV/c$^2$\,, \\
\indent which is 1.021\,$\cdot$\,m($\mu^\pm$)(\emph{exp}).

\bigskip

In the particles the mass of the electron neutrino is
\begin{itemize}
\item
\hspace{1.5cm}m($\nu_e$) = 1/136.74\,$\cdot$\,m($\nu_\mu$)
 $\cong$ $\alpha_f$m($\nu_\mu$)\,. 
\end{itemize}

\begin{itemize}
\item
\hspace{1cm}\emph{1/2 of the rest mass of the electron is approximately equal\\ 
\hspace{1cm}to the sum of the masses of the neutrinos in the electron}.\\
\indent\hspace{1cm} This is the so-called ``bare" part of the electron.
\end{itemize}

\hspace{2.5cm} m(e$^-$)/2 = $\Sigma_i$\,m($\nu_e$) = 
N/4\,$\cdot$\,m($\nu_e$). \hspace{2.6cm}\\

$\bullet$\hspace{1.5cm}\emph{the rest mass of a free electron or positron is}    
\begin{eqnarray} \mathrm{m(e^\pm)c^2}(\emph{theor}) =
\mathrm{N/2}\cdot{\mathrm{m}(\nu_e)\mathrm{c}^2}
= \mathrm{N/2}\cdot{\mathrm{m}(\bar{\nu}_e)\mathrm{c}^2}\nonumber\\ 
 = 0.5208\,\mathrm{MeV} = 1.019\,\mathrm{m(e}^\pm)\mathrm{c}^2(\emph{exp})\,.
\nonumber\end{eqnarray}

\newpage

The measured and calculated ratios of m($\mu^\pm$)/m(e$^\pm$) and \\
\indent m($\pi^\pm$)/m(e$^\pm$) provide
\begin{itemize}
\item
\emph{proof of the validity of our explanation of m($\pi^\pm$), 
m($\mu^\pm$) and m(e$^\pm$),\\ as well as of the validity of the 
relation m($\nu_e$) = $\alpha_f$m($\nu_\mu$)}.  
\end{itemize}

\medskip 

   Only photons, neutrinos, charge and the weak nuclear force are needed
to explain the rest masses of the electron, of the muon and of the stable 
mesons and baryons, and their spin. 

\bigskip

   {\bfseries Acknowledgments}. I gratefully acknowledge the contributions of\\ 
\hspace*{0.6cm}Dr. T. Koschmieder to this study. I thank Professor A. Gsponer \\ 
\indent for information about the history of the integer multiple rule.

\section*{Bibliography}

\noindent
[1] Gell-Mann,\,M. 1964. Phys.Lett.B {\bfseries111},1.

\smallskip
\noindent
[2] Nakamura, K. 2010. J.Phys.G {\bfseries 37}, 075021.

\smallskip
\noindent
[3] Skyrme,\,T.H.R. 1962. Nucl.Phys. {\bfseries 31},556.

\smallskip
\noindent
[4] El Naschie,\,M.S. 2002. Chaos,Sol.Frac. {\bfseries 14},649.

\smallskip
\noindent
[5] El Naschie,\,M.S.  2002. Chaos,Sol.Frac. {\bfseries 14},369.

\smallskip
\noindent
[6] El Naschie,\,M.S. 2003. Chaos,Sol.Frac. {\bfseries 16},353.

\smallskip
\noindent
[7] El Naschie,\,M.S. 2003. Chaos,Sol.Frac. {\bfseries 16},637.

\smallskip
\noindent
[8] Feynman,\,R.P. 1985. \emph{QED. The Strange Theory\\
\indent\,\,of Light  and Matter}. Princeton University Press, p.152.

\smallskip
\noindent
[9] Koschmieder,\,E.L. 2003. Chaos,Sol.Frac. {\bfseries18},1129.
 
\smallskip
\noindent
[10] Nambu,\,Y. 1952. Prog.Th.Phys. {\bfseries 7},595.

\smallskip
\noindent
[11] Fr\"ohlich,\,H. 1958. Nucl.Phys. {\bfseries 7},148. 

\smallskip
\noindent
[12] Koschmieder,\,E.L. and Koschmieder,\,T.H. 1999.\\
\indent \,\,Bull.Acad.Roy.Belgique {\bfseries X},289,\\
\indent \,\, http://arXiv.org/hep-lat/0002016.

\smallskip 
\noindent
[13] Wilson,\,K.G. 1974. Phys.Rev.D {\bfseries10},2445.

\smallskip
\noindent
[14] Born,\,\,M. and v.\,Karman,\,Th. 1912. Phys.Z. {\bfseries13},297.

\smallskip
\noindent
[15] Blackman,\,M. 1935. Proc.Roy.Soc.A {\bfseries148},365;384.

\smallskip
\noindent
[16] Blackman,\,M. 1955. Handbuch der Physik VII/1, Sec.12.

\smallskip
\noindent
[17] Born,\,\,M. and Huang,\,K. 1954. \emph{Dynamical Theory \\
\indent\,\, of Crystal Lattices}. Oxford.

\smallskip
\noindent
[18] Maradudin,\,A. et al. 1971. \emph{Theory of Lattice Dynamics \\
\indent\,\,in the Harmonic Approximation}. Academic Press,\\
\indent\,\, 2nd edition.

\smallskip
\noindent
[19] Ghatak,\,A.K. and Khotari,\,L.S. 1972. \emph{An introduction \\ 
\indent \,\, to Lattice Dynamics}. Addison-Wesley.

\smallskip
\noindent
[20] Rekalo,\,M.P., Tomasi-Gustafson,\,E. and Arvieux,\,J. 2002.\\
 \indent\,\, Ann.Phys. {\bfseries295},1.

\smallskip
\noindent
[21] Schwinger,\,J. 1962. Phys.Rev. {\bfseries128},2425.

\smallskip
\noindent
[22] Perkins,\,D.H. 1982. \emph{Introduction to High-Energy Physics}.\\
\indent \,\,Addison Wesley.

\smallskip
\noindent
[23] Koschmieder,\,E.L. 1989. Nuovo Cim. {\bfseries101 A},1017.

\smallskip
\noindent
[24] Born,\,\,M. 1940. Proc.Camb.Phil.Soc. {\bfseries36},160.

\smallskip
\noindent
[25] Koschmieder,\,E.L. 2000. http://arXiv.org/hep-lat\\
\indent\,\,/0005027.

\smallskip
\noindent
[26] Rosenfelder,\,R. 2000. Phys.Lett.B {\bfseries479},381.

\smallskip
\noindent
[27] Melnikov,\,K. and van Ritbergen,\,T. 2000. \\
\indent\,\,Phys.Rev.Lett. {\bfseries84},1673.

\smallskip
\noindent
[28] Liesenfeld,\,A. et al. 1999. Phys.Lett.B {\bfseries468},20.

\smallskip
\noindent
[29] Bernard,\,V., Kaiser,\,N. and Meissner,\,U-G. 2000. \\
\indent  \,\,Phys.Rev.C {\bfseries62},028021.

\smallskip
\noindent
[30] Mistretta,\,C. et al. 1969. Phys.Rev. {\bfseries184},1487.

\smallskip
\noindent
[31] Eschrich,\,\,I. et al. 2001. Phys.Lett.B {\bfseries522},233.

\smallskip
\noindent 
[32] Sommerfeld,\,A. 1952. \emph{Vorlesungen \"{u}ber \\
\indent\,\,Theoretische Physik}. Bd.V, p.56.

\smallskip
\noindent
[33] Debye,\,P. 1912. Ann.d.Phys. {\bfseries39},789.

\smallskip
\noindent
[34] Bose,\,S. 1924. Z.f.Phys. {\bfseries26},178.

\smallskip
\noindent
[35] Bethe,\,H. 1986. Phys.Rev.Lett. {\bfseries58},2722.

\smallskip
\noindent
[36] Bahcall,\,J.N. 1987. Rev.Mod.Phys. {\bfseries59},505.

\smallskip
\noindent
[37] Fukuda,\,Y. et al. 1998. Phys.Rev.Lett. \\
\indent\,\,{\bfseries81},1562;  2003. {\bfseries90},171302.

\smallskip
\noindent
[38] Ahmad,\,Q.R. et al. 2001. Phys.Rev.Lett. \\
\indent\,\,{\bfseries87},071301.

\smallskip
\noindent
[39] Lai,\,A. et al. 2003. Eur.Phys.J.\,C {\bfseries30},33.

\smallskip
\noindent
[40] Koschmieder,\,E.L. 2003. http://arXiv.org/physics\\
\indent\,\,/0309025.

\smallskip
\noindent
[41] Born,\,\,M. and Land\'{e},\,A. Verh.Dtsch.Phys.Ges. \\
\indent\,\,{\bfseries20},210. (1918)

\smallskip
\noindent
[42] Madelung,\,E. Phys.Z. {\bfseries19},524. (1918)

\smallskip
\noindent
[43] Badhuri,\,R.K. et al. Phys.Lett.B {\bfseries136},189. (1984)

\smallskip
\noindent
[44] M\"{u}ller,\,H-M. et al. http://arXiv.org:nucl-th\\
\indent\,\,/9910038. (1999)

\smallskip
\noindent
[45] Vretenar,\,D. et al. http://arXiv.org:nucl-th\\
\indent\,\,/0302070. (2003)

\smallskip
\noindent
[46] Dexheimer,\,V.A. et al. http://arXiv.org:\\
\indent\,\,0708.0131. (2007)

\smallskip
\noindent
[47] Born,\,\,M. and Stern,\,O. 1919. Sitzungsber.\\
\indent\,\,Preuss. Akad. Wiss. {\bfseries33},901.

\smallskip
\noindent
[48] Born,\,M. \emph{Atomtheorie des festen Zustandes}, \\
\indent\,\,Teubner. (1923)

\smallskip
\noindent
[49] Barut,\,A.O. 1979. Phys.Rev.Lett. {\bfseries 42},1251.

\smallskip
\noindent
[50] Gsponer,\,A. and Hurni,\,J-P. 1996. Hadr.J. {\bfseries 19},367.

\smallskip
\noindent
[51] Barut,\,A.O. 1978. Phys.Lett.B {\bfseries73},310.

\smallskip
\noindent
[52] Thomson,\,J.J. 1897. Phil.Mag. {\bfseries44},293.

\smallskip
\noindent
[53] Lorentz,\,H.A. 1903. \emph{Enzykl.Math.Wiss.} \\
\indent\,\,Vol.{\bfseries5},188.

\smallskip
\noindent
[54] Poincar\'{e},\,H. 1905. Compt.Rend. {\bfseries 140},1504. \\
\indent
Translated in:\\
\indent \,\,Logunov,\,A.A. 2001. \emph{On The Articles by \\ 
\indent\,\, Henri Poincare``On The Dynamics of the Electron"}.\\
\indent\,\, Dubna JINR.

\smallskip
\noindent
[55] Ehrenfest,\,P. 1907. Ann.Phys. {\bfseries 23},204.

\smallskip
\noindent
[56] Einstein,\,A. 1919. Sitzungsber.Preuss.Akad.Wiss.\\
\indent\,\,{\bfseries20},349.

\smallskip
\noindent
[57] Pauli,\,W. 1921. \emph{Relativit\"{a}tstheorie}. B.G. Teubner.\\ 
\indent
Translated in: \,\,1958. Theory of Relativity. \\
\indent\,\,Pergamon Press.

\smallskip
\noindent
[58] Poincar\'{e},\,H. 1906. Rend.Circ.Mat.Palermo {\bfseries21},129.

\smallskip
\noindent
[59] Uhlenbeck,\,G.E. and Goudsmit,\,S. 1925. Naturwiss. \\
\indent\,\,{\bfseries13},953 .

\smallskip
\noindent
[60] Dirac,\,P.A.M. 1928. Proc.Roy.Soc.London\,A \\
\indent\,\,{\bfseries117},610.

\smallskip
\noindent
[61] Gottfried,\,K. and Weisskopf,\,V.F. 1984. \emph{Concepts of\\ 
\indent
Particle Physics}. Vol.1,\,p.38. Oxford University Press.

\smallskip
\noindent
[62] Schr\"{o}dinger,\,E. 1930. Sitzungsber.Preuss.Akad.Wiss.\\ 
\indent\,\,{\bfseries24},418.

\smallskip
\noindent
[63] Dirac,\,P.A.M. 1962. Proc.Roy.Soc. {\bfseries268},57 

\smallskip
\noindent
[64] Mac\,Gregor,\,\,M.H. 1992. \emph{The Enigmatic Electron}.\\
\indent\,\,Kluwer.

\smallskip
\noindent
[65] Bender,\,D. et al. 1984. Phys.Rev.D {\bfseries30},515.

\smallskip
\noindent
[66] Koschmieder,\,E.L. 2005. http://arXiv.org\\
\indent\,\,/physics/0503206.

\smallskip
\noindent
[67] Koschmieder,\,E.L. 2003. Hadr.J. {\bfseries26},471,\\
\indent\,\,2003. http://arXiv.org/physics/0301060.

\smallskip
\noindent
[68] Jaffe,\,R.L. 2001. Phil.Trans.Roy.Soc.London.A \\
\indent\,\,{\bfseries359},391.

\smallskip
\noindent
[69] G\"ockeler,\,M. et al. 2002. Phys.Lett.B {\bfseries545},112.

\smallskip
\noindent
[70] Rivas,\,M. 2001. \emph{Kinematic Theory of \\ 
\indent\,\,Spinning Particles.} Kluwer.  

\smallskip
\noindent
[71] Poynting,\,J.H. 1909. Proc.Roy.Soc.A {\bfseries82},560.

\smallskip
\noindent
[72] Allen,\,J.P. 1966. Am.J.Phys. {\bfseries34},1185.

\smallskip
\noindent
[73] Koschmieder,\,E.L. 2003. http://arXiv.org\\
\indent\,\,/physics/0308069,\\
\indent\,\,2005. \emph{Muons: New Research},\,1. Nova.

\smallskip
\noindent
[74] Koschmieder,\,E.L. 2009. http://arXiv.org\\
\indent\,\,/0909.3681v3,\,\,2014.

\bigskip    

\section*{Appendix A}

\bigskip

\begin{center}

\large
{The lattice constant}

\normalsize
\end{center}

\noindent
An equation for the lattice constant in cubic lattices has been 
given by Born and Land\'{e} [41], Eq.(6) therein. Suppose 
there are N lattice points in a diatomic cubic crystal. There are then 
 N/2 masses m$_1$ and N/2 masses m$_2$  and the mass in
the crystal is N/2$\cdot$(m$_1$ + m$_2$). In each cell are eight
neighboring particles, i.e.\,\,there are N/8 such cells. The volume of 
each cubic cell is \emph{a}$^3$ and the total volume of the crystal 
is N\emph{a}$^3$/8. The density $\rho$ is then 
$\rho$ = N/2$\cdot$(m$_1$ + m$_2$)/(N\emph{a}$^3$/8),
from which follows that the lattice constant is given by
\begin{equation} \emph{a}^3 = 4\,(\mathrm{m}_1 + \mathrm{m}_2)/\rho\,.
\end{equation}

   We determine the lattice constant of the neutrino lattice with 
m($\nu_e$) = 0.365\,milli-eV/c$^2$ and m($\nu_\mu)$ = 
49.91\,milli-eV/c$^2$, from Eqs.(67,70), and with the density of  
the $\pi^\pm$\,mesons $\rho(\pi^\pm)$ = 
 m($\pi^\pm$)/Vol($\pi^\pm$) or with $\rho$ = 
139.57\,MeV/c$^2$Vol($\pi^\pm)$. The volume of the 
$\pi^\pm$\,mesons can be determined from the 
measured radius of the $\pi^\pm$\,mesons, 
r$_\pi$ = 0.880$\cdot$10$^{-13}$\,cm = r$_p$,
from Eq.(16). The third power of the lattice constant of  
the neutrino lattice of the $\pi^\pm$\,mesons is then    
\begin{equation} \emph{a}^3 = \frac{4\,(49.91 + 
0.365)\,\mbox{milli-eV}}{139.57\,\mathrm{MeV/Vol}(\pi^\pm)} = 
\frac{201.1\, \mbox{milli-eV}}{139.57\,\mathrm{MeV}}\times 
\mathrm{Vol}(\pi^\pm) \end{equation}
or \begin{equation} \emph{a}^3 = 1.44085\cdot 10^{-9}\cdot
\mathrm{Vol}(\pi^\pm)
\end{equation}
The volume of the $\pi^\pm$\,mesons is  
 4\,$\pi$/3\,$\cdot$\,r$_\pi^3$  with r$_\pi$ = 
0.88\,$\cdot$\,10$^{-13}$\,cm, Eq.(16). The measured
r$_\pi$  is, however, not equal to the sidelength d of the cubic
lattice, which is an integer multiple of the lattice distance \emph{a}.
We must therefore replace the measured r$_\pi$ by 
d/$\sqrt[3]{4\pi/3}$. The volume of the cubic $\pi^\pm$ meson is 
then equal to (r$_\pi$(exp))$^3$ and it follows that
   
\begin{equation} \emph{a} = 0.9939\cdot10^{-16}\,\mathrm{cm}\,.
\end{equation}

This agrees qualitatively with the neutrino lattice constant we use
\emph{a} = 1\,$\cdot$\,10$^{-16}$\,cm, which we postulated in Eq.(8),
the difference with \emph{a} = 1$\cdot$10$^{-16}$\,cm is well within  
the uncertainty of r$_\pi$. This 
agreement is, of course, a consequence of our determination of the
 neutrino masses m($\nu_e$)  and m($\nu_\mu$) with the help of our 
postulated \emph{a}. It is useful to know that lattice theory, expressed by 
Eq.(6) of B\&L or by Eq.(126), leads to the neutrino lattice constant. If the 
neutrino masses could be determined independently from our calculations, 
the lattice constant of a neutrino lattice could be calculated from Eq.(126) 
without making an assumption about \emph{a} in Eq.(17) for N, which we 
used for the determination of m($\nu_\mu$) and m($\nu_e$).
 
\vspace{0.5cm}

\section*{Appendix B}

\bigskip

\begin{center}

\large{The electron radius}

\normalsize

\end{center}

\noindent
The classical electron radius is given by [2]
\begin{equation} \mathrm{r(e)}_{c\,l} = 
\mathrm{e}^2/\mathrm{m(e^\pm)}\,\mathrm{c}^2
= 2.817940\cdot 10^{-13}\,\mathrm{cm}\,. \end{equation}
This equation is based on the premise that the electron has a symmetric 
spherical charge distribution and that the entire mass of m(e) is of 
electric origin. As mentioned in Section 11 the electron scattering
experiments do not confirm that the electron radius has this value, rather  
the charge radius of the electron has been found to be on the order of 
10$^{-16}$\,cm, instead of 10$^{-13}$\,cm.

   In Section 11 we have explained the mass of the electron with a cubic
lattice consisting to one half of electric oscillations and to the other 
half of electron neutrinos, Eqs.(77) and (83). 
The position of the charge elements of our model of the electron is shown 
in Fig.\,8, they are separated by the distance 2\emph{a} = 
2\,$\cdot\,10^{-16}$\,cm. 
With N/4 charge elements the volume of the cubic charge distribution 
is N/4\,$\cdot$\,(2\emph{a})$^3$ = 5.708\,$\cdot$\,10$^{-39}$\,cm$^3$, 
from which follows that the sidelength of the cubic charge distribution 
of the free electron is
\begin{equation} \mathrm{d(e)}_{cu} = 1.787\cdot10^{-13}\,\mathrm{cm}\,.
\end{equation}
The volume of the cubic charge distribution of a free electron with sidelength
d(e)$_{cu}$ corresponds to a charged sphere with the radius 
\begin{equation} \mathrm{r(e)}_{cu} = 1.10856\cdot10^{-13}\,\mathrm{cm}\,.
\end{equation}
For a comparison, the measured charge radius of the 
$\pi^\pm$\,mesons (Eq.16) is, with r$_\pi$ = r$_p$,
\begin{center} r$(\pi^\pm)$ = 0.88\,$\cdot$\,10$^{-13}$\,cm,
\end{center}
and the sidelength of the cubic lattice of the $\pi^\pm$\,mesons is
 
\begin{equation} \mathrm{d}(\pi^\pm) = \sqrt[3]{N}\cdot\mathrm{a} = 
1.4185\cdot10^{-13}\,\mathrm{cm} = 0.5034\cdot\mathrm{r(e)}_{c\,l}.
\end{equation}
   
   The apparent radius of a free cubic charge distribution, 
Eq.(131), is by 25\% larger than the radius of the 
$\pi^\pm$\,mesons. The dimension of the charge e$^\pm$ 
\emph{in} the $\pi^\pm$\,mesons must be equal to the measured 
charge radius of $\pi^\pm$. The measured charge radius r($\pi^\pm$)
is based solely on the 
interaction of electrons with $\pi^\pm$\,mesons, the non-interacting 
neutrinos in $\pi^\pm$ do not contribute to the scattering. The apparent 
radius of a free cubic electron, Eq.(131), is larger than the radius 
r($\pi^\pm$) of the same charge in the neutrino lattice of a $\pi^\pm$\,meson. 
The charge elements in the interior of the $\pi^\pm$\,mesons 
are closer together than in the free electron. When the charge e$^\pm$
is introduced into the neutral neutrino lattice of the $\pi^\pm$\,mesons 
the electric charge is compressed. That means that a binding 
energy must be involved, when an electron is added to the neutrino 
lattice of a $\pi^\pm$\,meson.  

   Comparing the classical electron radius Eq.(129) to the effective radius 
of a free cubic electron, Eq.(131), we find that 
\begin{equation} \mathrm{r(e)}_{c\,l} = 2.542\,\mathrm{r(e})_{cu} \,.
\end{equation}
Both radii differ, because the formula for the classical electron radius 
is based on the assumption, that the electron has a spherical charge 
distribution, whereas we deal with a cubic charge distribution.
The apparent contradiction between our theoretical effective charge  
radius of the cubic charge distribution of a free electron in Eq.(131), and 
the experimentally measured charge radius of the electron, which is on 
the order of $10^{-16}$\,cm [65], is a consequence of scattering 
formulas which assume that the electron is a point particle, not
a rigid cubic charge distribution of a finite size.

\vspace{1.5cm}   

\section*{Appendix C}

\begin{center}

\large{The magnetic moment of the muon}

\normalsize

\end{center}

\noindent
   The explanation of the magnetic moment of the electron in Section\,12
has to pass a critical test, namely it has to be shown that the same
considerations lead to a correct explanation of the magnetic moment of 
the muon $\mu_\mu$ = e$\hbar$/2m($\mu^\pm$)c,  which is 
about 1/200th of the magnetic  moment of the electron, but is known 
with nearly the same accuracy as $\mu_e$.  Both magnetic moments 
are related through the equation
\begin{equation} \frac{\mu_\mu}{\mu_e} = 
\frac{\mathrm{m(e}^\pm)}{\mathrm{m}(\mu^\pm)}
= \frac{1}{206.768}\,, \end{equation}
as follows from Eq.(96) applied to the electron and muon.
This equation agrees with the experimental results to the sixth decimal.
The muon has, as the electron, an anomalous magnetic moment which 
is too small to be considered here.  

   As shown in Section 9 the muons consist of a lattice of 
N/4 muon neutrinos $\nu_\mu$, respectively anti-muon 
neutrinos $\bar{\nu}_\mu$, of N/4 electron neutrinos and 
the same number of anti-electron neutrinos plus a charge
e$^\pm$. For the explanation of the magnetic moment of the muon we 
follow the same reasoning we have used for the explanation of the 
magnetic moment of the electron. We say that m($\mu^\pm)$  consists 
of two parts, one part which causes the magnetic moment and another 
part which does not contribute to the magnetic moment.  The part
of m($\mu^\pm$) which causes the magnetic moment must contain the 
charge, or circular electric oscillations, without which there would 
be no magnetic moment. It becomes immediately clear from the small
mass of the electron neutrinos, and from Eq.(79) for the energy of the 
electric oscillations in a free electron, that $\Sigma$\,m($\nu_e$) and 
E$_\nu$(e$^\pm$) are too small, as compared to the energy in the 
rest  masses of all neutrinos in the muons, to make up m($\mu^\pm)$/2.
If, however, the electric charge elements in the muon lattice bind to the 
muon neutrinos instead to the electron neutrinos, as in the case of the electron, 
then one obtains m($\mu^\pm$)/2 from the sum of the oscillation energy 
of the muon neutrinos, (E$_\nu(\mu^\pm$)/4), plus the sum of the 
energy in the masses of the muon neutrinos, plus the energy 
E$_\nu$(e$^\pm$) in the electric oscillations, and the charged part 
of the muon is
\vspace{0.2cm}
\begin{eqnarray}
\lefteqn {1/4\cdot\mathrm{E}_\nu(\mu^\pm) \,+\, 
\sum_i\,{\mathrm{m(\nu_\mu)c}^2} + 
\mathrm{E}_\nu(\mathrm{e}^\pm)  }\nonumber \\  
&  =  & 1/4\cdot\mathrm{E}_\nu(\mu^\pm) + 
 \mathrm{N}/4\cdot \mathrm{m}(\nu_\mu)\mathrm{c}^2 + 
\mathrm{m(e^\pm})\mathrm{c}^2/2 \nonumber\\   
  & = & 53.3125\,\mathrm{MeV} 
  =  0.50457\,\mathrm{m}(\mu^\pm)\mathrm{c}^2\,. 
\end{eqnarray}

\noindent
We have now used E$_\nu(\pi^\pm)$ = E$_\nu(\mu^\pm)$ according 
to Eq.(63), with E$_\nu(\pi^\pm$) = m($\pi^\pm)$c$^2$/2 = 
69.7851\,MeV  and also E$_\nu$(e$^\pm$) = m(e$^\pm$)c$^2$/2 from 
Eq.(84), as well as  m($\nu_\mu$)c$^2$ = 49.91\,milli-eV, (Eq.70), 
m($\mu^\pm$)c$^2$ = 105.6583\,MeV and N = 2.854\,$\cdot\,10^9$. 
Eq.(135) says that the part of m($\mu^\pm$) which carries the 
charge and causes the magnetic moment is $\approx$ 
m($\mu^\pm$)/2, provided that the charge elements \emph{bind to the
muon neutrinos} instead of the electron neutrinos.

  The remaining part of 
m($\mu^\pm$) which does not carry charge and does not contribute 
to the magnetic moment is given by
\medskip  
\begin{eqnarray}
\lefteqn{ 3/4\cdot\mathrm{E}_\nu(\mu^\pm) \,+\, 
\sum_i\,(\mathrm{m}(\nu_e) + \mathrm{m}(\bar{\nu}_e))\mathrm{c}^2 + 
\mathrm{m(e^\pm)}\mathrm{c}^2/2}\nonumber \\ 
&  =  & 3/4\cdot\mathrm{E}_\nu(\mu^\pm) 
+ 3/4\cdot\mathrm{N}\mathrm{m}(\nu_e)\mathrm{c}^2 \nonumber \\ 
 & = & 53.1201\,\mathrm{MeV} 
  =  0.50275\,\mathrm{m}(\mu^\pm)\mathrm{c}^2\,. 
\end{eqnarray}
The additional m($\mathrm{e}^\pm)$c$^2$/2 on the top line of Eq.(136) 
originates from the  energy in the mass of the N/4 electron neutrinos which 
make up the neutral part of the electron. For m($\nu_e)$  = m$(\bar{\nu}_e)$ 
we use the value 0.365\,milli-eV/c$^2$ as in Eq.(67). The sum of the 
mass of the charged part of m($\mu^\pm$) plus the neutral part of 
m($\mu^\pm$) is 1.0073\,m($\mu^\pm$). It is important to
note that Eqs.(135,136) depend critically on the validity of 
E$_\nu(\mu^\pm$) = E$_\nu(\pi^\pm$), from Eq.(63).  

   If the charge elements bind to the muon neutrinos in the muon
lattice, and if the charged part of the muon makes up 1/2 of the mass of 
the muon as in Eq.(135), then it follows from Eq.(96) that the magnetic 
moment of the muon is given by 
\begin{equation} \vec{\mu}_\mu = \frac{\mathrm{e}\hbar}
{2\mathrm{m}(\mu^\pm)/2\cdot\mathrm{c}}\cdot\vec{s}\,.  \end{equation}
With s = 1/2 we have 
\begin{equation}\mu_\mu = \mathrm{e}\hbar/
2\mathrm{m}(\mu^\pm)\mathrm{c}\,, \end{equation}
without the artificial g-factor.

We have thus shown that we can explain the magnetic moment of the 
muon with the same concept that we have applied to the explanation of  
the magnetic moment of the electron, namely that 1/2 of the mass of the 
electron does not contribute to the magnetic moment, because
this half of the mass does not carry charge. In the case of the muon the 
same is true, provided that the charge elements bind to the muon 
neutrinos in the muon lattice.

\end{document}